%% file: surfactant_v12.tex
\documentclass[11pt]{article}
\usepackage{verbatim,amsmath,amssymb}
\usepackage{epsfig,float,color}
\usepackage{geometry}
\usepackage{setspace}
\usepackage[authoryear,round]{natbib} 
\usepackage{xparse}                   
\usepackage{amsmath,amssymb,amsthm,mathrsfs,amsfonts,dsfont} 
\usepackage{longtable}
\usepackage{hyperref}
\usepackage{multirow}
\usepackage{tabularx}
\usepackage{titlesec}
\usepackage{scalerel}
\usepackage{tabularx}
\usepackage{textcomp}
\usepackage[shortlabels]{enumitem}
\usepackage[section]{placeins}		

\titleformat{\paragraph}{\normalfont\normalsize\itshape}{\theparagraph}{1em}{}
\titlespacing*{\paragraph}{\parindent}{3.25ex plus 1ex minus .2ex}{.75ex plus .1ex}

\definecolor{darkblue}{rgb}{0,0,1}
\hypersetup{colorlinks=true, breaklinks=true, linkcolor=darkblue, citecolor=darkblue, menucolor=darkblue, urlcolor=darkblue}

\newcommand{\tred}[1]{\textcolor{black}{#1}}

\input{neco_updated.tex}
\newcommand{\tss}{\textsuperscript}
\newcommand{\mdt}{\dif}
\DeclareMathOperator*{\ass}{\scalerel*{{\mathsf{A}}}{\sum}}

\ExplSyntaxOn
\seq_new:N \l_mariose_cites_seq
\seq_new:N \l_mariose_citecommands_seq
\cs_new_protected:Npn \mariose_citelist:nnn #1 #2 #3
 {
  \seq_set_split:Nnn \l_mariose_cites_seq { , } { #3 }
  \seq_clear:N \l_mariose_citecommands_seq
  \seq_map_inline:Nn \l_mariose_cites_seq
   {
    \seq_put_right:Nn \l_mariose_citecommands_seq { #2 { ##1 } }
   }
  \seq_use:Nnnn \l_mariose_citecommands_seq {~and~} {#1~} {~and~}
 }
\NewDocumentCommand{\citetlist}{m}
 {
  \mariose_citelist:nnn { , } { \citet } { #1 }
 }
\NewDocumentCommand{\citeplist}{O{}O{}m}
 {
  (#1 \mariose_citelist:nnn { ; } { \citeboth } { #3 } #2)
 }
\ExplSyntaxOff

\NewDocumentCommand{\citeboth}{m}{\citeauthor{#1},~\citeyear{#1}}

\begin{document}

\theoremstyle{definition}
\newtheorem{thm}{Theorem}[section]
\newtheorem{defn}[thm]{Definition}
\newtheorem{lem}[thm]{Lemma}
\newtheorem{cor}[thm]{Corollary}
\newtheorem{prop}[thm]{Proposition}
\newtheorem{rem}[thm]{Remark}
\newtheorem{note}[thm]{Note}

\begin{center}
\Large{\bf{A finite membrane element formulation for surfactants}}\\

\end{center}

\begin{center}
\large{Farshad Roohbakhshan\footnote{corresponding author, email: roohbakhshan@aices.rwth-aachen.de \\
Present address: Federal Institute for Materials Research and Testing (BAM), Unter den Eichen 87, 12205 Berlin, Germany}
and Roger A. Sauer}\\
\vspace{4mm}

\small{\textit{Aachen Institute for Advanced Study in Computational Engineering Science (AICES), RWTH Aachen
University, Templergraben 55, 52056 Aachen, Germany}}

\vspace{4mm}

Published\footnote{This pdf is the personal version of an article whose final publication is available at \href{https://doi.org/10.1016/j.colsurfa.2018.11.022}{www.sciencedirect.com}} 
in \textit{Colloids and Surfaces A: Physicochemical and Engineering Aspects}, \\
\href{https://doi.org/10.1016/j.colsurfa.2018.11.022}{DOI: 10.1016/j.colsurfa.2018.11.022} \\
Submitted on 16. August 2018, Revised on 4. November 2018, Accepted on 10. November 2018, Available online on 22. November 2018.
\end{center}

\rule{\linewidth}{.15mm}
{\bf Abstract:}
Surfactants play an important role in various physiological and biomechanical applications. An example is the respiratory system, where pulmonary surfactants facilitate the breathing and reduce the possibility of airway blocking by lowering the surface tension when the lung volume \tred{decreases} during exhalation. This function is due to the dynamic surface tension of pulmonary surfactants, which depends on the concentration of surfactants spread on the liquid layer lining the interior surface of the airways and alveoli. Here, a finite membrane element formulation for liquids is introduced that allows for the dynamics of concentration-dependent surface tension, as is the particular case for pulmonary surfactants. A straightforward approach is suggested to model the contact line between liquid drops/menisci and planar solid substrates, which allows the presented framework to be easily used for drop shape analysis. It is further shown how line tension can be taken into account. Following an isogeometric approach, NURBS-based finite elements are used for the discretization of the membrane surface. The capabilities of the presented computational model is demonstrated by different numerical examples -- such as the simulation of liquid films, constrained and unconstrained sessile drops, pendant drops and liquid bridges -- and the results are compared with experimental data. 


\vspace{-4mm}
\rule{\linewidth}{.15mm}
\section{Introduction}\label{s:intro}
Surface tension, which is due to the attraction of molecules at a fluid interface by the bulk of the fluid, leads to a minimization of the fluid surface area. However, if molecules of a surface active agent, usually called \emph{surfactant}, are added to the fluid interface, the surface tension decreases. Surfactants, which are amphiphilic compounds, have a hydrophobic tail, which is oil-soluble and allow them to stay on the surface, and a hydrophilic head, which is water-soluble and is able to reduce the surface tension by disrupting hydrogen bonds. This function is particularly crucial for the lung biomechanics as pulmonary surfactants facilitate the breathing and reduce the possibility of airways being blocked by lowering the surface tension as the lung volume decreases \citep{nkadi09}. By adding lung surfactants to pure water with a surface tension of about 70 mN/m, the surface tension can decrease to values lower than 2 mN/m \citeplist{goerke98,possmayer01}. Pulmonary surfactants are composed of roughly 80\% phospholipids (PL), 5-10\% neutral lipids (NL), mainly cholesterol, and 8-10\% surfactant proteins (SP) \citep{goerke98}. Around half of the PL content is dipalmitoylphosphatidylcholine (DPPC), which is the strongest surfactant compound in the mixture of the pulmonary surfactants and has the largest contribution to the reduction of surface tension. Although a solution of pure DPPC can reduce the surface tension to \tred{nearly} zero during monolayer compression, it does not respread properly during expansion and adsorbs very slowly to the interface \citeplist{hildebran79,ingenito99}; therefore, the DPPC molecules are lost during cyclic compression and expansion of the liquid-air interface. In fact, other components of the pulmonary surfactants also contribute to the dynamic characteristics of lung surfactants. Surfactant proteins, particularly \tred{SP-A}, SP-B and SB-C, among other functionalities, improve the adsorption rate of DPPC molecules during the expansion of the interface \citeplist{goerke98,ingenito99,veldhuizen00} and prevent the collapse of the \tred{surfactant film}. There are many references on the link between deficiencies in the components of the pulmonary surfactants and respiratory distress syndrome (RDS) in adults and infants \citeplist[e.g~]{gregory91,lewis93,hallman01,ma12}. The pulmonary surfactants have crucial contributions to the physiology and biomechanics of the respiratory system: They accelerate the oxygen diffusion through the air-water interface \citep{olmeda10}; they enhance lung immunity and its resistance against pulmonary infections \citep{han15} and foremost they play the key role in lung stability by increasing the stability of alveolar surfaces \citep{bachofen01}. 

Computational models of the respiratory system can provide new insights to these mechanisms and offer a path to developing better treatment of lung diseases such as RDS. Therefore, it is necessary to develop a computational framework to study the dynamics of pulmonary surfactants. With this motivation in mind, the objective of this paper is to present a new finite membrane element formulation that allows for concentration-dependent surface tension like in pulmonary surfactants. Although the constitutive models presented here are developed for pulmonary surfactants, other fluids with dynamic surface tension can be modeled similarly. 

Compared to earlier works, this paper presents a dynamic finite membrane element formulation for liquids that:
\vspace{-\topsep}
\bnumr[1)]
\setlength{\itemsep}{0pt}
\setlength{\parskip}{0pt}
\setlength{\parsep}{0pt}
\item allows for different concentration-dependent constitutive laws for dynamic surface tension of the liquid-gas interface; 
\item can be coupled with structural finite membrane, shell and solid elements; 
\item can model thin films, bubbles, liquid droplets and menisci;
\item is able to treat different external forces (like gravity and pressure), contact constraints and various boundary conditions like contact lines with specified contact angles;
\item can include the effect of line tension if needed.  
\enumr
\vspace{-\topsep}
Furthermore, in \citet{roohbakhshan18thesis}, the presented formulation is combined with the computational biological membrane model of \citet{biomembrane} in order to model surfactant-lined alveolar tissue.

The rest of this paper is organized as follows: In Sec.~\ref{s:dyn}, the existing empirical, theoretical and computational approaches towards describing the dynamic surface tension are briefly reviewed. Sec.~\ref{s:memtheo} summarizes the liquid membrane theory developed by \citetlist{membrane,droplet}, which includes the treatment of contact lines for quasi static droplets and menisci on the rigid planar substrate, and extends it by dynamic surface tension and contact line tension. In Sec.~\ref{s:models}, the two existing dynamic surface tension models, proposed by \citet{otis94} and \citet{saad10}, are adapted to general membranes. These models allow for dynamic changes of surface tension due to the variation of surfactant concentration. Sec.~\ref{s:FE} is devoted to the finite element solution, where the time discretization and integration of the dynamic model is discussed. Sec.~\ref{s:example} presents numerical examples, which show the performance and robustness of the formulation. Sec.~\ref{s:conc} concludes the paper.

\section{Dynamic surface tension}\label{s:dyn}
In this section, first a concise history of empirical methods for dynamic measurements of surface tension is presented. Then, the existing theories and computational models are briefly reviewed.

\subsection{Experimental methods}\label{s:dyn_exp}
In general, the existing techniques for dynamic measurement of surface tension can be grouped into three main categories: 1) Force method, 2) drop shape and pressure method and 3) flow method \citep{franses96}. In the force method, a solid is inserted in the liquid-gas interface, which creates a meniscus confined by a boundary with a specific contact angle on the solid surface. By balancing the forces at the contact line, the surface tension of the liquid-gas interface can be found. The typical examples for the force method are the \textit{Wilhelmy plate method} and the \textit{du No{\"u}y ring method} \citep{butt06}. In the shape and pressure method, the well known Young--Laplace equation, 
\eqb{l}
\Delta p = 2\,H\,\gamma~,
\label{e:YL}\eqe
which relates the pressure jump across the interface $\Delta p$ to the interface surface tension $\gamma$ and the interface mean curvature $H$, is inversely solved based on the captured shape of an interface. The contact angle is included as a boundary condition for Eq.~\eqref{e:YL}. If gravity is not neglected, $\Delta p$ and accordingly $H$ vary across the interface. The Bond number,
\eqb{l}
\mathrm{Bo} = \ds\frac{\Delta\rho\,g\,L^2}{\gamma}~,
\label{e:Bond}
\eqe
which defines the ratio of gravitational to interfacial forces, is usually used to evaluate the importance of gravity. Here $\Delta\rho$ is the density difference of the media on the two sides of the interface, $g$ is the gravitational constant and $L$ is a characteristic length, which can be for example the radius of a drop. In order to ignore gravity, $Bo \ll 1$.
In general, the drop shape method can be further grouped into \citep{saad16}
\vspace{-\topsep}
\bnumr[a)]
\setlength{\itemsep}{0pt}
\setlength{\parskip}{0pt}
\setlength{\parsep}{0pt}
\item Volume-radius limited, e.g.~\textit{pendant drop} (PD) and \textit{constrained sessile drop} (CSD) test;
\item Volume-angle limited, e.g.~\textit{sessile drop} (SD) and \textit{captive bubble} (CB) test;    
\item Volume-radius-radius limited, e.g.~\textit{2-edges-constrained liquid bridge}\footnote{Here, the terms liquid bridge and meniscus are used interchangeably.} (CLB) test, and 
\item Volume-radius-angle limited, e.g.~\textit{1-edge-constrained liquid bridge} (LB) test.
\enumr
\vspace{-\topsep}
In the flow method, the interface is controlled in a way that the shape analysis in not required to find the interface profile and numerical methods are not needed for solving the Young--Laplace equation \eqref{e:YL}. The well know examples for the flow method are the \textit{maximum bubble pressure method} (MBPM), \textit{growing drop method} and \textit{oscillating jet method}. For instance, in the MBPM and \textit{growing drop method}, the interface is restricted to be sections of a sphere with radius $R$, so that $H=2/R$ and Eq.~\eqref{e:YL} can be solved analytically as $\gamma = \Delta p\,R/2$. To keep the interface close to a spherical shape, the gravity effect should be negligible, which restricts these methods to the cases where the Bond number is small. As it is shown in Sec.~\ref{s:example}, the presented computational framework can simulate many of the existing empirical methods that are commonly used to measure dynamic surface tension.

Dynamic measurement of the surface tension of pulmonary surfactants goes back to the early work of \citet{clements57}, who used the \textit{Langmuir--Wilhelmy balance} for this purpose. Since then, the method \tred{is regularly used to express the isothermal relationship between surface tension and surface area} of lung extracts \citeplist[e.g.~]{hills85,sosnowski17}. Although this tool is very suitable to spread a well-defined monolayer of lipids and proteins on the air-liquid interface and the surface area can be varied simply by moving a barrier on the interface, it is susceptible to liquid leakage and it is slow compared to the dynamics of the respiratory system \citep{veldhuizen00}. Being a successful alternative of the \textit{Langmuir--Wilhelmy balance}, the \textit{pulsating bubble surfactometer} (PBS) works based on the formation of a gas bubble inside a chamber of liquid. The surface tension is found by solving the Young--Laplace equation assuming that the bubble is quasi spherical. Although it has been prevalently used since its introduction by \citet{enhorning77}, it has technical problems like liquid leakage and inaccuracy at low surface tensions \citep{veldhuizen00}, which are improved in a modified version, namely the \textit{captive bubble surfactometer} introduced by \citet{schurch89}. Recently, drop shape methods have become more popular for the dynamic measurement of pulmonary surfactants. As these methods are based on the inverse solution of the Young--Laplace equation \eqref{e:YL} by means of numerical techniques, they are more flexible and can include external forces like gravity and different boundary conditions like specified contact angles \citep{saad16}. Usually, the experiments are set up so that the drops are axially symmetric; thus, \textit{axisymmetric drop shape analysis} (ADSA) techniques are used, which either use regression methods to fit the experimental drop profile to a Laplacian curve \citep[e.g.~][]{rotenberg83} or solve an axisymmetric 2D variation of the Young--Laplace equation \eqref{e:YL}, which depends only on the arc length along the interface, through an iterative optimization procedure \citeplist[e.g.~]{saad11thesis,saad16}. The method is combined with the \textit{constrained sessile drop} (CSD) test \citep{saad10,saad12} and it is also used for the \textit{pendant drop} (PD) test by circular approximation of the droplet profile \citep{saad11design}. \tred{Recently, \citet{bangyozova17} employ the \textit{Brewster angle microscopy} (BAM) and ADSA technique for the assessment of healthy and diseased pulmonary surfactants. For thin films, alternative approaches like the \textit{pressure balance technique} can be used to directly measure the surface forces in terms of the disjoining pressure \citep[e.g.][]{todorov17}.}          

\subsection{Theoretical models}\label{s:dyn_theo}
Since the 1970s, several theoretical models have been introduced to explain the diffusion-adsorption process \citeplist{horn75,otis94,moris01,krueger00,saad10}, through which the molecules of surfactants are transferred from the bulk to the interface. All these models assume that the surface tension is determined only by the amount of surfactants at the interface. Surfactants are transfered to the interface in two phases: First, surfactants are conveyed from the liquid bulk to the layers adjacent to the interface through a diffusion process. Second, the surfactants are adsorbed from the layers to the interface. As diffusion and adsorption occur at different rates, most of the existing models consider either diffusion or adsorption as the key mechanism.
Thus, in general, the existing theoretical models can be grouped into \textit{diffusion-controlled} \citep[e.g.][]{loglio91} and \textit{adsorption/desorption-controlled} models \citeplist[e.g.~]{otis94,moris01,krueger00,saad10}. In the former approach, it is assumed that the diffusion of surfactant molecules through the bulk is the rate limiting process while, in the latter, the adsorption and desorption are assumed to be more dominant. In Sec.~\ref{s:models}, the \textit{adsorption/desorption-controlled} models of \citetlist{otis94,saad10} are introduced in detail and it is shown how they can be incorporated into the presented dynamic finite element formulation. 

\subsection{Computational models}\label{s:dyn_comp}
Although dynamic surface tension of pulmonary surfactants and lung extracts has been the subject of many empirical and theoretical studies in the last decades, there are few computational models that address the concentration-dependent surface tension of the pulmonary surfactants. In earlier approaches, the curves of surface tension vs.~surface area at constant temperature, obtained by experiments, are used to define the surface tension as a function of surface area. For instance, \citet{karakaplan80} propose an alveolar model discretized by triangular planar finite elements that combines the elastic properties of the alveolar wall and the surface tension of the pulmonary surfactants. 
They assume that the behavior of the surfactants is defined by two experimentally-fitted functions, one for inflation and another for deflation, that depend on the changes of local area. In the same fashion, \citet{kowe86} use an exponential form to add the triangular planar surface tension elements to the earlier model of \citet{dale80} who use elastic pin-joined bar elements to build a 3D alveolus model approximated by a truncated octahedron consisting of fibers only. To distinguish between inflation and deflation, two similar exponential functions with different parameters to be fitted are used. The single alveolus model of \citet{kowe86} is extended by \citet{denny95,denny97} to an alveolar model constructed by an assembly of 36 truncated octahedra. \citet{kojic06,kojic09} introduce a scaling approach to consider the hysteresis effect of surface tension in their proposed material model of biological membranes. Similar to other approaches, they also find the surface tension through a quasi static function derived from a single surface tension vs.~surface area hysteresis curve inferred from an experiment. However, as the experiments are restricted to a fixed range of area changes, they introduce a scaling method to be able to stretch their biological membranes to ranges that are different from a particular experiment. The computational models briefly introduced above are simple to implement but they are all time-independent; insensitive to the rate of changes in surface area; quasi-static rather dynamic and, excluding the models of \citet{kojic06,kojic09}, they are restricted to the surface area range dictated by the experiment, from which the parameters are identified. In contrast to these time-independent approaches, there are time-dependent computational models that explicitly consider the exchange of surfactants between the liquid bulk and the liquid-gas interface. The very first model of this kind is the mathematical model of \citet{archie73} that assumes a linear adsorption/desorption between the interface and the bulk. Accordingly, Archie assumes that the surfactant concentration is inversely proportional to the surface area and the surface tension is inversely proportional to the concentration. \citet{denny00,denny06} modify the earlier work of \citet{denny95,denny97} by using the dynamic surface tension model of \citet{otis94} instead of the quasi-static exponential model of surface tension. \citetlist{wiechert09,wiechert11thesis} also use the dynamic model of \citet{otis94} to add the interfacial energy of the surfactant layer to the surface of the alveolar wall modeled by finite solid elements.
\section{Liquid membrane theory}\label{s:memtheo}
The liquid membrane formulation considered here is following the finite element formulation of \citetlist{membrane,droplet}. This section briefly reviews the existing formulation and extends it to account for dynamic surface tension and constant line tension.

\subsection{Membrane kinematics}\label{s:kin}
In general, the membrane surface $\sS$ can be characterized by a mapping from a parametric domain $\sP$ with coordinates $\xi^1$ and $\xi^2\in[-1,~1]$ as
\eqb{l}
\bx = \bx\big(\xi^1,\xi^2,t\big)~,
\eqe
where $t\in[0,~T]$ denotes time. From this, follow the covariant tangent vectors to $\sS$,
\eqb{l}
\ba_\alpha = \ds\pa{\bx}{\xi^\alpha}~,\quad\alpha=1,2 ~.
\eqe
The associated contra-variant tangent vectors $\ba^\alpha = a^{\alpha\beta}\,\ba_\beta$ are obtained from the metric $[a^{\alpha\beta}] = [a_{\alpha\beta}]^{-1}$ and $a_{\alpha\beta}=\ba_\alpha\cdot\ba_\beta$. Accordingly, an infinitesimal surface area in the deformed configuration is related to the parametric domain as $\dif a = J_a\,\dif\xi^1\,\dif\xi^2$ with $J_a=\sqrt{\det a_{\alpha\beta}}$ and similarly $\dif A = J_A\,\dif\xi^1\,\dif\xi^2$ with $J_A=\sqrt{\det A_{\alpha\beta}}$ in the reference configuration. Therefore, the surface stretch between the undeformed surface $\sS_0$ and the deformed surface $\sS$ is
\eqb{l}
J := \dfrac{\dif a}{\dif A} = \dfrac{J_a}{J_A} ~.
\label{e:J}\eqe 
Further, the material time derivative of the surface stretch is \citeplist{sauer18CISM}
\eqb{l}
\dot{J} := \dfrac{\mdt J}{\mdt t} := J\,\divz_{\mkern-4.5mu\mrs}\,\bv = J\,\ba^\alpha\cdot\dot\ba_\alpha
= \tred{\dfrac{J}{2}\,a^{\alpha\beta}\,\dot{a}_{\alpha\beta} = -\dfrac{J}{2}\,\dot{a}^{\alpha\beta}\,a_{\alpha\beta}}~, 
\label{e:dotJ}\eqe
where $\bv := \dot\bx$ is the material velocity. The surface divergence follows from $\divz_{\mkern-4.5mu\mrs}\,\bv = \tr\nabla_{\!\mrs}\,\bv $, where $\nabla_{\!\mrs} := \big( \ba_\alpha\otimes\ba^\alpha\big)\cdot\nabla_{\!\mrx} $ is the surface gradient operator and $\nabla_{\!\mrx}$ is the regular gradient operator in the current configuration. The surface normal is
\eqb{l}
\bn = \dfrac{\ba_1\times\ba_2}{\norm{\ba_1\times\ba_2}} 
\label{e:bn}\eqe 
and the components of the curvature tensor $\bb = b_{\alpha\beta}\,\ba^\alpha\otimes\ba^\beta$ are given by
\eqb{l}
b_{\alpha\beta}=\bn\cdot\ba_{\alpha,\beta}=-\bn_{,\beta}\cdot\ba_\alpha~,
\label{e:bab}\eqe
where $\ba_{\alpha,\beta} = \partial\ba_\alpha/\partial\xi^\beta$. The mean and Gaussian curvature of surface are, respectively, 
\eqb{l}
H := \dfrac{1}{2}\tr{\bb} = \dfrac{1}{2}\,b^\alpha_\alpha = \dfrac{1}{2}\,a^{\alpha\beta}\,b_{\alpha\beta}~,
\label{e:meanH}\eqe
and
\eqb{l}
\kappa := \det{\bb} = \dfrac{b}{a}~,
\label{e:kappa}\eqe
where
\eqb{l}
a = \det[a_{\alpha\beta}]~, \quad
b = \det[b_{\alpha\beta}]~.
\eqe

\subsection{Constitution of liquid membranes}\label{s:const}
The stress state in liquid membranes has a hydrostatic component that is governed by the surface tension $\gamma$ and a viscous component that is assumed here to follow a simple linear (i.e.~Newtonian) viscosity model \citep{sahu17}. In this case, the  surface stress tensor can be written as
\eqb{l}
\bsig = \sigma^{\alpha\beta}\,\ba_\alpha\otimes\ba_\beta~,
\eqe
where
\eqb{l}
\sigma^{\alpha\beta} = \gamma\,a^{\alpha\beta} - \tred{\eta\,\Big(\dot{a}^{\alpha\beta} + \dfrac{\dot{J}}{J}\,a^{\alpha\beta}\Big)}
\label{e:sigab1}\eqe
are the contra-variant stress components and $\eta$ is the kinematic surface viscosity. It is treated as constant here. $\gamma$ on the other hand can change with time. This is different to the formulation of \citet{membrane} that treats $\gamma$ constant. \tred{The $\dot J/J$ term in Eq.~\eqref{e:sigab1} is included to ensure that the viscous stress is purely deviatoric and does not affect the surface tension $\gamma := \tfrac{1}{2}\, \sigma^{\alpha\beta}\,a_{\alpha\beta}$. This can be seen from Eq.~\eqref{e:dotJ}.}

For many applications, such as pulmonary surfactants, surface tension shows hysteresis. Hence, to determine $\gamma(t)$ at the current time step $t$, an evolution equation is needed to find the surface tension changes over time.  This evolution law has the form of an ordinary differential equation (ODE) e.g.~$\dot\gamma = f(\gamma,J,t)$. For such a formulation, one can use the dynamic {CR} and {AR} models introduced in Secs.~\ref{s:ALM} and \ref{s:CRM}, respectively. Later in Sec.~\ref{s:time}, a time integration scheme of these models is introduced.
\tred{
\begin{rem}
The current modeling framework is based on a single deformation measure and does not consider a decomposition of the surface deformation into multiple sources as has been recently presented by \citet{sauer18multi}.
\end{rem}}

\subsection{Strong form}\label{s:sf}
According to Cauchy's formula, the membrane traction on a surface normal to $\ba^\alpha$ is 
\eqb{l}
\bt^\alpha = \bsig\,\ba^\alpha~.
\label{e:cuachy}\eqe
Furthermore, the balance of linear momentum provides us with the strong form of the equilibrium equation, which governs the membrane, as
\eqb{l}
\bt^\alpha_{;\alpha} + \bff = \mathbf{0}~,
\label{e:sf}\eqe
where the inertia effects are neglected since only problems at small length scale are considered here. Further, $\bff$ is a distributed body force, which can be decomposed into in-plane and out-of-plane components as $\bff = f^\alpha\,\ba_\alpha + p\,\bn $, where $p$ is the net pressure exerted on the membrane surface. For the case that there is fluid only on one side, e.g.~for a droplet, $p$ equals to the difference of the fluid pressure $p_\mrf$ and the external pressure $p_\mathrm{ext}$, acting on the membrane surface as
\eqb{l}
p = p_\mrf - p_\mathrm{ext}~.
\label{e:p}\eqe
The fluid pressure
\eqb{l}
p_\mrf = p_\mrv + p_\mrh 
\label{e:pf}\eqe
is composed of the hydrostatic pressure
\eqb{l}
p_\mrh = \rho\,\bg\cdot\bx ~,
\label{e:ph}\eqe
where $\rho$ is the fluid density and where $\bg = [0,~0,~\tred{-}g]^\mrT$ is the gravity vector, and the capillary pressure, $p_\mrv$, across the liquid-gas interface $\sS_\mathrm{LG}$, can be interpreted as the Lagrange multiplier for the volume constraint
\eqb{l}
g_\mrv = g_\mrv(\bx) := V_0 - V = 0~,
\label{e:gv}\eqe
in case of incompressible droplets \citeplist{membrane,droplet}. Here, $V_0$ is the prescribed volume.
For liquid droplets and menisci, where the internal fluid flow is not significant, it is computationally more efficient to model the internal behavior by the  scalar Eq.~\eqref{e:gv}, instead of discretizing the interior and solving a flow problem there.  
The solution of Eqs.~\eqref{e:sf}~and~\eqref{e:gv} is found by applying the Dirichlet and Neumann boundary conditions
\eqb{llll}
\bu \is \bar\bu & $on$~\partial_u\sS~, \\[1mm]
\bt \is \bar\bt & $on$~\partial_t\sS~,
\eqe
on the membrane boundary $\partial\sS=\partial_u\sS\,\cup\,\partial_t\sS$. An example for the traction $\bar\bt$ is the distributed contact line force described in Sec.~\ref{s:cont}.

\subsection{Weak form}\label{s:wf}
As shown by \citet{membrane}, the weak form of Eq.~\eqref{e:sf} is given by
\eqb{l}
G_{\mathrm{int}} - G_{\mathrm{ext}}=0\quad\forall\bw\in\sW~,
\label{e:wf_2}\eqe
where
\eqb{lll}
G_\mathrm{int} \dis \ds\int_{\sS}\bw_{;\alpha}\cdot\sig^{\alpha\beta}\,\ba_\beta\, \dif a ~, \\[4mm]
G_\mathrm{ext} \dis \ds\int_\sS w_\alpha\,f^\alpha\,\dif a
+ \int_\sS w\,p\,\dif a
+ \int_{\partial_t\sS} \bw\cdot\bar\bt\,\dif s ~,
\label{e:Gintext}\eqe
are the internal and external virtual work contributions, respectively. Here, $\bw\in\sW$ is a kinematically admissible variation of $\bx\in\sS$ and $\sW$ is a suitable space for $\bw = w_\alpha\,\ba^\alpha + w\,\bn$, where $w_\alpha := \bw\cdot\ba_\alpha$ and $w:=\bw\cdot\bn$. 

\begin{rem}
For the extension of Eq.~\eqref{e:wf_2} to liquid films with bending resistance, e.g.~for lipid bilayers, see \citet{liquidshell}.
\end{rem}

\subsection{Contact line}\label{s:cont}
\begin{figure}[ht]
\centering
\includegraphics[height=40mm]{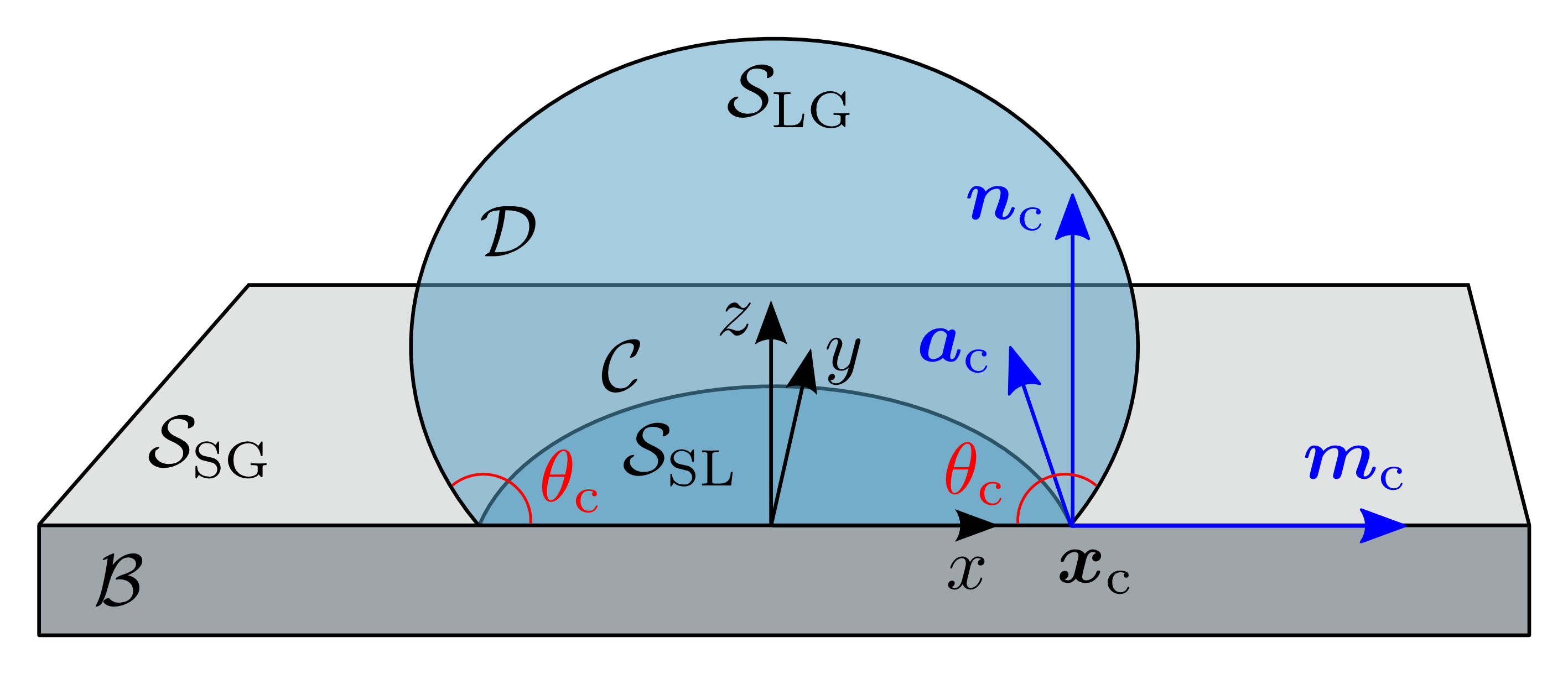}
\caption{Droplet contact model: The solid substrate, the liquid droplet and the contact line are denoted by  $\sB$, $\sD$ and $\sC$, respectively.}
\label{f:droplet}
\end{figure}

Any computational model for the simulation of droplets and liquid menisci should be able to handle the contact line. The numerical enforcement of equilibrium at the contact line in the context of nonlinear finite elements is explored in detail in the work of \citet{droplet}, which also presents a general droplet contact model. The model is extended by \citet{frictdroplet} to allow for contact angle hysteresis, using a frictional sliding algorithm, arbitrary meniscus shapes and arbitrary substrate roughness, heterogeneity and compliance. As shown in Fig.~\ref{f:droplet}, the droplet contact model consists of two bodies: The solid substrate $\sB$ and the liquid droplet $\sD$. The droplet and solid substrate have separate interfaces with the surrounding gas, denoted by $\sS_\mathrm{SG}$ and $\sS_\mathrm{LG}$, respectively and they share the $\sS_\mathrm{SL}$ interface.
The contact line $\sC $ is the common intersection of the three mentioned interfaces. Along the contact line, the droplet surface meets the substrate at a contact angle $\theta_\mrc = \theta_\mrc(\bx_\mrc)$, which can depend on the position $\bx_\mrc \in \sC$ if the substrate is heterogeneous or if contact angle hysteresis occurs. Here, we assume that the substrate is homogeneous and the contact angle is constant. For the behavior of droplets on chemically heterogeneous substrate surfaces see \citet{luginsland17}. 

\begin{figure}[ht]
\begin{center} \unitlength1cm
\begin{picture}(15.0,4.0)
\put(1.0,0.5){\includegraphics[height=40mm]{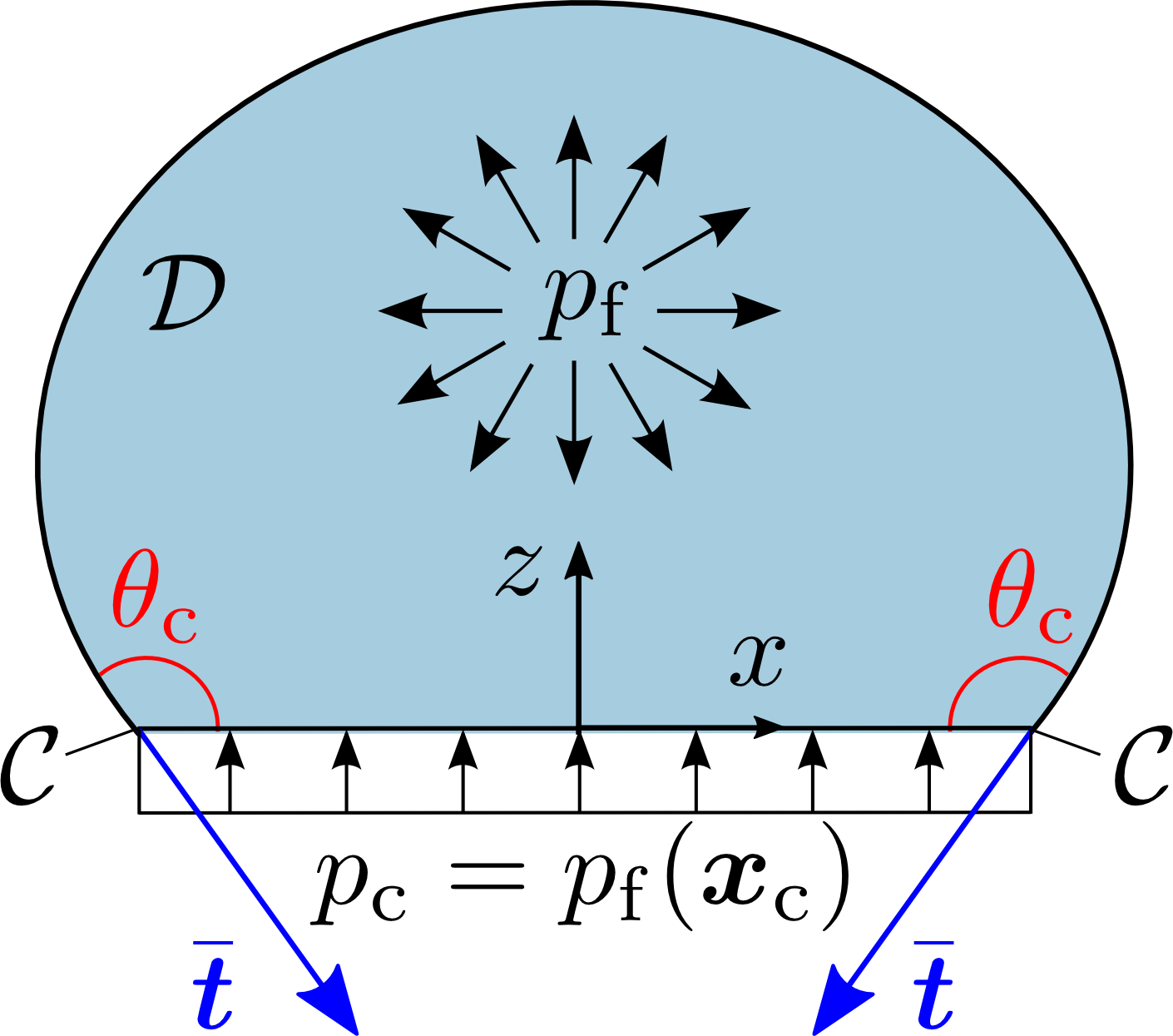}}
\put(0.0,0.5){(a)}
\put(8.0,0.0){\includegraphics[height=40mm]{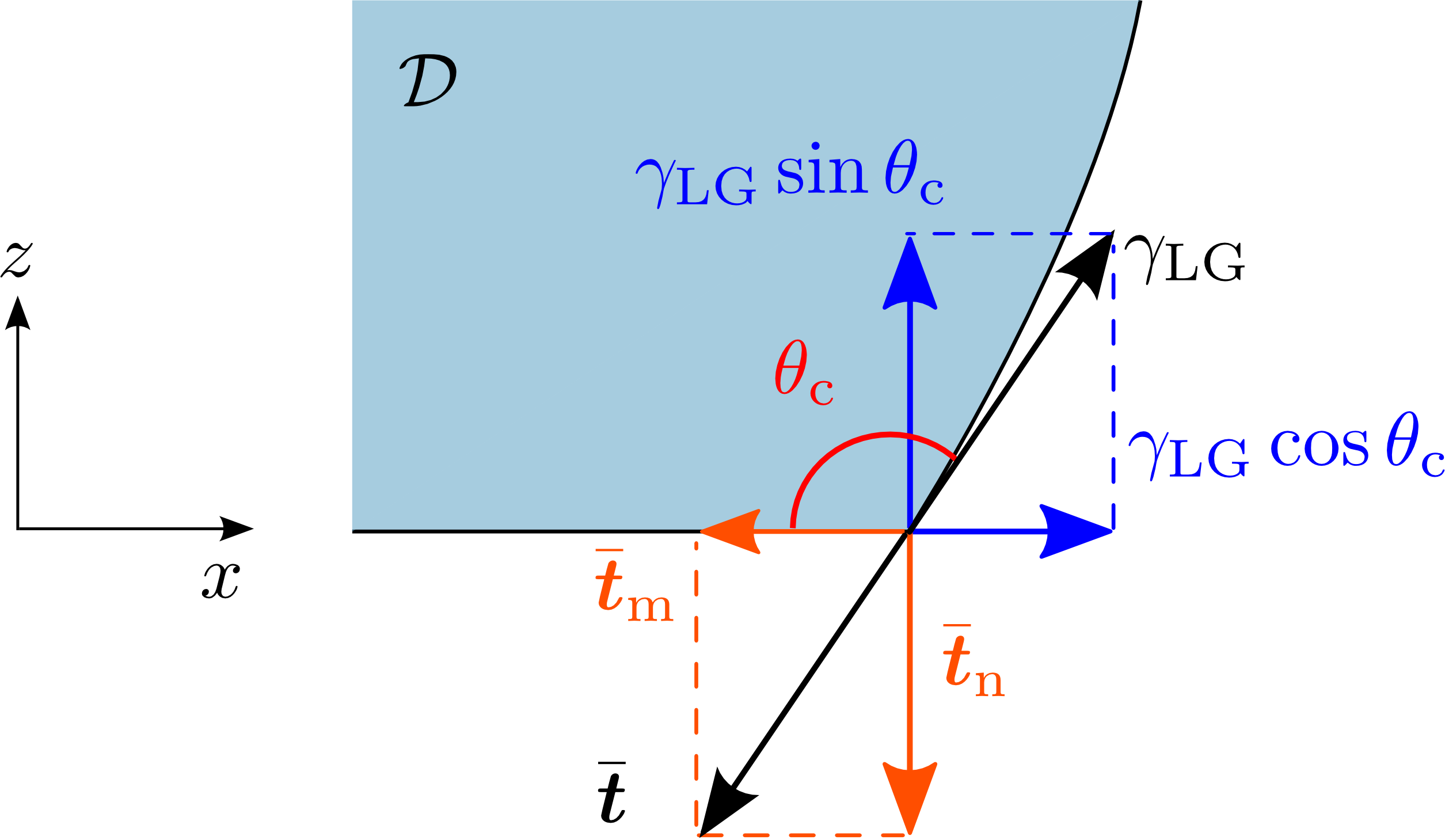}}
\put(8.0,0.5){(b)}
\end{picture}
\caption{Forces acting on the droplet: (a) Line force $\bar\bt$ and contact pressure $p_\mrc$ acting on the solid-liquid interface $\sS_\mathrm{SL}$ and (b) equilibrium along the contact line.}
\label{f:cline}
\end{center}
\end{figure}

In contrast to \citet{droplet,memtheo,frictdroplet} and \citet{sauer18mono} who consider \textit{closed} membrane models that explicitly account for the solid-liquid interface $\sS_\mathrm{SL}$, on which the contact constraints for non-penetration and sticking are enforced point-wise, here, $\sS_\mathrm{SL}$ is not included in the model as the contact conditions at this interface are assumed to be homogeneous. Without $\sS_\mathrm{SL}$ the model becomes an \textit{open} membrane model. In this case, the contact impenetrability constraint is replaced by the following choice of boundary conditions at the contact line: 1) a pure Dirichlet boundary condition, where $\bx_\mrc$ is fixed, 2) a pure Neumann boundary condition, where $\bar\bt$ is given (see Figs.~\ref{f:cline}.a) and 3) a mixed Dirichlet--Neumann boundary condition. Here, the first approach is followed for pinned droplets and menisci and the third one is adopted for droplets and menisci that have fixed contact angles. 

As shown in Fig.~\ref{f:cline}.b, the Neumann traction at the contact boundary is given by
\eqb{l}
\bar\bt = \bar{t}_\mrm\,\bm_\mrc + \bar{t}_\mrn\,\bn_\mrc ~,
\label{e:qc}\eqe
with the components
\eqb{lll}
\bar{t}_\mrm \is -\gamma_\mathrm{LG}\,\cos \theta_\mrc ~, \\
\bar{t}_\mrn \is -\gamma_\mathrm{LG}\,\sin \theta_\mrc ~. 
\label{e:yl}
\eqe
Further, $\bar{t}_\mrm = \gamma_\mathrm{SG} - \gamma_\mathrm{SG}$. 
These component are defined in the basis $\{\bm_\mrc,~\ba_\mrc~,\bn_\mrc\}$ illustrated in Fig.~\ref{f:droplet}.
The substrate normal $\bn_\mrc$ is considered fixed here (e.g.~$\bn_\mrc\,\hat{=}\,[0, 0, 1]^\mrT$); $\ba_\mrc$ is the tangent to the contact line at $\bx_\mrc$ and the outward unit normal of contact line $\sC$ is given by 
\eqb{l}
\bm_\mrc = \ds\frac{\ba_\mrc \times \bn_\mrc}{\norm{\ba_\mrc \times \bn_\mrc}}~.
\label{e:mc}\eqe

In the following, we discuss the application of mixed Dirichlet--Neumann boundary conditions. By imposing a Dirichlet boundary condition in normal direction $\bn_\mrc$, the normal component of the contact force, $\bar{t}_\mrn$, is found as the reaction force corresponding to this boundary condition. Here, two approaches are proposed to define an appropriate value for the tangential component of the contact line force, $\bar{t}_\mrm$, corresponding to a given contact angle $\theta_\mrc$.

\subsubsection{General shapes}\label{s:cont_gen} 
According to Eq.~\eqref{e:yl}, the contact line force $\bar{t}_\mrm $ depends explicitly on the surface tension of the liquid-gas interface $\gamma_\mathrm{LG}$. Even though the surface tension is not constant, i.e.~$\gamma_\mathrm{LG} = \gamma(\bx_\mrc,t)$, Eq.~\eqref{e:yl} can still be used to enforce the contact angle at $\bx_\mrc$ for droplets and menisci.

\subsubsection{Droplets}\label{s:cont_drop}
\begin{figure}[ht]
\begin{center} \unitlength1cm
\begin{picture}(15.0,5.0)
\put(0.0,0.5){\includegraphics[height=45mm]{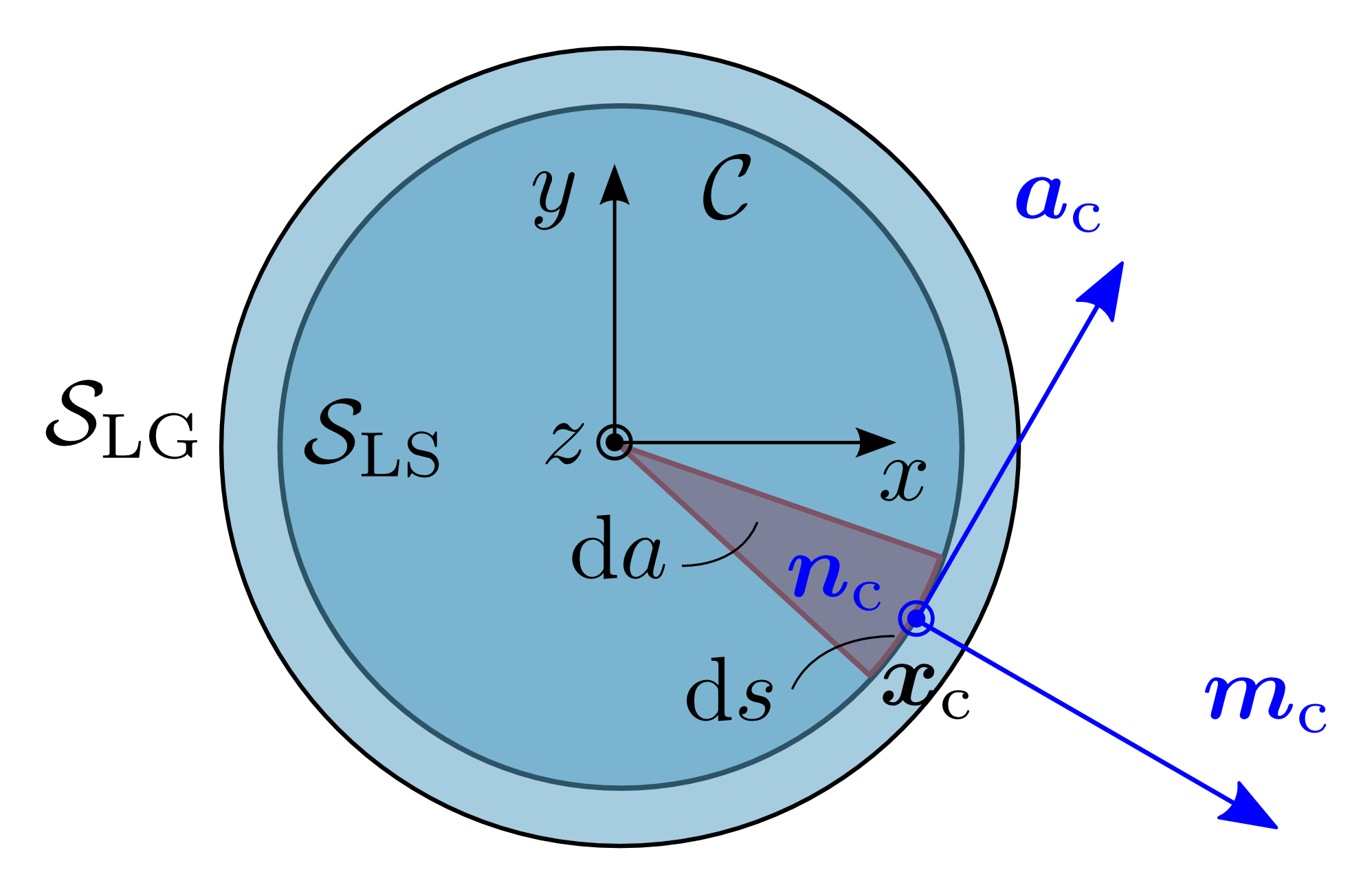}}
\put(1.0,0.5){(a)}
\put(7.0,0.0){\includegraphics[height=50mm]{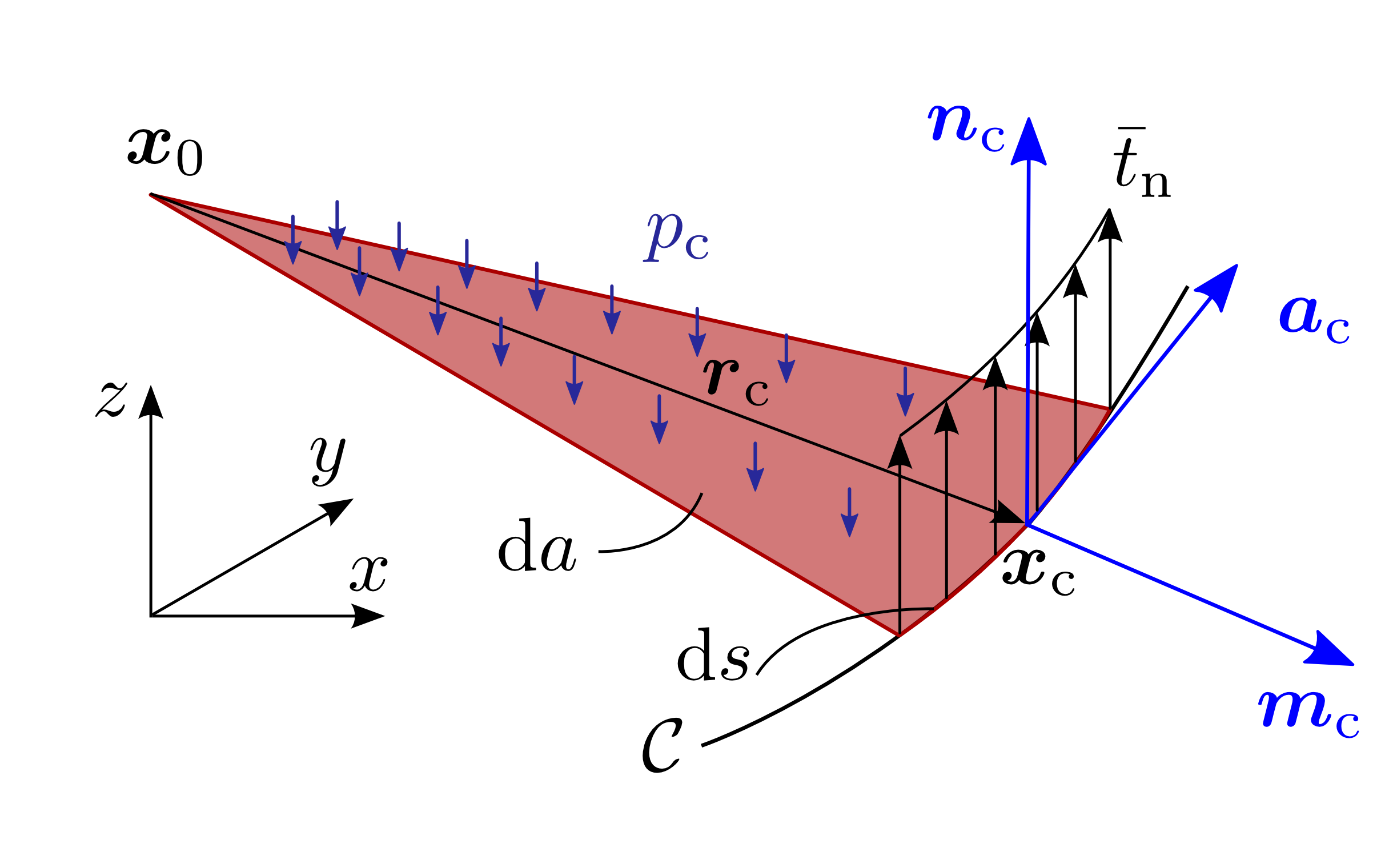}}
\put(8.0,0.5){(b)}
\end{picture}
\caption{Balance of forces acting on the substrate in the normal direction: (a) Top view of an infinitesimal area $\dif a$ on the solid-liquid interface $\sS_\mathrm{SL}$ and (b) forces acting on $\dif a$ in the normal direction.}
\label{f:balance}
\end{center}
\end{figure}

Alternatively, Eq.~\eqref{e:yl} can be reformulated as 
\eqb{l}
\bar{t}_\mrm = \bar{t}_\mrn\,\cot \theta_\mrc ~,
\label{e:qm1}\eqe
such that $\bar{t}_m$ now depends on the unknown normal force $\bar{t}_\mrn$. In the contact-based models of \citet{droplet,memtheo,frictdroplet}, $\bar{t}_\mrn$ corresponds to the normal contact force at $\bx_\mrc$ (and it is termed $q_\mrn$ instead of $\bar t_\mrn$). Here, it corresponds to the reaction force at the Dirichlet boundary, which can be determined analytically for planar rigid substrates. As shown in Fig.~\ref{f:cline}.a, if the contact surface is horizontal and planar, the contact pressure $p_\mrc$ is equal to the internal pressure of the fluid \eqref{e:pf}
\eqb{l}
p_\mrf(\bx_\mrc) = p_\mrv + \rho\,\bg\cdot\bx_\mrc ~.
\eqe
As schematically shown in Fig.~\ref{f:balance}, considering an infinitesimal surface area $\dif a = \dfrac{1}{2}\,\norm{\br_\mrc}\,\dif s$ on the substrate, $\bar{t}_\mrn$ can be found by balancing the forces in the normal direction $\bn_\mrc$ as
\eqb{l}
\Big(\ds\frac{1}{2}\,p_\mrc\,\norm{\br_\mrc} - \bar{t}_\mrn \Big) \dif s = 0~,
\label{e:pcqn}\eqe
which gives
\eqb{l}
\bar{t}_\mrn = \ds\frac{1}{2}\,p_\mrc\,\norm{\br_\mrc} ~.
\label{e:qn}\eqe
Here, $\br_\mrc := \bx_\mrc - \bx_0$, where $\bx_0$ denotes the center of the circular contact surface.
Plugging Eq.~\eqref{e:qn} into Eq.~\eqref{e:qm1}, the tangential force along the contact line is obtained as 
\eqb{l}
\bar{t}_\mrm =  \ds\frac{1}{2}\,p_\mrc\,\norm{\br_\mrc}\,\cot \theta_\mrc ~,
\label{e:qm2}\eqe
which is applied as a displacement-follower load distributed along the contact line to enforce the specified contact angle $\theta_\mrc$. It should be noted that the latter expression for $\bar{t}_\mrm$ cannot be used for the particular case of liquid menisci that are modeled with two Dirichlet boundaries (see Fig.~\ref{f:comp_model}.d)\footnote{For this special case, there are two unknown normal contact forces along two separated contact lines. In order to find them analytically, one needs more balance equations like Eq.~\eqref{e:pcqn}, which are not considered here.}. 

To summarize, the droplet contact model introduced here is based on three assumptions:
\vspace{-\topsep}
\bnumr[1)]
\setlength{\itemsep}{0pt}
\setlength{\parskip}{0pt}
\setlength{\parsep}{0pt}
\item As the solid substrate is rigid, horizontal and planar and accordingly the solid-liquid interface $\sS_\mathrm{SL}$ remains planar, there is no need to explicitly consider this interface and the droplet can be modeled as an \textit{open} membrane. 
\item To model the contact between the droplet and the rigid substrate, a Dirichlet boundary condition is imposed on the contact line in the direction of the substrate normal $\bn_\mrc$.
\item To enforce a given contact angle $\theta_\mrc$, a distributed force $\bar{\bt}_\mrm = \bar{t}_\mrm\,\bm_\mrc$ is applied along the contact line, where $\bar{t}_m$ is obtained according to Secs. \ref{s:cont_gen} or \ref{s:cont_drop}.  
\enumr 
\vspace{-\topsep}

\subsection{Line tension}
The contact angle introduced in Sec.~\ref{s:cont} is called an \textit{intrinsic} contact angle \citep{marmur97} when the solid substrate has a smooth surface with homogeneous chemical and interfacial properties. As already mentioned, the Young equation 
\eqb{l}
 \cos \theta_\mrc = \dfrac{\gamma_\mathrm{SG}-\gamma_\mathrm{SL}}{\gamma_\mathrm{LG}}
\label{e:yl2}\eqe
is usually used to measure the contact angle. However, it has been experimentally observed for small systems that the contact angle predicted by Eq.~\eqref{e:yl2} is different from the empirically measured contact angle, due to the effect of line tension in $\sC$. Therefore, Eq.~\eqref{e:yl2} is modified as \citeplist{duncan95,marmur97} 
\eqb{l}
 \cos \theta_\mrc = \cos \theta_\mrc^\infty - \dfrac{\lambda\kappa_\mrg}{\gamma_\mathrm{LG}}~,
\label{e:yl3}\eqe
where $\lambda$ (with the unit of force) is the line tension along the contact line $\sC$ and $\kappa_\mrg$ is the geodesic curvature of $\sC$ at point $\bx_\mrc\in\sC$. For the special case of axially symmetric droplets and menisci, $\kappa_\mrg = 1/r_\mrc$. Here, $r_\mrc = \norm{\br_\mrc}$ is the radius of the circular three-phase contact line in the plane of the substrate surface, described in Sec.~\ref{s:cont_drop}. Further, $\theta_\mrc^\infty$ is the contact angle as $r_\mrc\rightarrow\infty$, given by Eq.~\eqref{e:yl2}.  
In analogy to surface tension, line tension can be associated with the free line energy
\eqb{l}
W_\lambda := \ds\int_\sC \lambda\,\dif s ~.
\label{e:Wl}\eqe
As shown in Appendix~\ref{s:wf_lt}, the spatial variation of the free energy $W_\lambda$ gives the contribution of the line tension to weak form (\ref{e:Gintext}.1) as 
\eqb{l}
G_\lambda := \delta W_\lambda = \ds\int_{-1}^{+1} \bw_{;\alpha}\cdot\lambda\,\ba_\mrc\,\norm{\ba_\mrc}^{-1}\,\dif\xi~.
\label{e:Gextt}\eqe

\section{Material models for surface tension}\label{s:models}
In Sec.~\ref{s:dyn_theo}, the existing literature on the theoretical models for the surfactant-dependent dynamics of surface tension is briefly reviewed. Here, the adsorption-limited (AL) model of \citet{otis94} and the compression-relaxation (CR) model of \citet{saad10}, which are in good agreement with experimental studies and can be efficiently implemented within a finite membrane element formulation, are discussed. The implementation is then discussed in Sec.~\ref{s:FE}.

\subsection{Adsorption-limited ({AL}) model}\label{s:ALM}
\citet{otis94} introduce a dynamic model for \emph{surfactant {TA}}\textsuperscript{\textregistered}\footnote{Surfactant TA\textsuperscript{\textregistered} is an artificial surfactant, produced by the Tokyo Tanabe company, which is widely used for clinical treatment of respiratory distress syndrome (RDS).} based on an experiment with a \textit{pulsating bubble surfactometer} ({PBS}). For the adsorption and desorption processes, it is assumed that the behavior of surfactant concentration $\Gamma$ at the liquid-gas interface can be described in three different regimes. These regimes are distinguished by two specific concentration values: The maximum equilibrium interfacial concentration $\Gamma^*$ and the maximum interfacial concentration $\Gamma_\mathrm{max}$, which are constants to be determined from experiments. \citet{otis94} assume that if the maximum surfactant concentration is reached and the surface area is reduced beyond the corresponding level of $\Gamma_\mathrm{max}$, surfactants are squeezed out of the interface so that the concentration of surfactants remains at $\Gamma_\mathrm{max}$. Furthermore, for the concentration values between $\Gamma^*$ and $\Gamma_\mathrm{max}$, it is assumed that there is neither adsorption nor desorption, which implies that the surfactant molecules remain at the interface. For the regime below $\Gamma^*$, adsorption and desorption processes are governed by Langmuir kinetics \citep{miller94}. Therefore, three governing equations for the dynamics of the surface concentration can be postulated as \citep{otis94}
\eqb{l} 
\ds\frac{\dif(\phi\,J)}{\dif t} =
\left\{
\begin{array}{lll}
J\,\left[K_1\,\big(1-\phi\big) - k_2\,\phi\right]	\quad\quad &$if$& 0\leq\phi \leq 1 \\[5mm]
0 \hfill &$if$& 1 < \phi < \phi_\mathrm{max} \\[3mm]
-\phi_\mathrm{max}\,\ds\frac{\dif J}{\dif t} \hfill &$if$& \phi = \phi_\mathrm{max}
\end{array} \right. ~,
\label{e:ALMc}\eqe
where $\dif(...)/\dif t$ is the material time derivative, $J$ is the local stretch and $K_1 := k_1\,C$. The coefficients of adsorption onto and desorption from the surface are $k_1$ and $k_2$, respectively. The bulk concentration $C$ is assumed to be constant as diffusion effects, which occur at different rates, are neglected w.r.t.~the adsorption/desorption process. Thus, $k_1$ and $C$ are merged into one parameter $K_1$. Here, for the sake of simplicity, the normalized concentration $\phi := \Gamma/\Gamma^*$ and the normalized maximum concentration $\phi_\mathrm{max} := \Gamma_\mathrm{max}/\Gamma^*$ are introduced. 
\begin{rem} 
Here, following continuum theory, the model of \citet{otis94} is defined locally in terms of local stretch $J$ rather than the total surface area of the membrane as it is done in \citet{otis94}.
\end{rem}
\begin{rem}
\citet{moris01} extend the formulation of \citet{otis94} by including diffusion effects. In this sense, $C$ varies with time and position within the bulk. However, for thin liquid membranes, such sophisticated formulations are not efficient. Furthermore, \citet{krueger00} modify the model of \citet{otis94} by including a secondary collapse layer, which leads to a more complicated set of state equations.  
\end{rem}
From Eq.~\eqref{e:ALMc}, one can derive the surface tension in form of an equation of state, which relates the surfactant interfacial concentration $\Gamma$ to the surface tension $\gamma$. For this  purpose, \citet{otis94} suggest two straight lines that meet at $\Gamma = \Gamma^*$, which give  
\eqb{l}
\gamma = 
\left\{
\begin{array}{lll}
\gamma_0 - m_1\,\phi \quad\quad &$if$& \phi \leq 1 \\[4mm]
\gamma^* -m_2\,(\phi - 1) \hfill &$if$& 1 < \phi \leq \phi_\mathrm{max}
\end{array} \right. ~,
\label{e:ALMs}\eqe
where $\gamma_0$ is the temperature-dependent surface tension of water, e.g.~$\approx 70~\mrm\mrN/\mrm$ at $25^\circ~\mrC$, and $\gamma^*$ is the minimum equilibrium surface tension, corresponding to the maximum equilibrium surfactant concentration $\Gamma^*$. The parameters $K_1$, $k_2$ and $m_2$ are the only unknowns that need to be identified by fitting model \eqref{e:ALMc}-\eqref{e:ALMs} to experimental results. On the other hand, $m_1$ can be found by knowing $\gamma^*$ and using Eq.~(\ref{e:ALMs}.1). If there are no surfactants on the interface, i.e.~$\phi = 0$, the surface tension should be that of pure water, i.e.~$\gamma=\gamma_0$. Therefore,  
\eqb{l}
m_1 = \gamma_0 - \gamma^* ~.
\eqe 
If normalized concentration values are used, $\Gamma^*$ is not needed; however, it can be assumed as $3~\mrm\mrg/\mrm^2$ \citep{moris01}. Two other parameters, namely $\gamma^*$ and $\gamma_\mathrm{min}$, need to be determined directly from experiments. The minimum equilibrium surface tension $\gamma^*$ can be determined by finding the lowest surface tension obtained for repeated measurements that are in equilibrium and have considerably high surfactant concentration in the bulk. For example, it can be 22.2 mN/m \citep{ingenito99} or 25 mN/m \citep{moris01}. Similarly, the minimum surface tension $\gamma_\mathrm{min}$, can be obtained from cyclic experiments, which usually give $1 \sim 2~\mrm\mrN/\mrm$ \citep{moris01}. Finally, from Eq.~(\ref{e:ALMs}.2), 
\eqb{l}
\phi_\mathrm{max} = 1 + \dfrac{\gamma^* - \gamma_\mathrm{min}}{m_2} ~.
\eqe        
Thus, the {AL} model needs six independent parameters, i.e. $K_1$, $k_2$, $m_2$, $\gamma_0$, $\gamma^*$, and $\gamma_\mathrm{min}$, among which the first three ones are found by curve fitting and the last three ones are directly determined from an experiment.

\subsection{Compression-relaxation ({CR}) model}\label{s:CRM}
\citet{saad10} propose a model that does not explicitly depend on the concentration values. Instead, the instantaneous surface tension and surface area determine the rate of surface tension change w.r.t.~time. The model is developed by superposition of two cases, where either the surface stretch or concentration is fixed. Assuming that the  surface stretch $J$ is constant, Eq.~(\ref{e:ALMc}.1) becomes
\eqb{l} 
\ds\frac{\dif \Gamma}{\dif t} =
K_1\,\big(\Gamma^*-\Gamma\big) - k_2\,\Gamma ~.
\label{e:ALMc_1}\eqe
If the maximum equilibrium interfacial concentration $\Gamma^*$ is known, the instantaneous equilibrium interfacial concentration $\Gamma_\mathrm{eq}$, depends on the instantaneous bulk concentration $C$ as \citep{otis94}
\eqb{l} 
\ds\frac{\Gamma_\mathrm{eq}}{\Gamma^*} = \dfrac{K_1}{K_1 + k_2} ~.
\label{e:Gam_eq}\eqe 
If the bulk concentration $C$ is fixed, by plugging Eq.~\eqref{e:Gam_eq} into Eq.~\eqref{e:ALMc_1}, one can derive
\eqb{l} 
\ds\frac{\dif \Gamma}{\dif t} = k\,\big(\Gamma_\mathrm{eq}-\Gamma\big) ~,
\label{e:dGam_dt}\eqe
where $k := K_1 + k_2$ is a coefficient for the dynamic adsorption and desorption. Assuming a linear relationship between surface tension and interfacial concentration (see Eq.~\eqref{e:ALMs}), the rate of changes in surface tension can be obtained as
\eqb{l} 
\ds\frac{\dif \gamma}{\dif t} = k\,\big(\gamma_\mathrm{eq}-\gamma\big) ~,
\label{e:dgam_dt}\eqe 
where the equilibrium surface tension $\gamma_\mathrm{eq}$ corresponds to the specified equilibrium interfacial concentration $\Gamma_\mathrm{eq}$. For the adsorption or spreading process, where $\gamma \geq \gamma_\mathrm{eq}$, the adsorption coefficient $k=k_\mra$ defines the adsorption rate. Similarly, for the desorption or relaxation, where $\gamma < \gamma_\mathrm{eq}$, the desorption rate is given by the desorption coefficient $k=k_\mrr$. Both $k_\mra$ and $k_\mrr$ are parameters to be found by fitting the model to experimental results.

The changes of surface tension $\gamma$ w.r.t.~the surface stretch $J$ can be characterized by the elasticity parameter
\eqb{l}
\epsilon = \ds\dfrac{\dif \gamma}{\dif \big(\ln J\big)}
\label{e:e_0}\eqe
that is assumed constant here. It corresponds to the elastic surface bulk modulus at $J\approx1$ \citep{liquidshell}. Rearranging and dividing by $\dif t$ leads to
\eqb{l}
\ds\frac{\dif\gamma}{\dif t} = \epsilon\,\dfrac{1}{J}\dfrac{\dif J}{\dif t}~.
\label{e:dgam_dt_s2}\eqe
The elasticity coefficient $\epsilon>0$ should also be obtained through a parameter identification process for compression ($\epsilon = \epsilon_\mrc$) and expansion ($\epsilon = \epsilon_\mre$). 
Furthermore, according to experiments \citep[e.g.][]{otis94}, there is a maximum surfactant concentration, which is the limit where the layers of surfactants collapse. Therefore, the surface tension cannot be reduced below the minimum surface tension $\gamma_\mathrm{min}$ by further compression.

Based on the four main processes that affect the dynamic response of surface tension -- namely 1) adsorption or spreading, 2) desorption or relaxation, 3) elasticity during compression and 4) elasticity during expansion -- \citet{saad10} postulate a unified formulation that combines all cases. In the framework of continuum mechanics, it can be formulated as     
\eqb{l}
\ds\frac{\dif\gamma}{\dif t} = 
\left\{
\begin{array}{lll}
k\,\big(\gamma_\mathrm{eq}-\gamma\big) + \epsilon\,\ds\frac{1}{J}\,\ds\frac{\dif J}{\dif t} \quad\quad &$if$& \gamma \geq \gamma_\mathrm{min} \\[4mm]
0	\hfill &$if$& \gamma < \gamma_\mathrm{min}
\end{array} \right. ~,
\label{e:CRM}
\eqe
with
\eqb{l} 
k =
\left\{
\begin{array}{lll}
k_\mra	\quad\quad &$if$& \gamma \geq \gamma_\mathrm{eq} \\[2mm]
k_\mrr	\hfill &$if$& \gamma < \gamma_\mathrm{eq}
\end{array} \right. ~,
\eqe
\eqb{l} 
\epsilon =
\left\{
\begin{array}{lll}
\epsilon_\mre \quad\quad &$if$& {\dif J}/{\dif t} > 0 \\[2mm]
\epsilon_\mrc \hfill &$if$& {\dif J}/{\dif t} < 0
\end{array} \right.~. 
\eqe

Here, it is assumed that the adsorption/desorption and elasticity processes can happen at the same time. The {CR} model requires six different parameters: $k_\mra$, $k_\mrr$, $\epsilon_\mrc$ and $\epsilon_\mre$ are set by curve fitting of experimental results. $\gamma_\mathrm{min}$ and $\gamma_\mathrm{eq}$ are  directly measured by experiments.

\begin{rem}\label{r:marangoni}
From a computational point of view, surface tension can be defined as follows: 
\vspace{-\topsep}
\bnumr[1)]
\setlength{\itemsep}{0pt}
\setlength{\parskip}{0pt}
\setlength{\parsep}{0pt}
\item Point-wise, where the surface tension varies locally (i.e.~$\gamma = \gamma(\bx,t)$), as it is used here; 
\item Element-wise, where the surface tension is assumed to be uniform over each finite element $\Omega^e$ (i.e.~$\gamma = \gamma(\Omega^e,t)$), see e.g.~\citetlist{wiechert09,wiechert11thesis}, or
\item Globally, where the surface tension is considered uniform over the whole membrane surface (i.e.~$\gamma = \gamma(t)$) as is done in \citetlist{otis94,saad10}.
\enumr
\vspace{-\topsep}

From a physical point of view, local variation of concentration results in a surface tension gradient, which in turn induces an interfacial flow known as \textit{Marangoni flow} \citeplist{scriven60,velarde02}. \tred{Among different applications, Marangoni flow can be used to transport particles at fluid interfaces for example for pulmonary drug delivery \citep{sharma17}}. If the bulk viscosity is significant, the Marangoni flow will cause an internal flow in the bulk of the fluid through viscous forces. However, in the examples presented in this work, the bulk viscosity is neglected. Generally, the Marangoni effect occurs due to temperature or concentration gradients and also at interfaces between different liquids. As the {AL} and {CR} models, introduced in Sec.~\ref{s:models}, are explicitly or implicitly dependent on the surfactant concentration, Marangoni flow can be expected if the point-wise approach is followed. The Lagrangian description of the membrane presented here can capture such flows provided that they are not too large, as shown in the example of Sec.~\ref{s:film_rel}.

If the surface flow becomes large, an arbitrary Lagrangian--Eulerian (ALE) formulation can be used \citeplist{yang07,ganesan09}, which is not considered here. The ALE approach requires expanding the material time derivatives into a spacial and a convective part \citep{stone90}
\eqb{l}
\ds\pa{\Gamma}{t} + \nabla_\mrs\Gamma\cdot\dot{\bx} = f\big(\Gamma,C\big) + D_\mrs\,\nabla_\mrs^2\Gamma~,
\label{e:conv-dif}\eqe
where the surface gradient operator, $\nabla_{\!\mrs}$, is given in Sec.~\ref{s:kin}. The rear term in Eq.~\eqref{e:conv-dif} represents surface diffusivity assuming Fick's laws of diffusion, where $D_\mrs$ is the diffusivity. $f$ is a surfactant source term representing the surface adsorption/desorption process, which can be derived following the approach of \citet{moris01}. The interfacial flows, governed by Eq.~\eqref{e:conv-dif}, are not considered in this study.
\end{rem}

\section{Finite element solution}\label{s:FE}
The finite element method is used to solve the governing equations. Thus, the membrane surface $\sS$ is approximated by the discretized surface $\sS^h$, which is constructed from elements $\Omega^e$. In this section, first the \tred{finite element (FE)} interpolation is introduced and then the discretized weak form is summarized following \citetlist{membrane,droplet}. Here also the contribution of line tension to the discretized weak form is derived. Third and foremost, the evolution equations for surface tension according to the {AL} and {CR} models are discretized within an implicit time integration scheme. As the {BVP} \eqref{e:wf_2} is highly nonlinear, it should be linearized and solved by an iterative algorithm such as the Newton--Raphson method.

\subsection{Finite element interpolation}\label{s:int}
Any point on the surface $\bx \in \sS$ is approximated by $\bx^h \in \sS^h$ as
\eqb{l}
\bx \approx \bx^h = \mN \, \mx_e ~,
\label{e:xh}\eqe
where $\mx_e$ is a vector that contains all nodal (or control point) positions belonging to $\Omega^e$ and the shape functions are arranged as $\mN:= [N_1\bone,\, N_2\bone,\, ...,\, N_{n_e}\bone]$, with $n_e$ being the number of nodes of element $\Omega^e$. Similarly in the reference configuration, 
\eqb{l}
\bX \approx \bX^h = \mN \, \mX_e ~.
\eqe
Due to Eq.~\eqref{e:xh}, we have
\eqb{lll}
\ba_\alpha \ais \mN_{,\alpha} \, \mx_e ~, \\
\ba_{\alpha,\beta} \ais \mN_{\alpha,\beta} \, \mx_e ~,
\eqe
where 
\eqb{lll}
\mN_{,\alpha} \dis [N_{1,\alpha}\bone,\, N_{2,\alpha}\bone,\, ...,\, N_{{n_e},\alpha}\bone]~,\\
\mN_{,\alpha\beta} \dis [N_{1,\alpha\beta}\bone,\, N_{2,\alpha\beta}\bone,\, ...,\, N_{n_e,\alpha\beta}\bone]
\eqe
and
\eqb{l}
N_{A,\alpha} = \dfrac{\partial N_A}{\partial \xi^\alpha}~,\quad N_{A,\alpha\beta} = \dfrac{\partial^2 N_A}{\partial \xi^\alpha\,\partial \xi^\beta} ~, \quad (A=1, ..., n_e)~.
\eqe
Following a Galerkin approach, the variation $\bw$ is approximated like the deformation, which gives
\eqb{l}
\bw \approx \mN \, \mw_e ~.
\eqe
Here, for the finite element interpolation, quadratic {NURBS}-based shape functions
\eqb{lll}
 N_A\big(\xi^1,\xi^2\big) = \ds\frac{w_A\,\hat{N}_A^e\big(\xi^1,\xi^2\big)}{\sum_{A=1}^{n_e} w_A\,\hat{N}_A^e\big(\xi^1,\xi^2\big)},
\label{e:shpfct}\eqe
with $n_e = 9$ are used. Here, $\{\hat{N}_A^e\}_{A=1}^{n_e}$ is the B-spline basis function expressed in terms of Bernstein polynomials using the B{\'e}zier extraction operator \citep{borden11}. However, as shown by \citet{membrane}, classical finite elements with quadratic Lagrange interpolation polynomials are also suitable for the modeling of liquid drops and menisci.

\subsection{FE force vectors}\label{s:fe_vect}
Following the FE setting introduced in Sec.~\ref{s:int}, the discretized version of weak form \eqref{e:wf_2} is
\eqb{l}
\ds\sum_{e=1}^{n_\mathrm{el}}\,\mw_e^\mrT\,\big(\mf_\mathrm{int}^e - \mf_\mathrm{ext}^e\big) = 0
\label{e:wf_3}\eqe
with the internal FE force vector $\mf_\mathrm{int}^e := \mf_{\mathrm{int}\tau}^e + \mf_{\mathrm{int}\lambda}^e$, where
\eqb{l}
\mf_{\mathrm{int}\tau}^e = \ds\int_{\Omega^e_0}\mN^\mrT_{,\alpha}\,\tau^{\alpha\beta}\,\ba_\beta\,\dif A ~, 
\label{e:fint}\eqe
\eqb{l}
\mf_{\mathrm{int}\lambda}^e := \ds\int_{-1}^{+1} \mN_{\mrt,\xi}^\mrT\,\lambda\,\ba_\mrc\,\norm{\ba_\mrc}^{-1}\,\dif\xi ~,
\label{e:fintt}\eqe
and the external FE force vector
\eqb{l}
\mf_\mathrm{ext}^e = \mf_{\mathrm{ext}f}^e + \mf_{\mathrm{ext}p}^e + \mf_{\mathrm{ext}t}^e + \mf_\mathrm{extc}^e ~, 
\label{e:fext1}\eqe
where
%
%
\eqb{l}
\mf_{\mathrm{ext}f}^e = \ds\int_{\Omega^e_0} \mN^\mrT\,\bff_{\!0}\,\dif A ~,\quad
\mf_{\mathrm{ext}t}^e = \ds\int_{\Gamma^e_t}\mN^\mrT_\mrt\,\bar\bt\,\dif s ~, \quad
\mf_{\mathrm{ext}p}^e = \ds\int_{\Omega^e} \mN^\mrT\,p\,\bn\,\dif a
\label{e:fext2}\eqe
and
\eqb{l}
\mf_\mathrm{extc}^e = \cos\theta_\mrc\ds\int_{\Gamma^e_\mrc} \mN^\mrT_\mrt\,\gamma\,\bm_\mrc\,\dif s 
\label{e:fextc2}\eqe
if the contact line model of Sec.~\ref{s:cont_gen} is used and
\eqb{l}
\mf_\mathrm{extc}^e = \ds\frac{1}{2}\,\cot\theta_\mrc\ds\int_{\Gamma^e_\mrc} \mN^\mrT_\mrt\,p_\mrc\,\bm_\mrc\,\norm{\br_\mrc}\,\dif s
\label{e:fextc1}\eqe
if the contact line model of Sec.~\ref{s:cont_drop} is used.
Here, it is assumed that the external distributed load per surface is $\bff = \bff_{\!0}/J + p\,\bn$, where $\bff_{\!0}$ is a dead load per reference area and $p$ is a live pressure as described in Sec.~\ref{s:sf}. However, in the examples of Sec.~\ref{s:example}, dead loading on $\sS$ is not considered. Besides, $\mf_{\mathrm{ext}t}^e$ is the contribution of a prescribed distributed load $\bar\bt$ on the Neumann boundaries $\Gamma^e_t \subset \partial_t\sS$ other than the contact boundary; $\mf_\mathrm{extc}^e$ is the contribution of the tangential force $\bq_\mrm$ along the contact line element $\Gamma^e_\mrc \subset \sC$ and $\mN_\mrt$ is the array of shape functions corresponding to the contact line.  

\subsection{The Newton--Raphson iteration}\label{s:NR}
As the discretized weak form \eqref{e:wf_3} holds for any admissible $\mw_e$, the governing equation can be reduced to the system of ODEs
\eqb{l}
\mf = \mf(\mx,\dot\mx,p_\mrv) = \mathbf{0} ~,
\label{e:f}\eqe
where $\mf$ is the global residual force vector formed by the assembly 
\eqb{l}
\mf := \ds\ass_{e=1}^{n_\mathrm{el}}\big(\mf^e_\mathrm{int}-\mf^e_\mathrm{ext}\big)~,
\eqe
where $\mf_\mathrm{int}^e$ and $\mf_\mathrm{ext}^e$ are the FE force vectors introduced in Sec.~\ref{s:fe_vect} and $\ass$ denotes the classical FE assembly operator. In Eq.~\eqref{e:f}, $\mx$ is a vector that collects all nodal positions (e.g.~in a global Cartesian coordinate system), $\dot\mx:=\dif\mx/\dif t$, and $p_\mrv$ is a single scalar Lagrange multiplier (see Sec.~\ref{s:sf}). $\dot\mx$ is eliminated by a time discretization scheme (see Sec.~\ref{s:time}) such that $\mf = \mathbf{0}$ becomes an algebraic equation for $\mx$ and $p_\mrv$.   
Following the introduced FE setting, the volume $V$, enclosed by the discretized surface, is \citep{membrane}
\eqb{l}
V \approx V^h = \dfrac{1}{3}\ds\sum_{e=1}^{n_\mathrm{el}} \int_{\Omega^e} \bn^\mrT\,\mN\,\dif a \, \mx_e ~.
\eqe
This equation is derived for closed droplets, but it also applies to open droplets, if the origin of the coordinate system lies on the contact surface.
Thus, the complete system of equations to be solved by the finite element method is
\eqb{l}
\mF = \mF(\mx,p_\mrv) := 
\begin{bmatrix}
\mf(\mx,p_\mrv) \\[2mm]
g_\mrv(\mx)
\end{bmatrix} \tred{ = \mathbf{0}}~.
\eqe
As $\mF = \mF(\mx,p_\mrv)$ is nonlinear, it can be solved by the Newton--Raphson method, which is an iterative algorithm. This requires the linearization of $\mF$ w.r.t.~$\mx$ and $p_\mrv$ as
\eqb{l}
\Delta \mF = \mK\, \Delta\mU ~,
\label{e:Df}\eqe
where $\mK$ is the global tangent matrix and
\eqb{l}
\Delta\mU :=  
\begin{bmatrix}
\Delta\mx \\[2mm]
\Delta p_\mrv
\end{bmatrix}
\eqe
includes all degrees of freedom. The global tangent matrix can be arranged as
\eqb{l}
\mK :=
\begin{bmatrix}
\mk & - \ml_\mathrm{ext} \\[2mm]
\mh_\mrv & 0
\end{bmatrix}
\label{e:Kglobal}\eqe
where $\mk := \mk_\mathrm{int} - \mk_\mathrm{ext}$ and
\eqb{l}
\mk_\mathrm{int} := \ds\ass_{e=1}^{n_\mathrm{el}}\mk^e_\mathrm{int}~,\quad \mk_\mathrm{ext} := \ds\ass_{e=1}^{n_\mathrm{el}}\mk^e_\mathrm{ext}~,\quad \ml_\mathrm{ext} := \ds\ass_{e=1}^{n_\mathrm{el}}\ml^e_\mathrm{ext}~,\quad \mh_\mrv := \ds\ass_{e=1}^{n_\mathrm{el}}\mh^e_\mrv~,
\eqe
with 
\eqb{l}
\mk^e_\mathrm{int} := \ds\pa{\mf^e_\mathrm{int}}{\mx_e}~,\quad \mk^e_\mathrm{ext} := \ds\pa{\mf^e_\mathrm{ext}}{\mx_e}~,\quad \ml^e_\mathrm{ext} := \ds\pa{\mf^e_\mathrm{ext}}{p_\mrv}~,\quad \mh^e_\mrv := \ds\pa{g_\mrv}{\mx_e} ~.
\label{e:tangents}\eqe
The element tangent matrices $\ml^e_\mathrm{ext}$ and $\mh^e_\mrv$ can be found in \citet{membrane}. The internal stiffness matrices $\mk^e_{\mathrm{int}\tau}$ and $\mk^e_{\mathrm{int}\lambda}$ and the tangent matrix $\mk^e_\mathrm{extc}$ for the contact line force vector are derived in Appendix~\ref{s:fe_tangent} in detail. 
\begin{rem}
For cases where the volume is not constrained, e.g.~for free films, Eq.~\eqref{e:Df} reduces to $\Delta \mf = \mk\, \Delta\mx$.
\end{rem} 
\begin{rem}
It should be noted that, for the constitutive laws presented in Sec.~\ref{s:models}, the internal stiffness tangent $\mk_\mathrm{int}$ is affected by the nonlinearity of the surface tension $\gamma$; therefore, it has extra terms compared to \citet{membrane} (see Appendix~\ref{s:fe_tangent}).   
\end{rem} 
\begin{rem}
All the tangents $\mk^e_\mathrm{int}$, $\mk^e_\mathrm{ext}$ and $\ml^e_\mathrm{ext}$ are derived at the current time step, $t_n$. Hence, the history quantities coming from the previous time steps ($t_{n-1}$, $t_{n-2}$, ...) are treated as constants in the Newton--Raphson iteration and thus do not need to be linearized. 
\end{rem} 

\subsection{Time integration}\label{s:time}
In the introduced dynamic setup, the surface tension and stress are time dependent. Thus, they should be updated with a discretized time integration scheme. Here, the backward Euler scheme is used, i.e.
\eqb{l}
\dot{\gamma} = \ds\frac{\dif\gamma}{\dif t} \approx \ds\frac{\Delta\gamma}{\Delta t} = \ds\frac{1}{\Delta t}\big(\gamma_n - \gamma_{n-1}\big)~, 
\label{e:BaEu_gam}\eqe  
\tred{
for the CR model \eqref{e:CRM},
\eqb{l}
\dfrac{\dif\big(J\,\phi\big)}{\dif t} \approx \ds\frac{1}{\Delta t}\big(J_n\,\phi_n - J_{n-1}\,\phi_{n-1}\big)~, 
\label{e:BaEu_phi}\eqe  
for the AL model \eqref{e:ALMc},} and
\eqb{l}
\dot{a}^{\alpha\beta} \approx \ds\frac{1}{\Delta t}\,\big(a^{\alpha\beta}_n-a^{\alpha\beta}_{n-1}\big) ~,
\label{e:BaEu_aab}\eqe 
\tred{for the viscosity model \eqref{e:sigab1}. Further, $\dot{J}$ in Eq.~\eqref{e:sigab1} is computed from Eqs.~\eqref{e:dotJ} and \eqref{e:BaEu_aab}. Here,} $\Delta t$ is the time step between $t_{n-1}$ and $t_n$, \tred{which is considered to be constant.} The contribution of the dynamic concentration-dependent surface tension $\gamma_n$, which varies by $J_n$, to the internal tangent matrix is given in Secs.~\ref{s:tiALM} and \ref{s:tiCRM}. The solution algorithm for the introduced dynamic problem is summarized in Table~\ref{t:algr}.   
\begin{table}[h!]
\centering
\fbox{\parbox{0.75\textwidth}{
\bnumr[1),leftmargin=*]
\setlength{\itemsep}{0pt}
\item Retrieve relevant quantities from $t_{n-1}$, i.e. \tred{$J_{n-1}$, $\gamma_{n-1}$ or $\phi_{n-1}$}.
\item Update the current time step: \tred{$t_n = t_{n-1} + \Delta t$}.
\item Newton--Raphson iteration: \\
\fbox{\parbox{0.7\textwidth}{
\bnumr[i),leftmargin=*]
\setlength{\itemsep}{0pt}
  \item Update $\gamma_n$ for the corresponding {dynamic models} of Sec.~\ref{s:models} \tred{(see Secs.~\ref{s:tiALM} and \ref{s:tiCRM})}.
  \item Compute the residual force vector $\mf_n := \mf_\mathrm{int} - \mf_\mathrm{ext}$.
  \item Update \tred{the} tangent matrices \tred{(\ref{e:Kglobal}-\ref{e:tangents})}. 
  \item Add the contribution of $\gamma_n$ to the tangent matrices (see Secs.~\ref{s:tiALM} and \ref{s:tiCRM}).
  \item Solve Eq.~\eqref{e:Df} for unknowns $\Delta\mx_n$ and $\Delta p_\mrv$.
  \item Update \tred{$\mx_n := \mx_n + \Delta\mx_n$} and $J_n$ \tred{\eqref{e:J}}.
  \item Check for convergence, \tred{e.g.~$|\Delta\mU^\mrT\,\mF| \leq \mathrm{Tol}$}. 
\enumr}}
\enumr}}
\caption{Solution algorithm for the dynamic model}
\label{t:algr}
\end{table}
\tred{
\begin{rem}
It should be noted that an explicit integration scheme offers no advantage because the Newton--Raphson iteration is needed anyway due to the other nonlinearities. Furthermore, as the system is strongly nonlinear, an explicit scheme would require prohibitively small time steps to run stable. Implicit time schemes generally run stable for much larger time steps.
\end{rem}
\begin{rem}
%
Here, all quantities are understood to be evaluated at the current time step $t_n$ unless stated otherwise.
\end{rem}
\begin{rem}
In the present FE formulation, the only degrees of freedom are displacements $\mx$ (and the capillary pressure $p_\mrv$ if the bulk incompressibility is taken into account). Therefore, only the geometry and displacement are interpolated by the finite element shape functions. 
In the CR model, the unknown $\phi$ and, in the AL model, the unknown $\gamma$ do not appear in the FE system as they are expressed in terms of unknown $J$ and are eliminated locally. 
\end{rem}
\begin{rem}
At time step $t_n$, the surface tension $\gamma_n$, the surfactant concentration $\phi_n$, the surface stretch $J_n$ and the components of metric tensor $a^{\alpha\beta}_n$ are stored at each Gaussian quadrature point to be retrieved at the next time step $t_{n+1}$.  
\end{rem}
}

\subsubsection{{AL} model}\label{s:tiALM}
Following Eq.~\eqref{e:BaEu_phi}, a time-discretized evolution equation for the concentration $\phi_n$ is obtained from Eq.~\eqref{e:ALMc} as
\eqb{l} 
\ds\frac{\Delta(J\,\phi)}{\Delta t} = \ds\frac{J_n\,\phi_n - J_{n-1}\,\phi_{n-1}}{\Delta t} = 
\left\{
\begin{array}{lll}
J_n\,\big[K_1\,(1-\phi_n) - k_2\,\phi_n\big] \quad &$if$& \phi_{n} \leq 1 \\[5mm]
0 \hfill &$if$& 1 < \phi_{n} < \phi_\mathrm{max} \\[3mm]
-\phi_\mathrm{max}\,\ds\frac{J_n - J_{n-1}}{\Delta t} \hfill &$if$& \phi_{n} = \phi_\mathrm{max}
\end{array} \right.~,
\label{e:ALMcbE1}\eqe
which gives the time-discretized version of the {CR} model as
\eqb{l} 
\phi_n = 
\left\{
\begin{array}{lll}
\big[1 + \Delta t\,(K_1+k_2)\big]^{-1}\Big(\Delta t\,K_1 + \ds\frac{J_{n-1}}{J_n}\phi_{n-1}\Big) \quad\quad &$if$& \phi_{n} \leq 1 \\[6mm]
\ds\frac{J_{n-1}}{J_n}\phi_{n-1} \hfill &$if$& 1 < \phi_{n} < \phi_\mathrm{max} \\[6mm]
\ds\frac{J_{n-1}}{J_n}\big(\phi_{n-1} + \phi_\mathrm{max}\big) - \phi_\mathrm{max} \hfill &$if$& \phi_{n} = \phi_\mathrm{max}
\end{array} \right.~.
\label{e:ALMcBEpw}\eqe

Then, the surface tension is updated according to the equation of state Eq.~\eqref{e:ALMs}, which gives
\eqb{l}
\gamma_n = 
\left\{
\begin{array}{lll}
\gamma_0 - m_1\,\phi_n \quad\quad &$if$& \phi_n \leq 1 \\[4mm]
\gamma^* -m_2\,\big(\phi_n - 1\big) \hfill &$if$& 1 < \phi_n \leq \phi_\mathrm{max}
\end{array} \right. ~.
\label{e:ALMsbE}\eqe

The linearization of $\gamma = \gamma(\bx_n,t_n)$ leads to
\eqb{l}
\Delta_\mrx\gamma = \dfrac{\partial \gamma}{\partial J}\,\Delta_\mrx J ~,
\label{e:Dgamx}\eqe
where
\eqb{l}
\dfrac{\partial \gamma}{\partial J} = -k_\mrs\,k_\mrc\,\ds\frac{1}{J^2}~,
\label{e:dgam_ALM}\eqe
with
\eqb{l} 
k_\mrc := \ds\dfrac{\partial \phi_n}{\partial (1/J_n)} =  
\left\{
\begin{array}{lll}
\big[1 + \Delta t\,(K_1+k_2)\big]^{-1}\,\phi_{n-1}\,J_{n-1} \quad\quad &$if$& \phi_{n-1} \leq 1 \\[6mm]
\phi_{n-1}\,J_{n-1} \hfill &$if$& 1 < \phi_{n-1} \leq \phi_\mathrm{max} \\[6mm]
\big(\phi_{n-1} + \phi_\mathrm{max}\big)\,J_{n-1} \hfill &$if$& \phi_{n-1} = \phi_\mathrm{max}
\end{array} \right.
\label{e:kc}\eqe
and
\eqb{l} 
k_\mrs = \ds\frac{\partial \gamma}{\partial \phi} = 
\left\{
\begin{array}{lll}
-m_1 \quad\quad &$if$& \phi_n \leq 1 \\[4mm]
-m_2 \hfill &$if$& 1 < \phi_n \leq \phi_\mathrm{max}
\end{array} \right. ~.
\label{e:ks}\eqe
The contribution of $\Delta_\mrx\gamma$ to the material tangent matrix is shown in Appendix~\ref{s:int_tang}. 

\subsubsection{{CR} model}\label{s:tiCRM}
Plugging Eq.~\eqref{e:CRM} into Eq.~\eqref{e:BaEu_gam}, the time-discretized version of the {CR} model is obtained as
\eqb{l}
\gamma_n = \ds\frac{1}{1+k\,\Delta t}\,\Big[ \gamma_{n-1} + k\,\gamma_\mathrm{eq}\,\Delta t + \epsilon\Big(1 - \ds\frac{J_{n-1}}{J_n}\Big)\Big]~, 
\label{e:CRMbEpw}\eqe
where
\eqb{l} 
k =
\left\{
\begin{array}{lll}
k_\mra	\quad\quad &$if$& \gamma_{n-1} \geq \gamma_\mathrm{eq} \\[2mm]
k_\mrr	\hfill &$if$& \gamma_{n-1} < \gamma_\mathrm{eq}
\end{array} \right. ~,
\eqe
\eqb{l} 
\epsilon =
\left\{
\begin{array}{lll}
\epsilon_\mre \quad\quad &$if$& J_n \geq J_{n-1} \\[2mm]
\epsilon_\mrc \hfill &$if$& J_n < J_{n-1}
\end{array} \right.~. 
\eqe

The linearization of $\gamma$ is also given by Eq.~\eqref{e:Dgamx}. From Eq.~\eqref{e:CRMbEpw}, we now have
\eqb{l}
\dfrac{\partial \gamma}{\partial J} = \ds\frac{\epsilon}{1+k\,\Delta t}\,\frac{J_{n-1}}{\big(J_n\big)^2} ~.
\label{e:dgam_CLM}\eqe

\section{Numerical examples}\label{s:example}
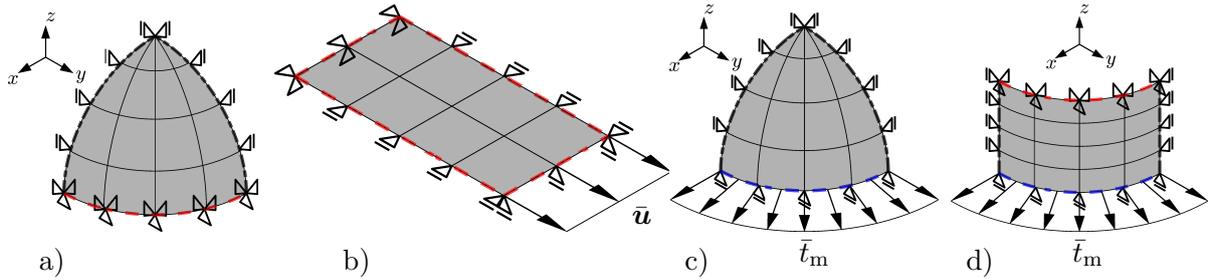
\begin{figure}[ht]
\begin{center} \unitlength1cm
\unitlength1cm
\begin{picture}(0,4.5)
\put(-7.9,0.5){\includegraphics[height=29mm]{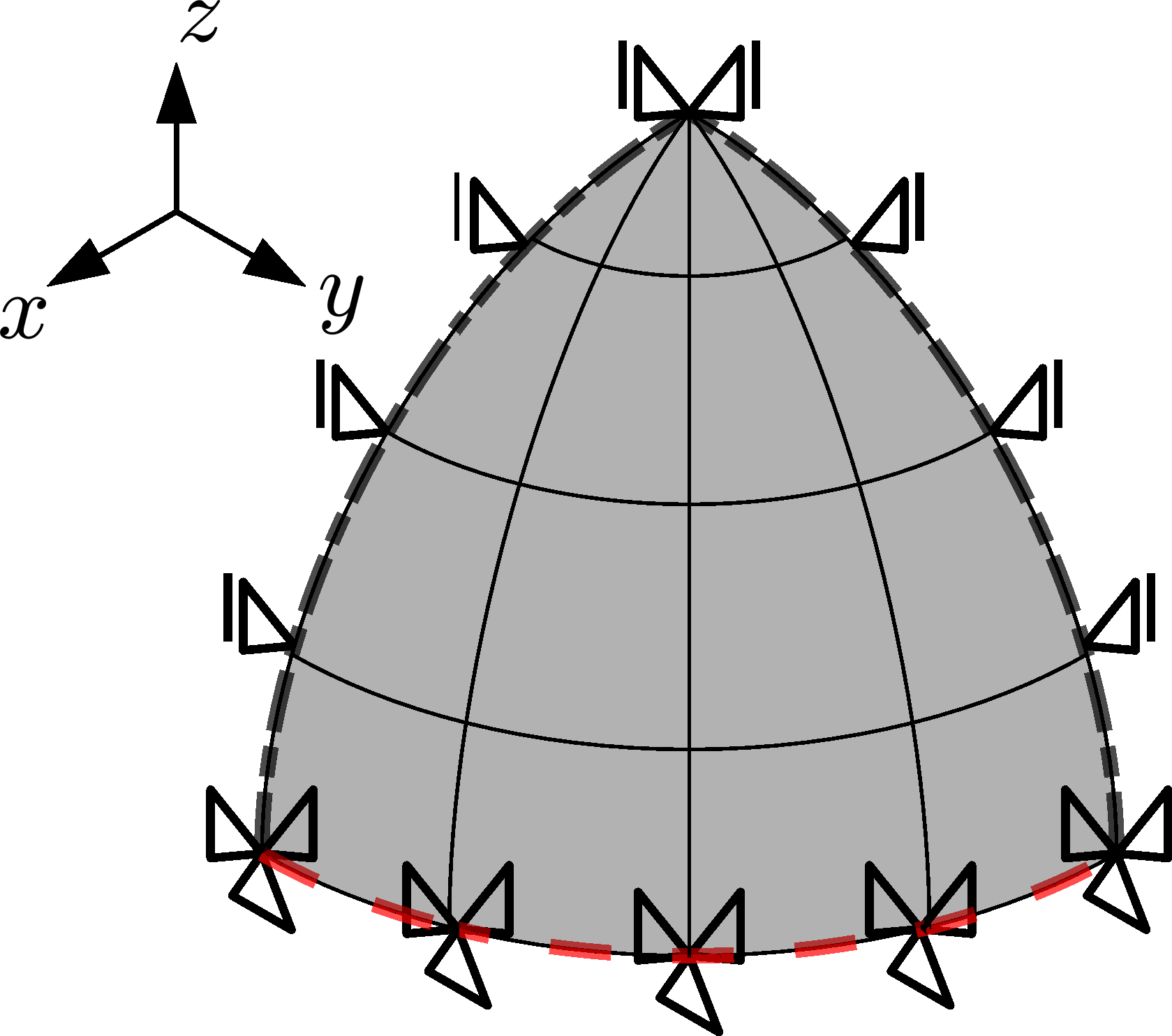}}
\put(-7.5,0.0){a)}
\put(-4.4,0.5){\includegraphics[height=30mm]{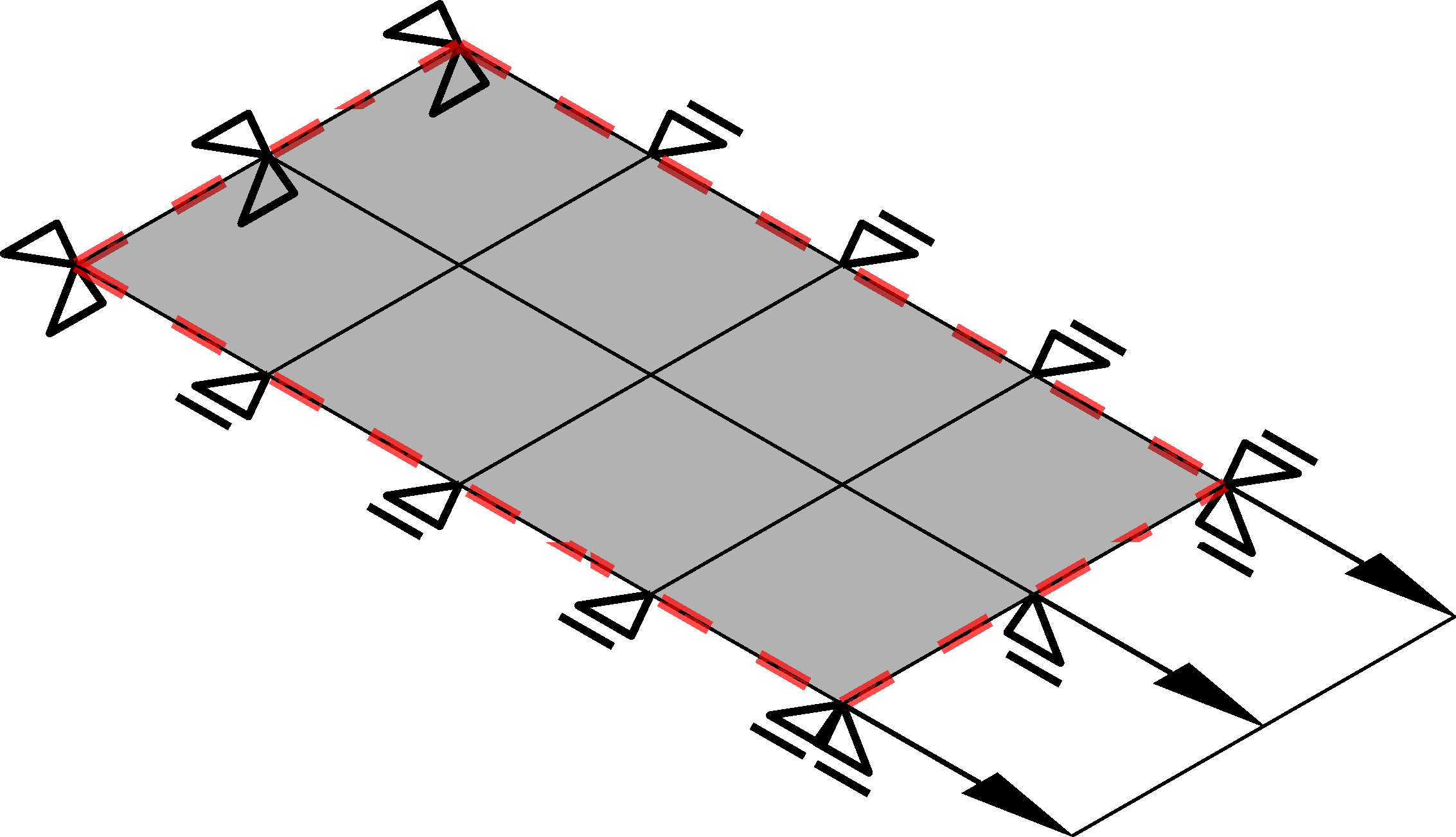}}
\put( 0.3,0.6){$\bar\bu$}
\put(-3.5,0.0){b)}
\put( 0.8,0.5){\includegraphics[height=30mm]{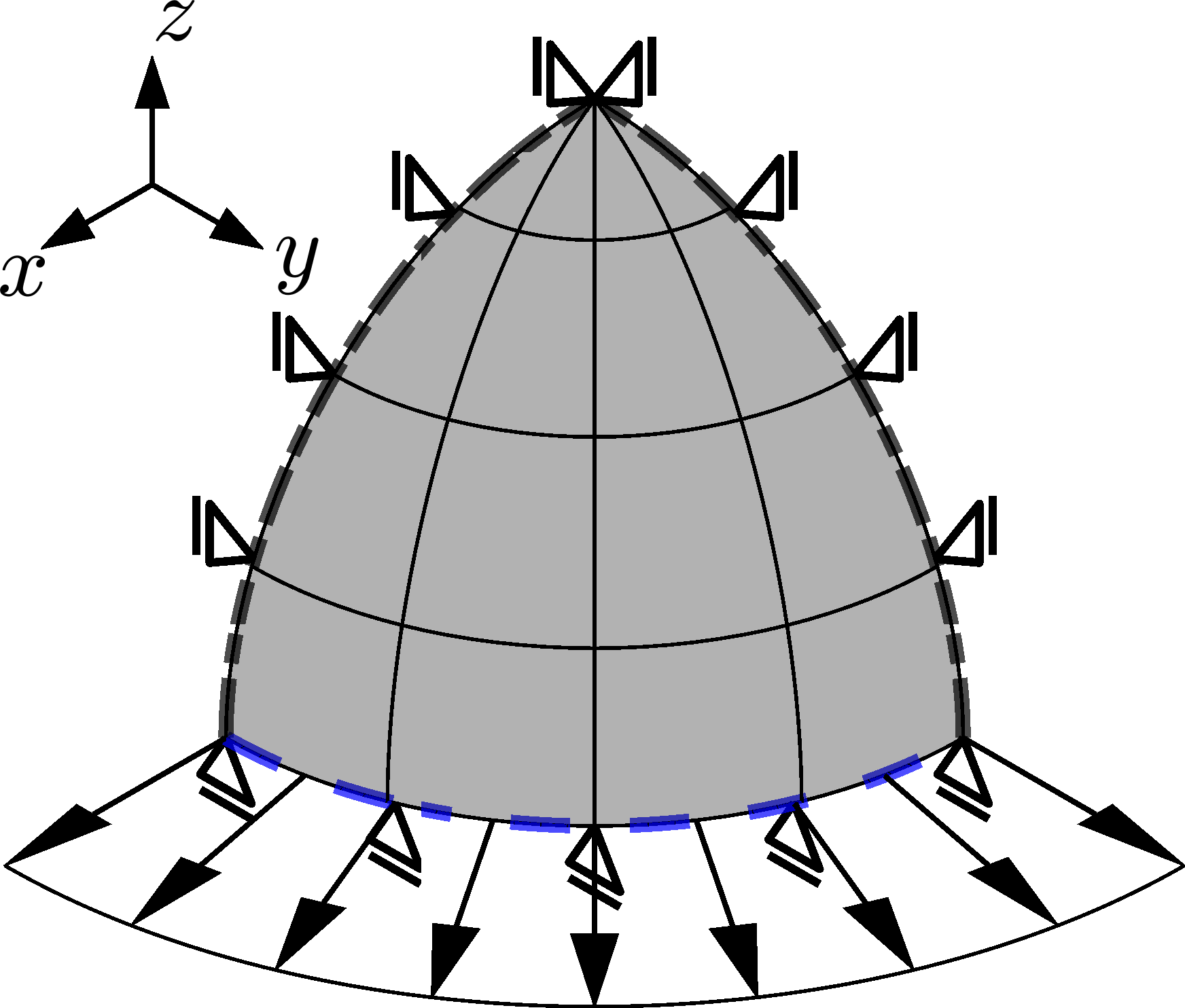}}
\put( 2.5,0.1){$\bar{t}_\mrm$}
\put( 1.0,0.0){c)}
\put( 4.5,0.5){\includegraphics[height=30mm]{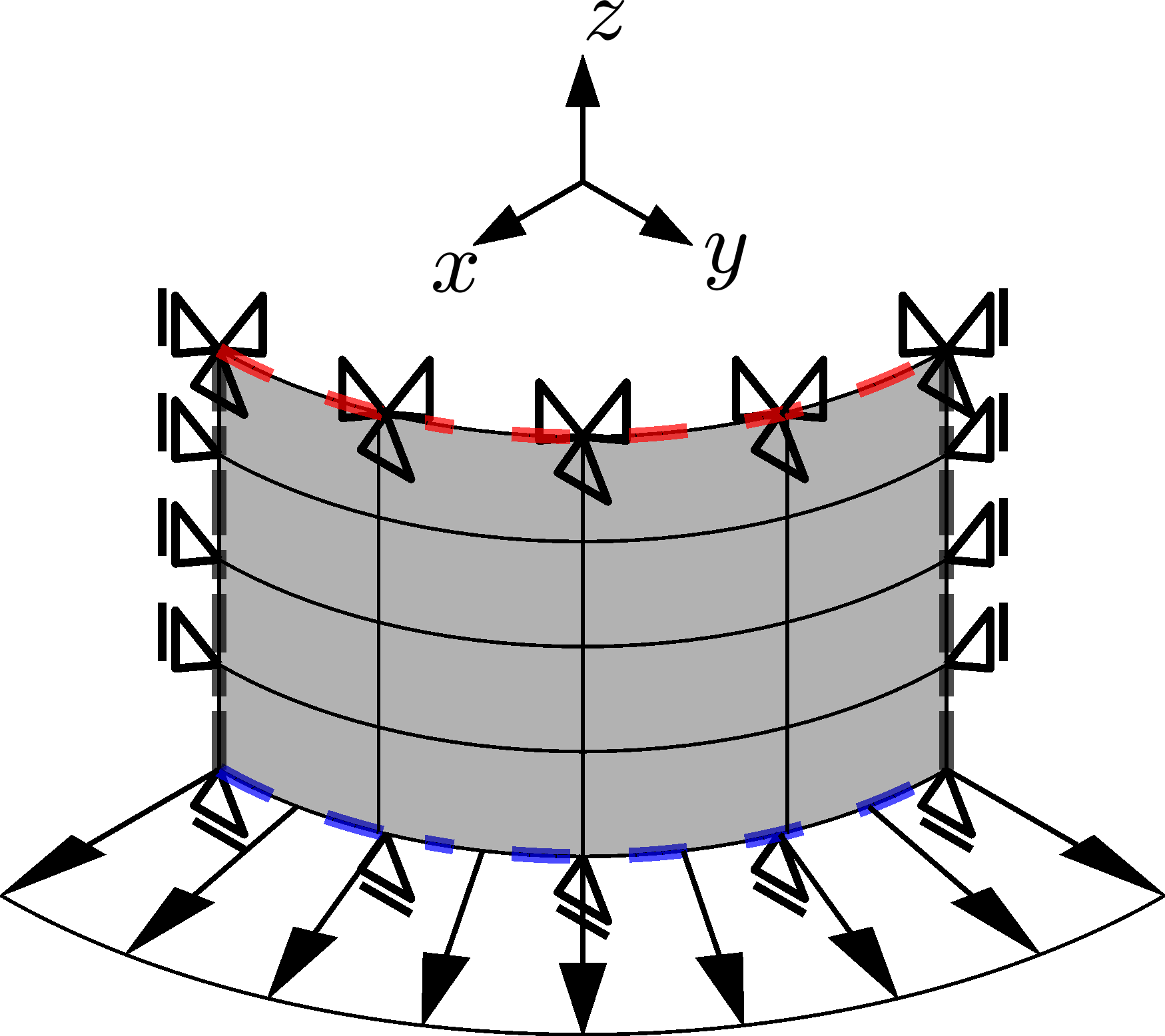}}
\put( 6.1,0.1){$\bar{t}_\mrm$}
\put( 4.7,0.0){d)}
\end{picture}
\caption{Computational models with boundary conditions: a) Constrained sessile drop (1/4 droplet, $n_\mathrm{el}=16$), b) thin film ($n_\mathrm{el}=8$), c) sessile drop (1/4 droplet, $n_\mathrm{el}=16$) and d) liquid bridge (1/4 meniscus, $n_\mathrm{el}=16$). The drops and meniscus are fluid filled and the thin film is a free surface without bulk. The symmetry boundaries are denoted by black dash-dot lines and the pure Dirichlet and the mixed Dirichlet--Neumann boundaries are denoted by red and blue dashed lines, respectively.}
\label{f:comp_model}
\end{center}
\end{figure}

In Sec.~\ref{s:intro}, the typical approaches for dynamic surface tension measurement -- namely the \textit{Langmuir--Wilhelmy balance} \citep{clements57}, \textit{captive bubble surfactometer} (CBS) \citep{schurch01} and \textit{constrained sessile drop} (CSD) \citep{saad10} -- are introduced. Here, the capabilities of the presented membrane formulation to simulate similar cases are demonstrated. The main focus of the simulations is on the drop shape methods. First, the CSD test is examined following the experimental setup and the CR material model of \citet{saad10}. Then, the behavior of a thin liquid film in cyclic compression/expansion and relaxation without external loads is analyzed. Next, the \textit{pendant drop} (PD), \textit{sessile drop} (SD) and \textit{liquid bridge} (LB) tests are simulated. Finally, the influence of the line tension on the contact angle is studied. Fig.~\ref{f:comp_model} shows the computational models with the corresponding boundary conditions of all the examples. As discussed by \citetlist{alonso04,hermans15}, in addition to the adsorption/desorption mechanisms, surface viscosity has also important contribution to the interfacial behavior of pulmonary surfactants. The corresponding values of the surface viscosity are not reported for the experiments from which the material parameters are taken here. Thus, the surface viscosity is taken as $\eta = 1~\mrm\mrN\mrs/\mrm$, which is within the ranges reported by \citetlist{alonso04,rudiger05} for similar surfactants. Furthermore, in all the examples, except the thin film, gravity is included as a hydrostatic pressure, according to Eq.~\eqref{e:ph}, where the droplet is assumed to have the density of water at $37^\circ$, i.e.~$\rho = 993~\mrk\mrg/\mrm^3$, and $g=\tred{9.8}~\mrN/\mrm^2$. For all the examples, the time step is $\Delta t = 0.03\,\mrs$ unless otherwise specified. \tred{With such a time step, a good Newton--Raphson performance is obtained for the loading conditions and the meshes used here.} The geometries are discretized by quadratic {NURBS} meshes, where the number of elements $n_\mathrm{el}$ is specified in Fig.~\ref{f:comp_model}.  

\subsection{Constrained sessile drop (CSD) test}\label{s:csd}
As the first example, the \textit{constrained sessile drop} (CSD) test, which is one of the most popular methods for dynamic measurement of surface tension, is simulated. The method is also used by \citet{saad10} for axisymmetric drop shape analysis. As the droplet is constrained, it is modeled as a pinned droplet, where the contact angle is not fixed. Therefore, the contact line formulation presented in Sec.~\ref{s:cont}, is not needed for this problem. The droplet is modeled as a hemisphere with initial radius of 1.5 mm, whose opening radius is fixed on a rigid pedestal and its volume is decreased and increased cyclically. As shown in Fig.~\ref{f:comp_model}.a, only 1/4 of the droplet is modeled to exploit the symmetry of the problem in order to reduce the computational cost. 
\begin{figure}[ht]
\begin{center} \unitlength1cm
\unitlength1cm
\begin{picture}(0,4.5)
\put(-7.5,0.0){\includegraphics[height=45mm,trim={330px 0 550px 0},clip]{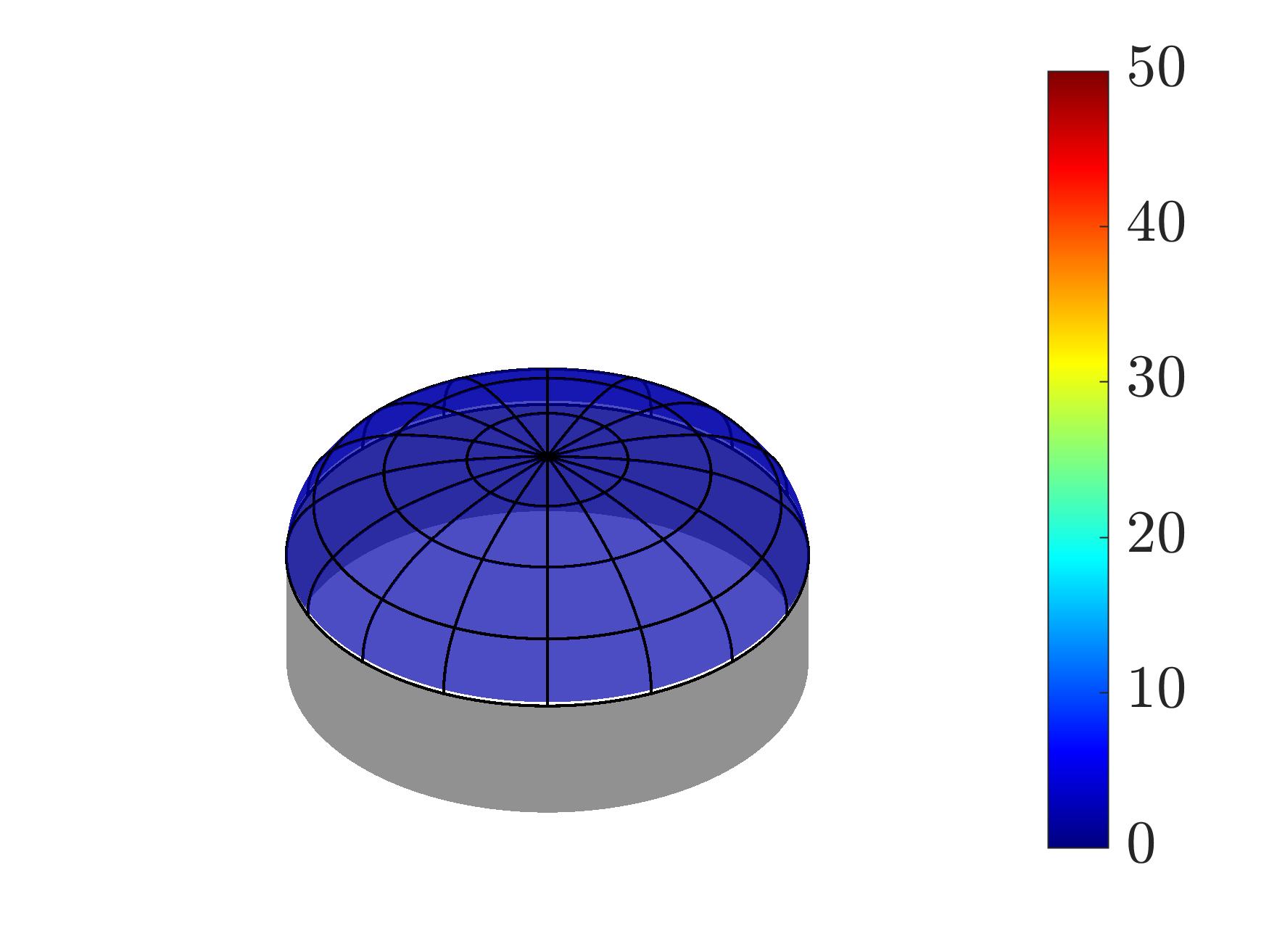}}
\put(-7.5,0.0){a)}
\put(-4.0,0.0){\includegraphics[height=45mm,trim={330px 0 550px 0},clip]{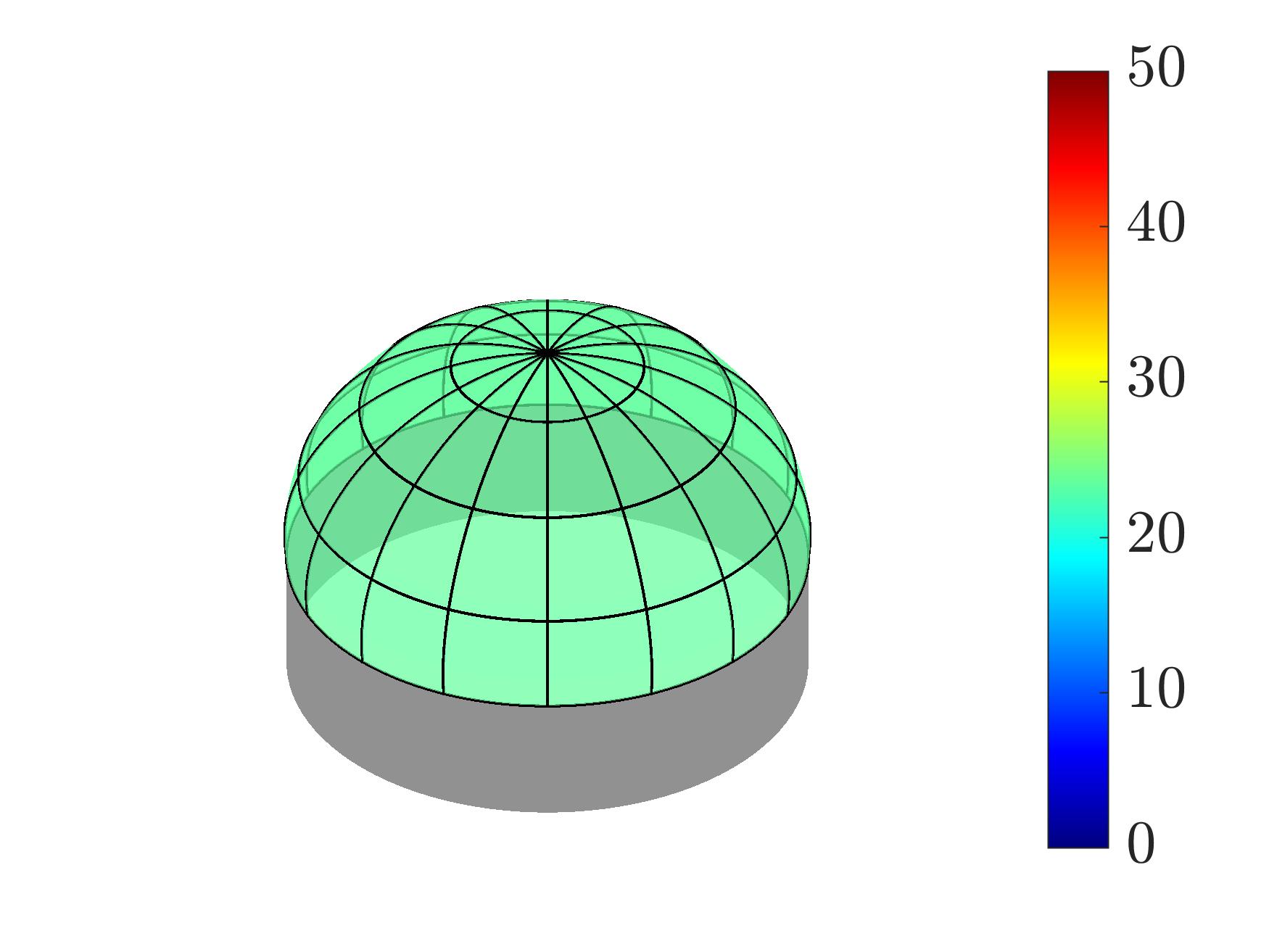}}
\put(-4.0,0.0){b)}
\put(-0.5,0.0){\includegraphics[height=45mm,trim={330px 0 550px 0},clip]{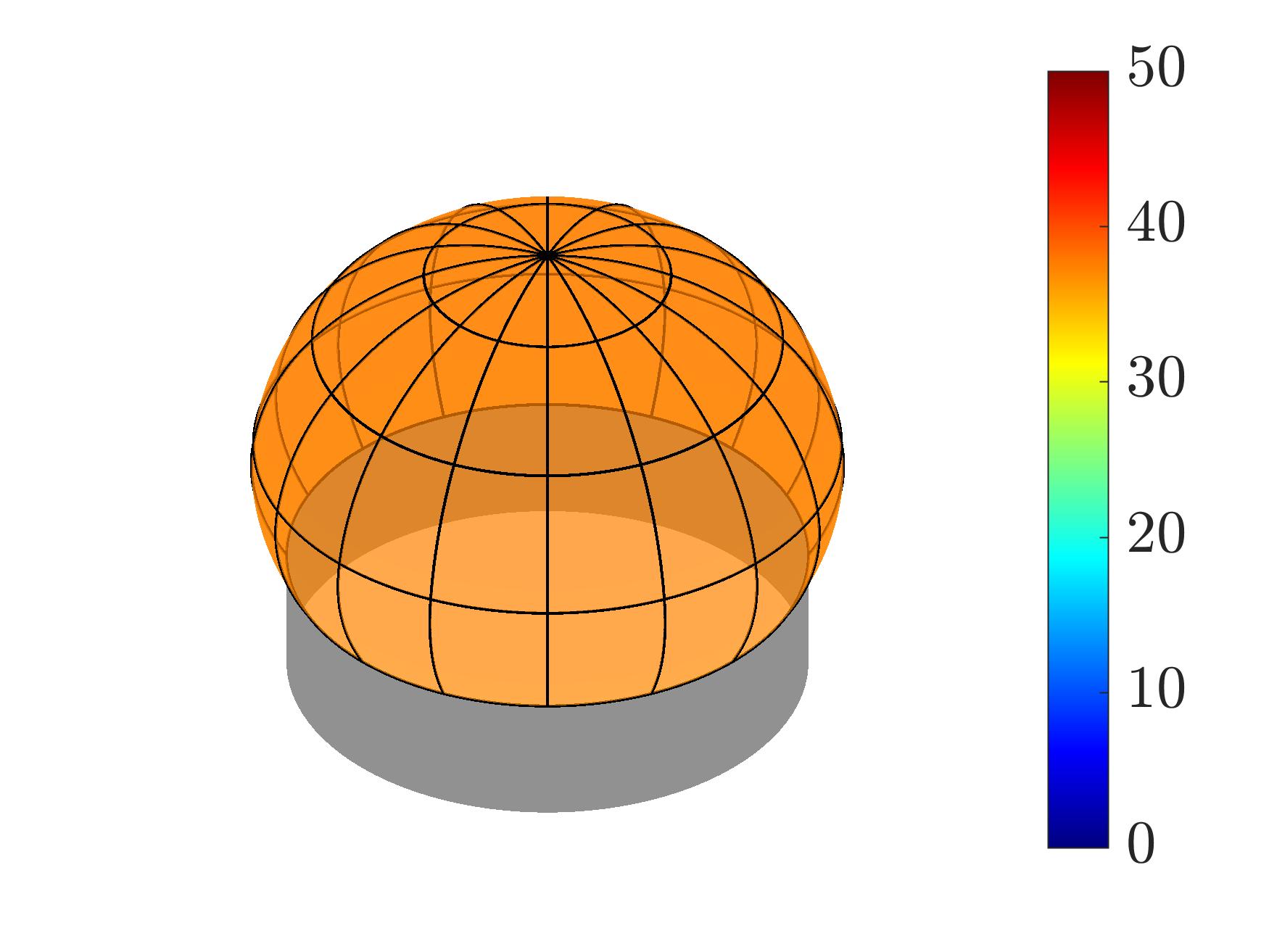}}
\put(-0.5,0.0){c)}
\put( 3.0,0.0){\includegraphics[height=45mm,trim={250px 0 30px 0},clip]{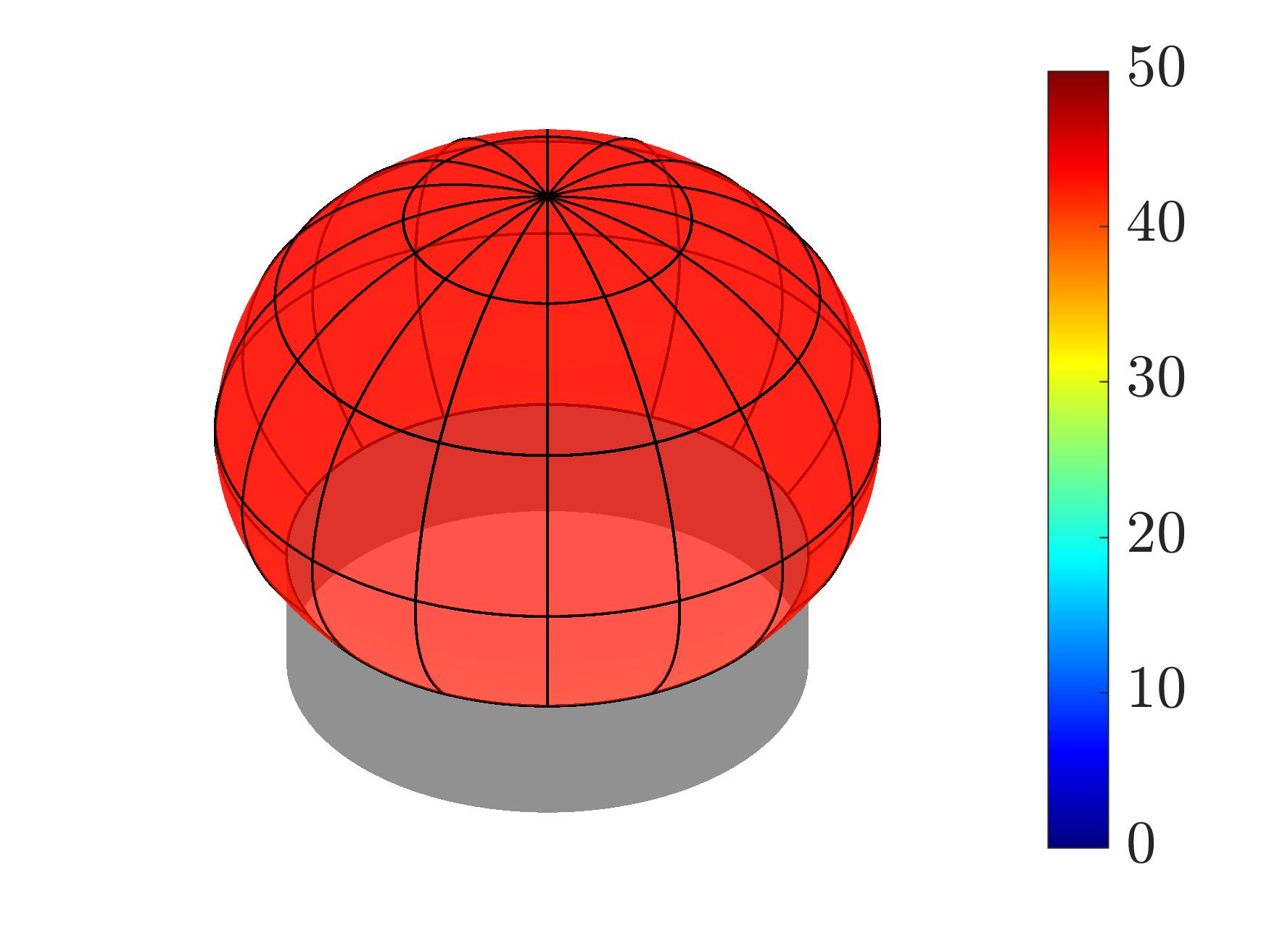}}
\put( 3.0,0.0){d)}
\end{picture}
\caption{The CSD test: Full droplet, modeled with the parameters of case 1 in Tab.~\ref{t:CRM_par}, at a) $V=0.5\,V_0$, b) $V=V_0$, c) $V=2\,V_0$ and d) $V=3\,V_0$ colored by the surface tension $\gamma~[\mrm\mrN/\mrm]$.}
\label{f:csd_drops}
\end{center}
\end{figure}
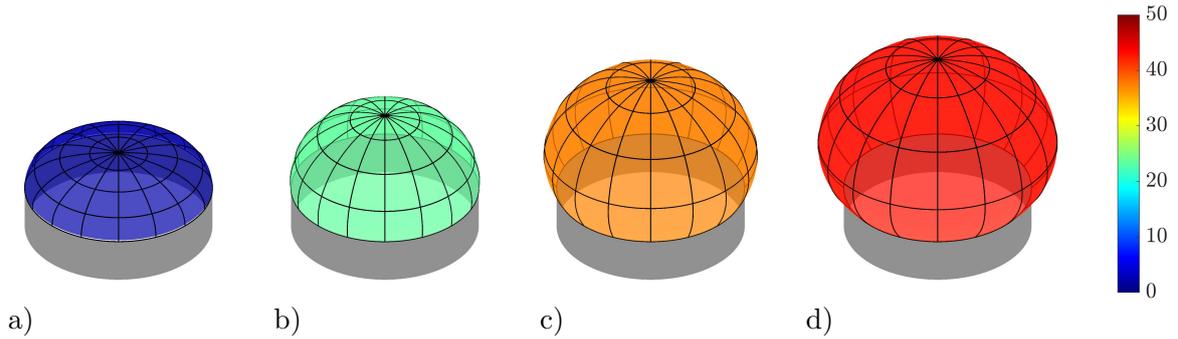

Following \citep{saad10}, for each test, two different concentrations of BLES\footnote{BLES\textsuperscript{\textregistered} (bovine lipid extract surfactant) is a commercial natural surfactant made of lung surfactant lavaged from cows. It is produced by BLES Biochemicals Inc., Ontario, Canada.} in humid and dry conditions are considered. As already mentioned, the {CR} model requires the elasticity of compression $\epsilon_\mrc$, the elasticity of expansion $\epsilon_\mre$, the relaxation coefficient $k_\mrr$, the adsorption coefficient $k_\mra$, the equilibrium surface tension $\gamma_\mathrm{eq}$ and the minimum surface tension $\gamma_\mathrm{min}$. The corresponding parameters are adopted from \citet{saad10} as listed in Tab.~\ref{t:CRM_par}. The droplet volume is controlled by the volume constraint \eqref{e:gv}. Fig.~\ref{f:csd_drops} show the droplet, modeled with the parameters of case 1 in Tab.~\ref{t:CRM_par}, at different volumes colored by the surface tension $\gamma~[\mrm\mrN/\mrm]$.
\begin{table}[ht!]
\centering
{\renewcommand{\arraystretch}{1.2}
\begin{tabular}{ rcccccccc }
  \hline
  & & & $\epsilon_\mrc$ & $\epsilon_\mre$ & $k_\mrr$ & $k_\mra$ & ${\gamma_\mathrm{min}}^\dagger$ & ${\gamma_\mathrm{eq}}^\dagger$ \\
  & Content & Humidity & [mN/m] & [mN/m] & [s\tss{-1}] & [s\tss{-1}] & [mN/m] & [mN/m] \\
  \hline
  1)& BLES 0.5 mg/ml & Humid & 125.1 & 157.8 & 0.547 & 2.474 & 2 & 24 \\
  2)& BLES 0.5 mg/ml & Dry   & 112.7 & 120.0 & 0.001 & 2.783 & 2 & 22 \\
  3)& BLES 2.0 mg/ml & Humid & 126.3 & 136.0 & 3.751 & 4.991 & 2 & 23 \\
  4)& BLES 2.0 mg/ml & Dry   & 123.1 & 129.9 & 0.006 & 0.712 & 2 & 23 \\
  \hline
  \multicolumn{6}{l}{$^\dagger$ The values are estimated by the authors.}
\end{tabular}
\caption{Parameters of the {CR} model \citep{saad10}}
\label{t:CRM_par}}
\end{table}

Fig.~\ref{f:csd_time} shows the changes of mean surface tension and surface area of the droplet as a function of time. As the effect of gravity is considered, the droplet does not remain spherical. Accordingly, as the volume of droplet changes, small deviations in the surface tension are expected. Thus, to compare the results, the average surface tension,
\eqb{l}
\bar\gamma(t) = \dfrac{1}{S}\ds\int_\sS\gamma(t,\bx)\,\dif a ~,
\label{e:mean_gam}\eqe
is examined. Here, $S$ is the total droplet surface area, in the deformed configuration, given as
\eqb{l}
S = \ds\int_\sS \dif a = \ds\int_{\sS_0} J\,\dif A ~.
\eqe
Similarly, the initial droplet surface is given as
\eqb{l}
S_0 = \ds\int_{\sS_0} \dif A 
\eqe
in the reference configuration. As shown in the example of Sec.~\ref{s:film_rel}, if enough time is given to the system to relax, it \tred{finds} a new configuration, where the surface tension is more homogeneous. Fig.~\ref{f:csd_gam} compares the standard deviation of the surface tension, $\sigma(\gamma)$, and its mean value $\bar\gamma$. As the standard deviation of the surface tension is small w.r.t.~the mean value for all the examples except the liquid film of Sec.~\ref{s:film_rel}, it is only reported here to keep the paper short.   
\begin{figure}[ht!]
\begin{center} \unitlength1cm
\unitlength1cm
\begin{picture}(0,9.0)
\put(-6.8,4.5){\includegraphics[height=45mm]{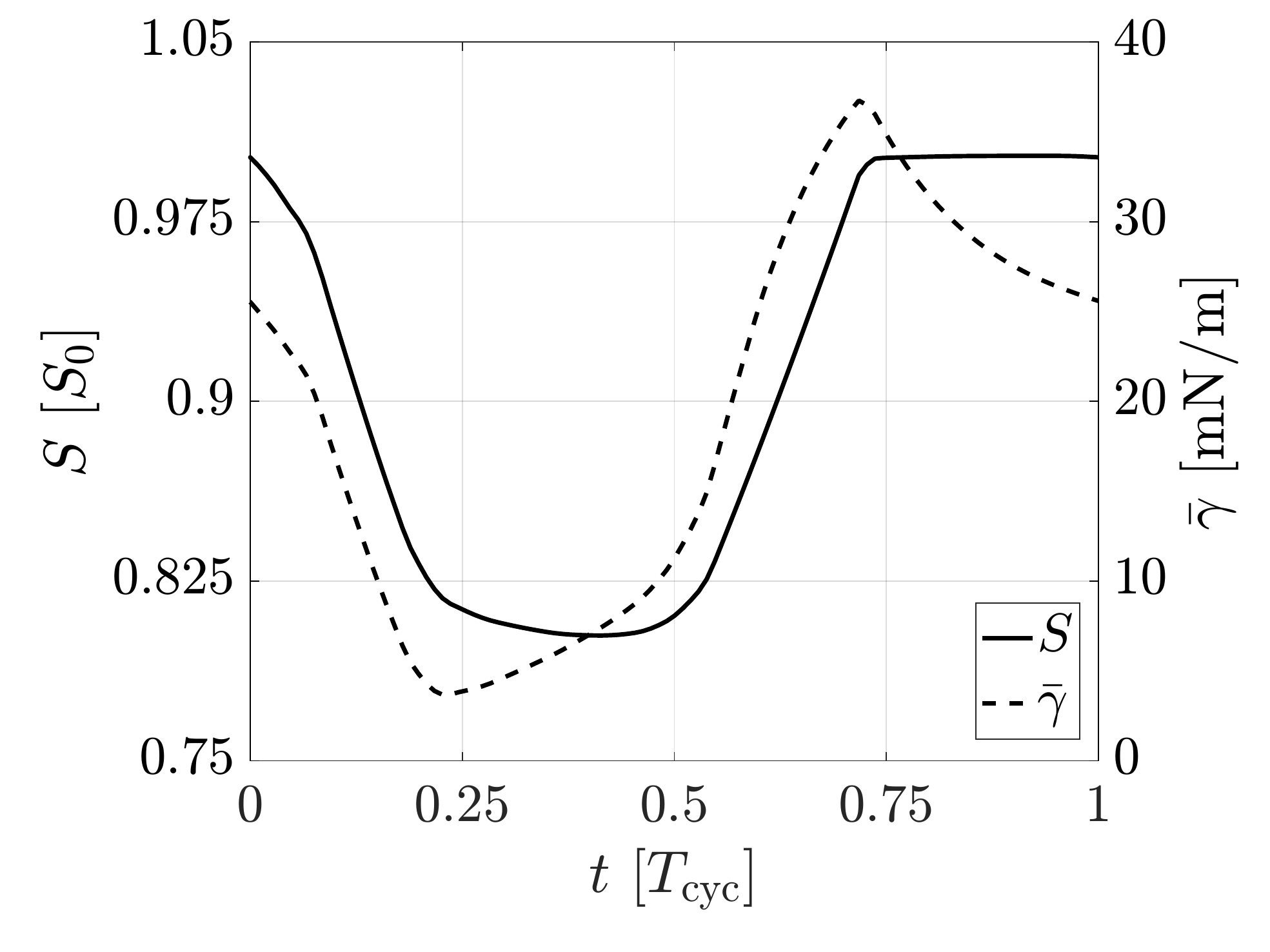}}
\put(-7.0,5.1){a)}
\put( 0.3,4.5){\includegraphics[height=45mm]{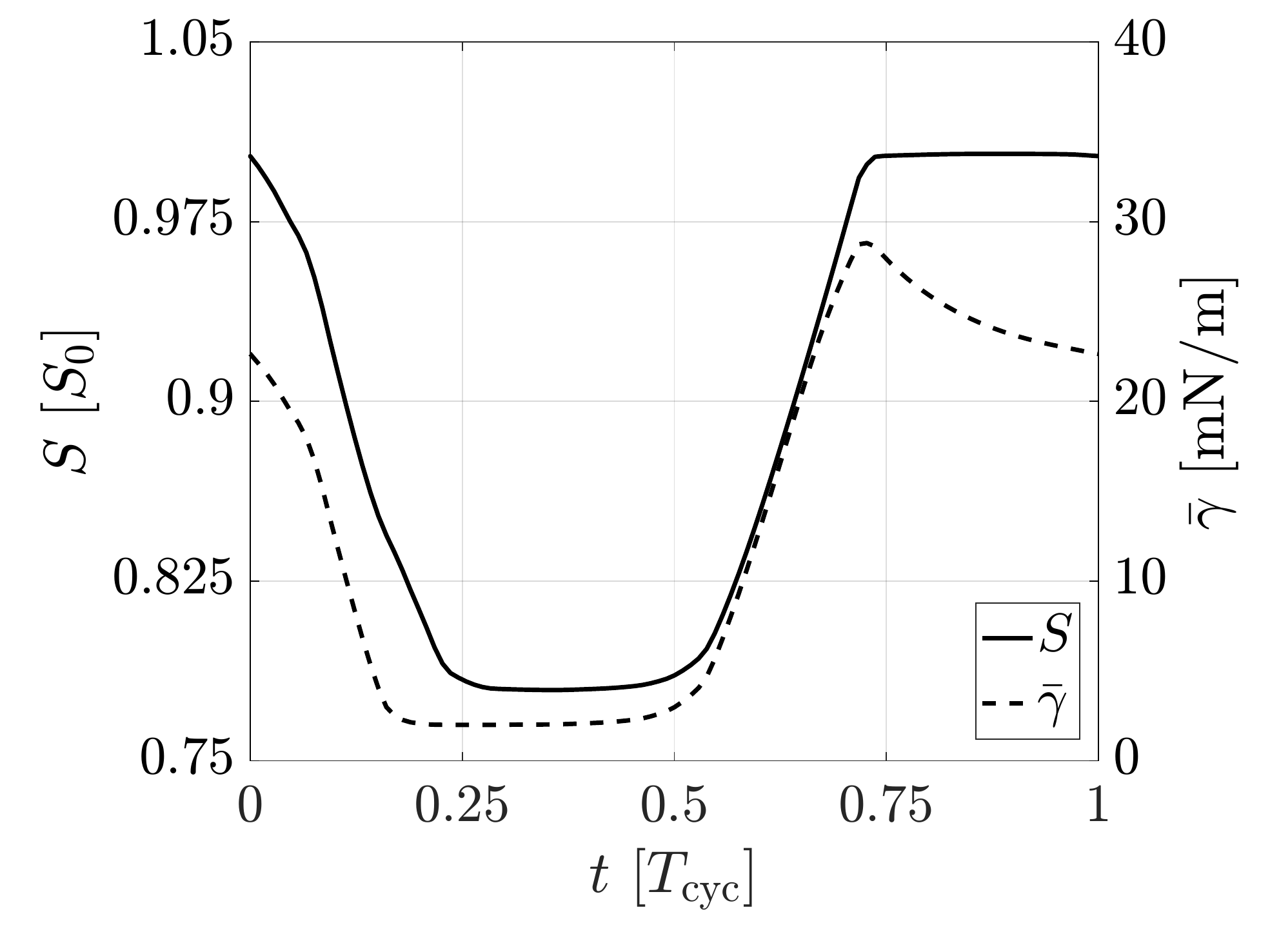}}
\put( 0.1,5.1){b)}
\put(-6.8,0.0){\includegraphics[height=45mm]{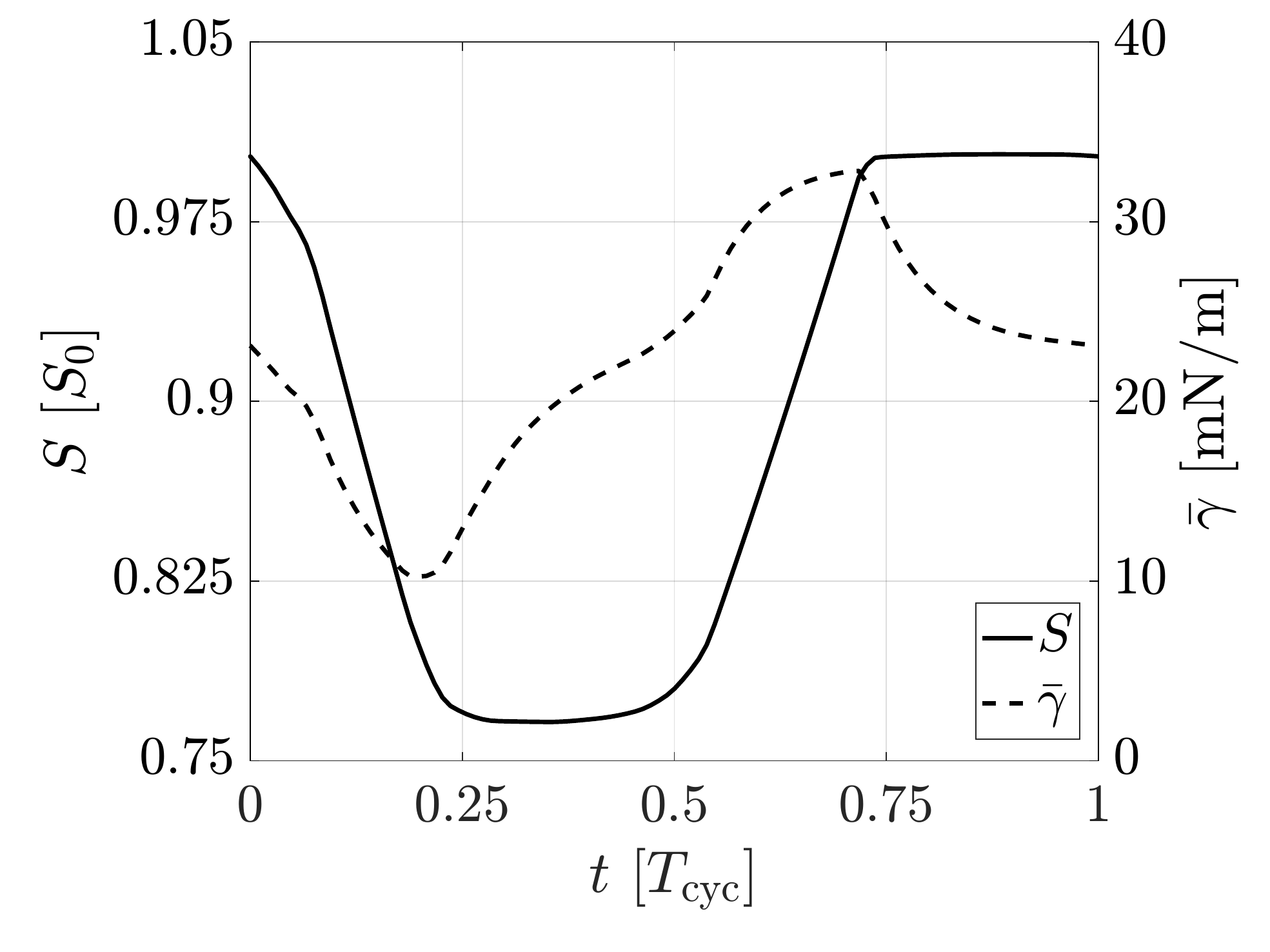}}
\put(-7.0,0.6){c)}
\put( 0.3,0.0){\includegraphics[height=45mm]{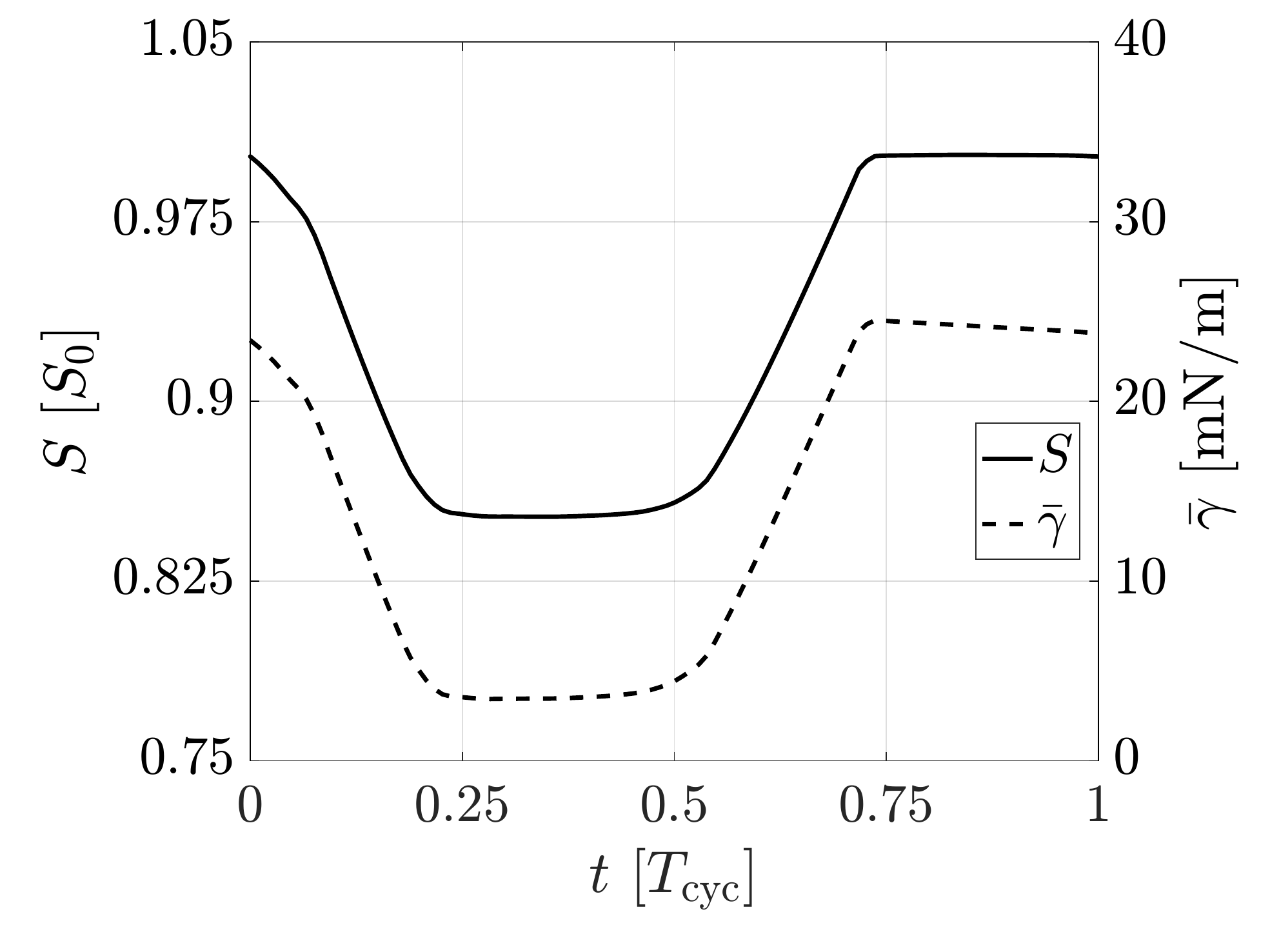}}
\put( 0.1,0.6){d)}
\end{picture}
\caption{The CSD test: FE results of the CR model showing the change of the average surface tension $\bar\gamma$ and the relative surface area $S/S_0$ vs.~time. Figs. a-d correspond to the cases 1-4 in Tab.~\ref{t:CRM_par}, respectively.}
\label{f:csd_time}
\end{center}
\end{figure}
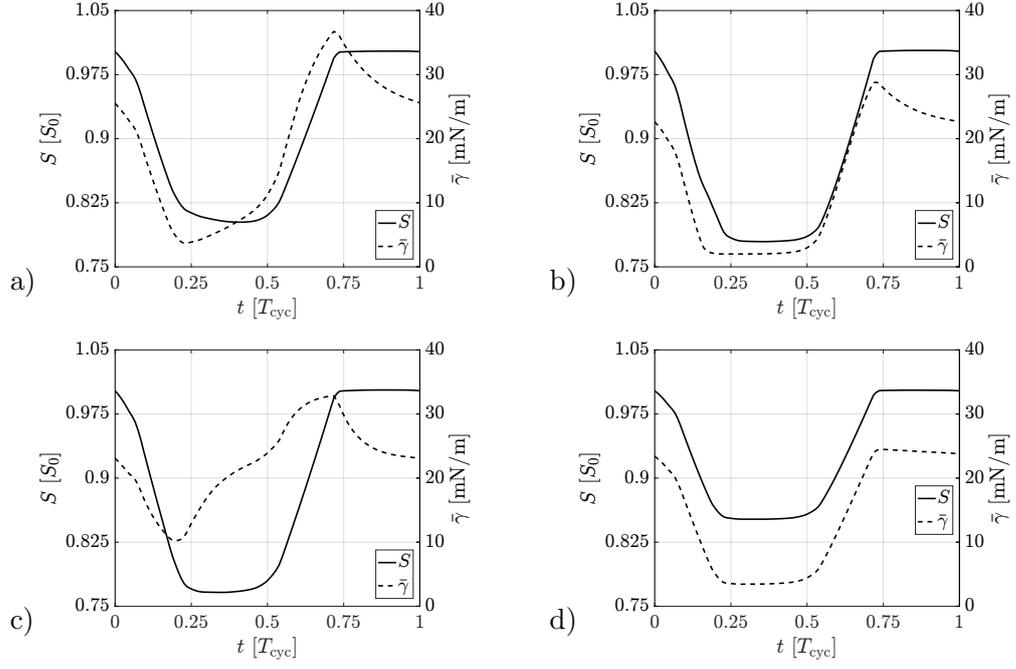
\begin{figure}[ht!]
\begin{center} \unitlength1cm
\unitlength1cm
\begin{picture}(0,9.0)
\put(-6.8,4.5){\includegraphics[height=45mm]{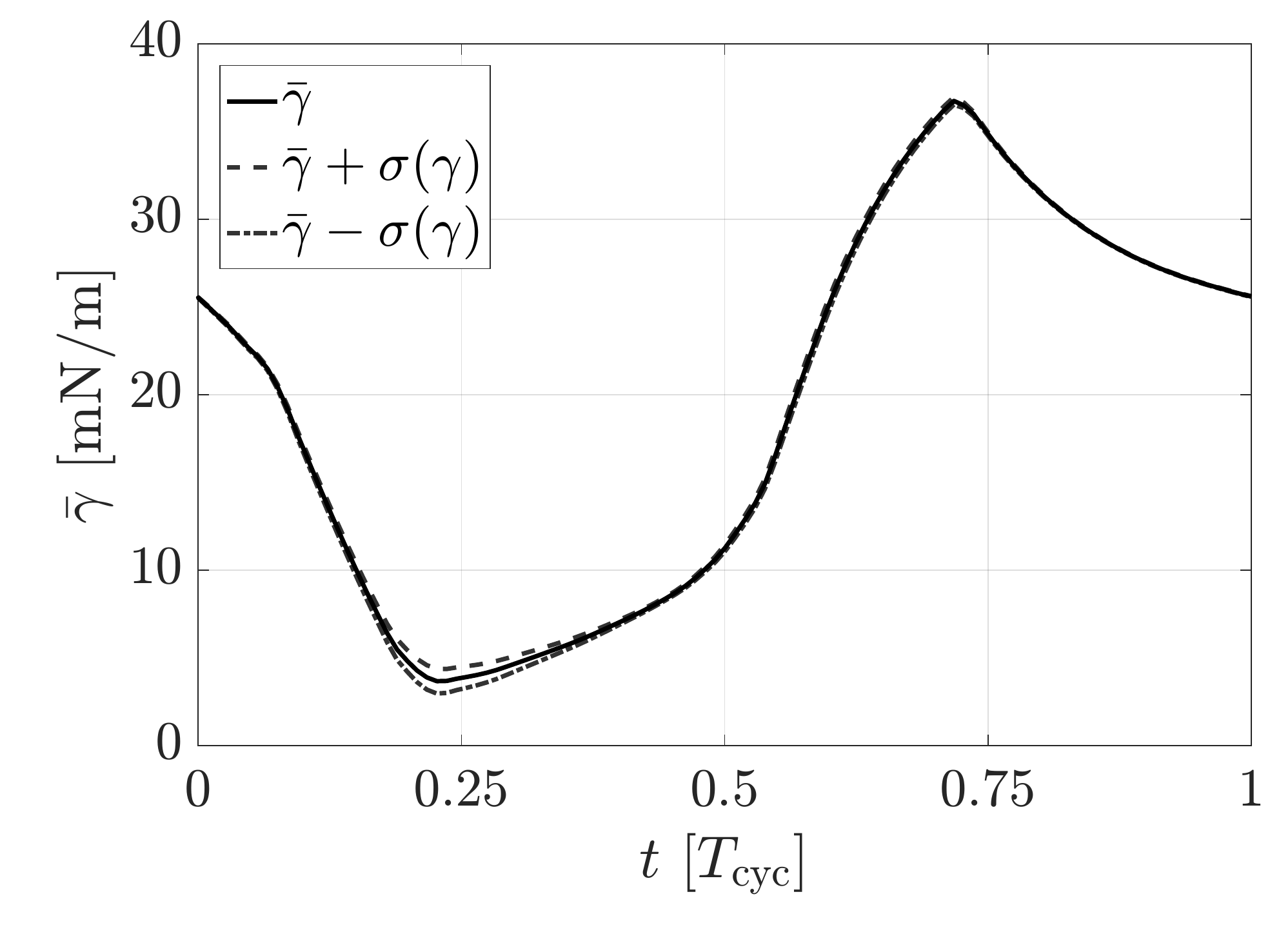}}
\put(-7.0,5.1){a)}
\put( 0.3,4.5){\includegraphics[height=45mm]{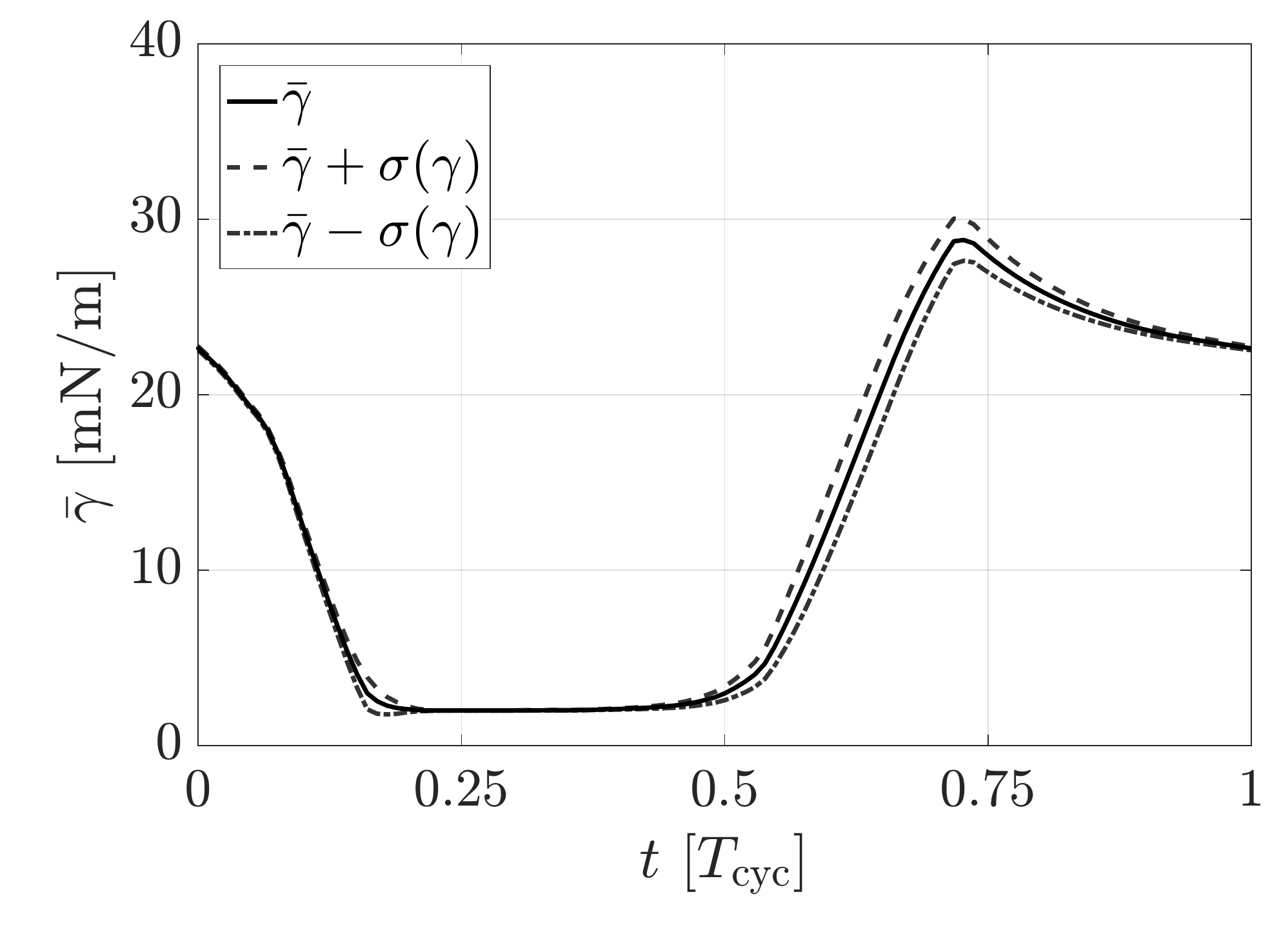}}
\put( 0.1,5.1){b)}
\put(-6.8,0.0){\includegraphics[height=45mm]{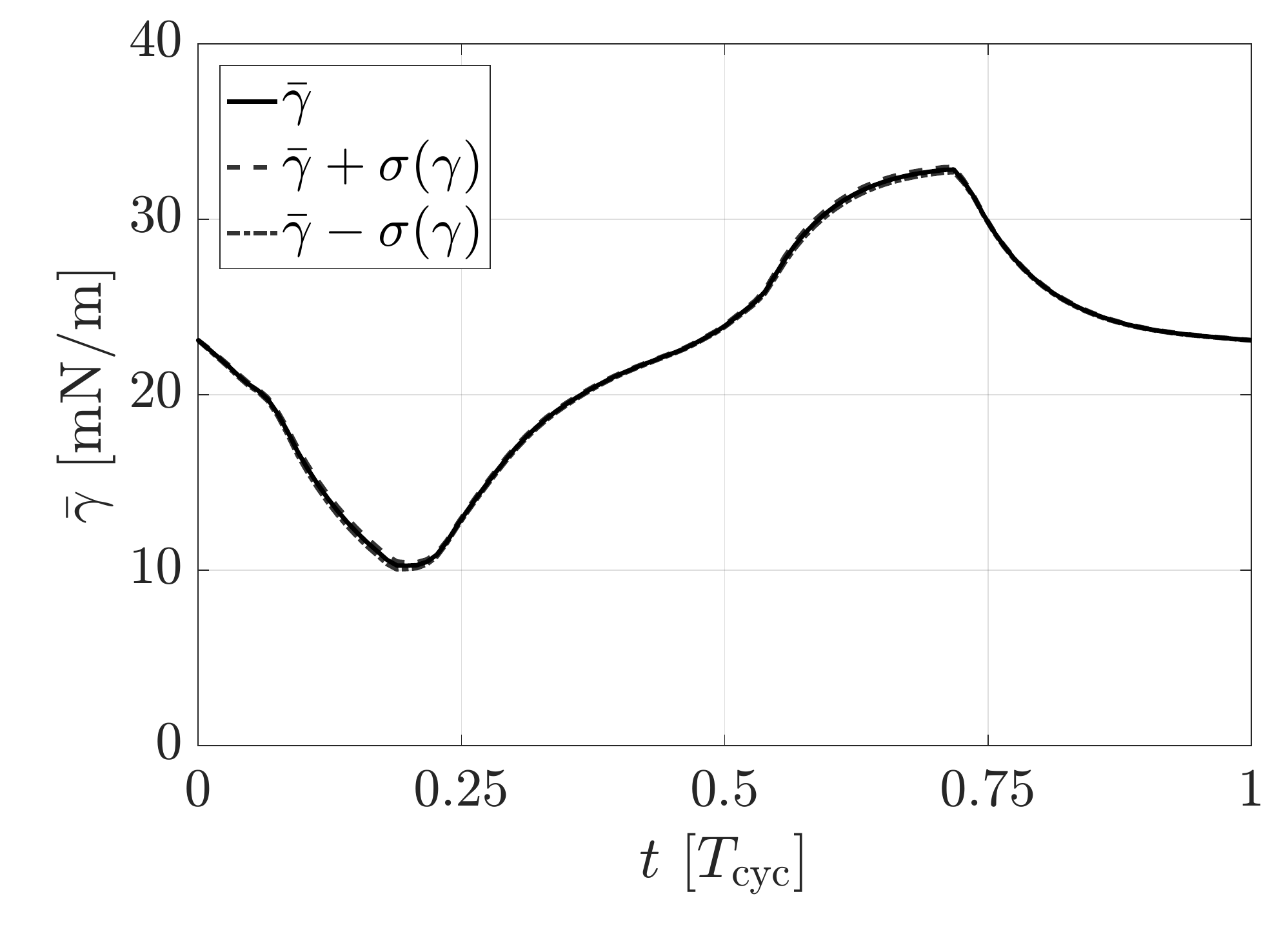}}
\put(-7.0,0.6){c)}
\put( 0.3,0.0){\includegraphics[height=45mm]{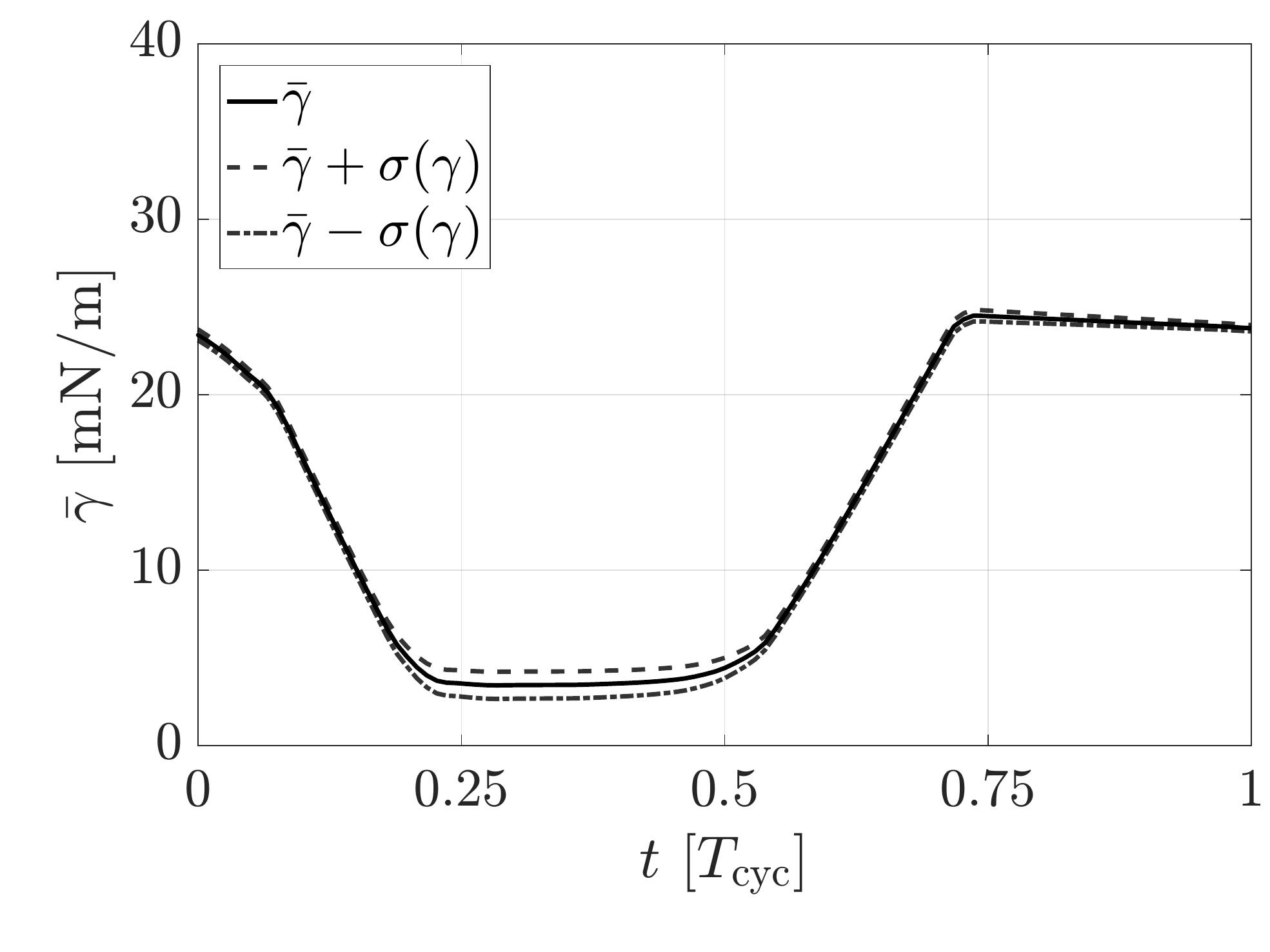}}
\put( 0.1,0.6){d)}
\end{picture}
\caption{The CSD test: FE results of the CR model showing the change of the average surface tension $\bar\gamma$ and the standard deviation of the surface tension $\sigma(\gamma)$ vs.~time. Figs. a-d correspond to the cases 1-4 in Tab.~\ref{t:CRM_par}, respectively.}
\label{f:csd_gam}
\end{center}
\end{figure}
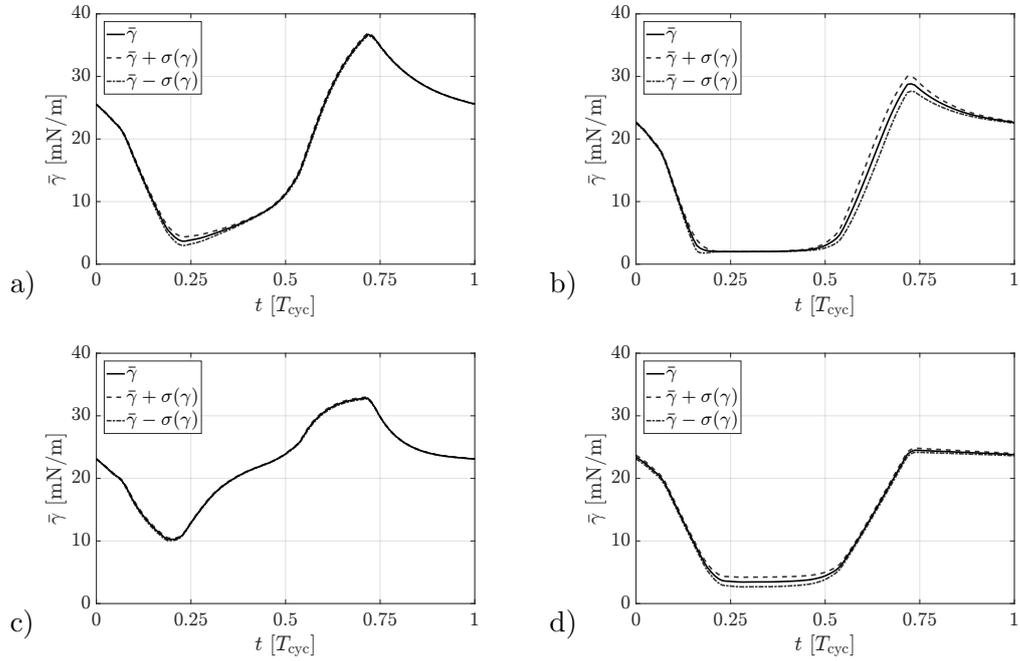

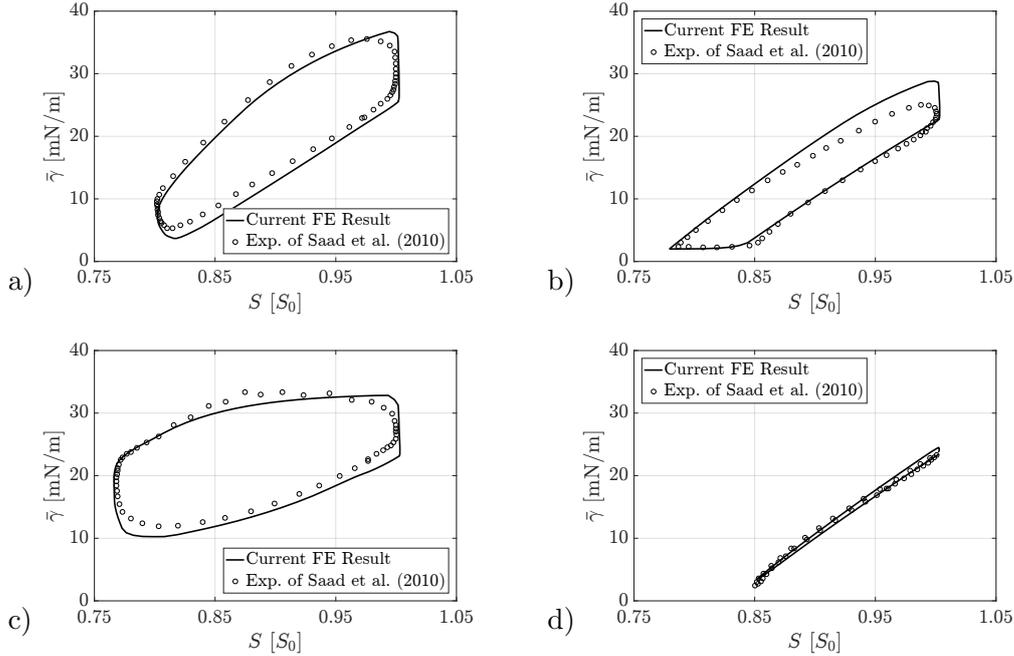
\begin{figure}[ht!]
\begin{center} \unitlength1cm
\unitlength1cm
\begin{picture}(0,9.0)
\put(-6.8,4.5){\includegraphics[height=45mm]{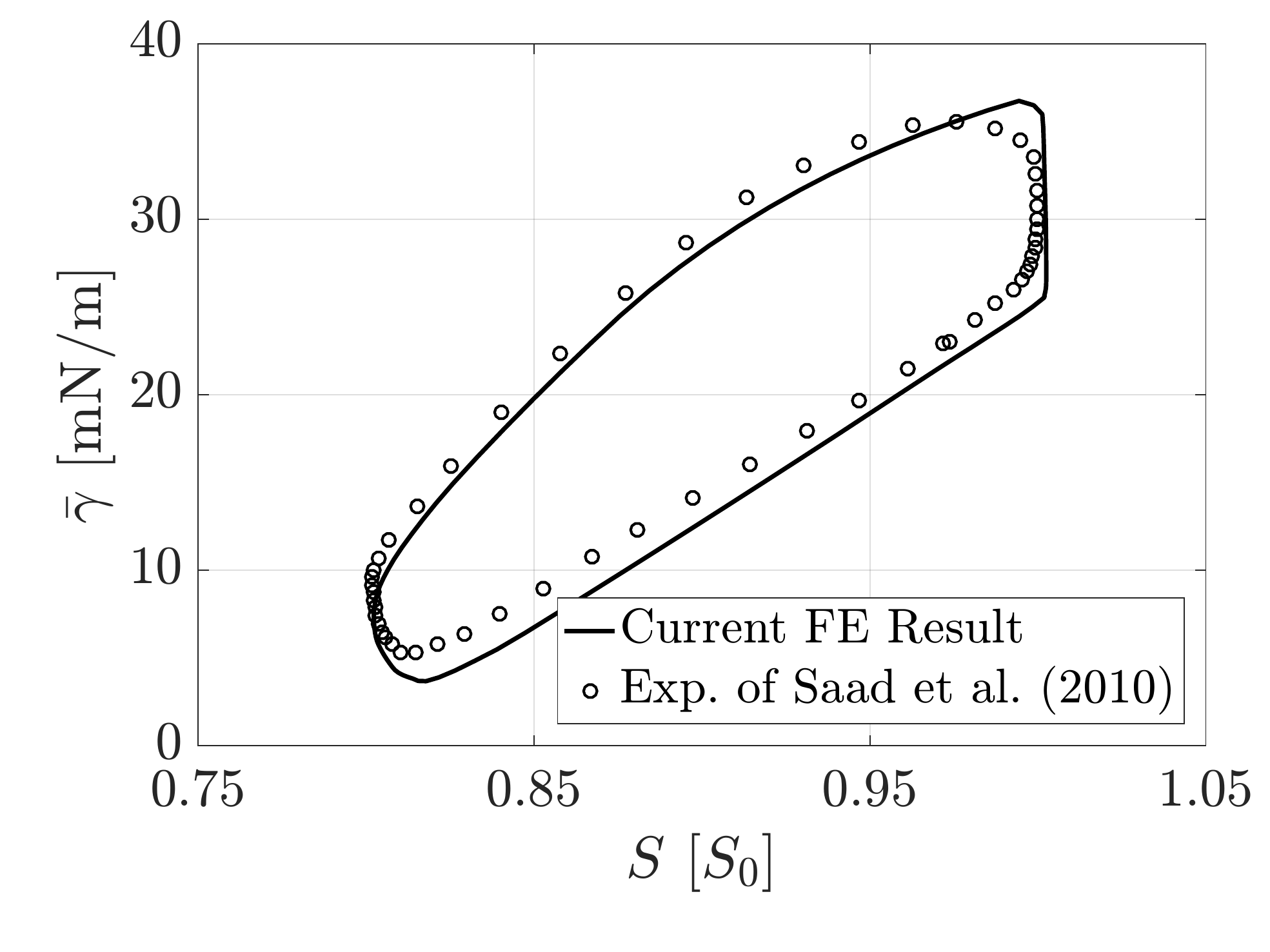}}
\put(-7.0,5.1){a)}
\put( 0.3,4.5){\includegraphics[height=45mm]{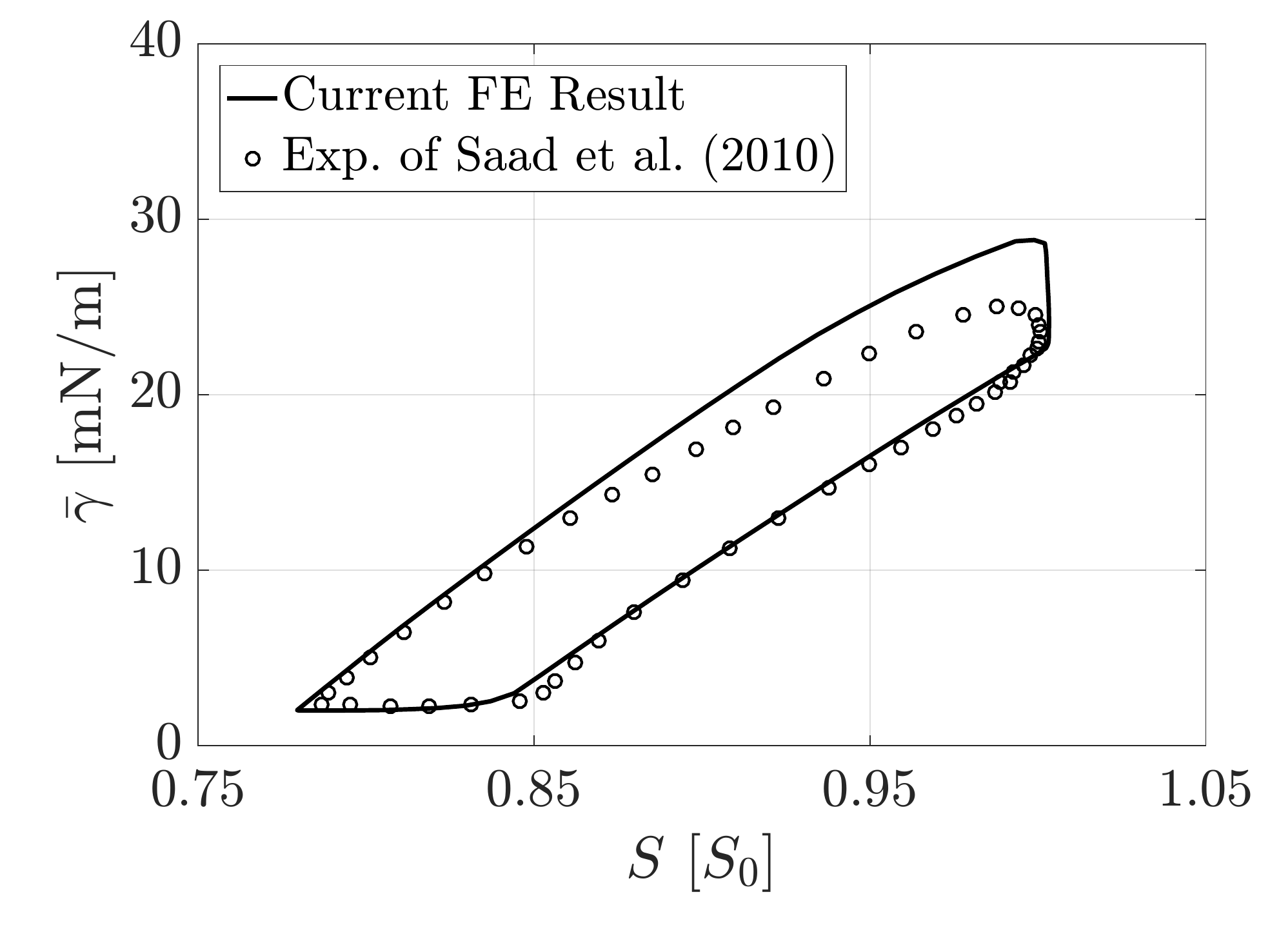}}
\put( 0.1,5.1){b)}
\put(-6.8,0.0){\includegraphics[height=45mm]{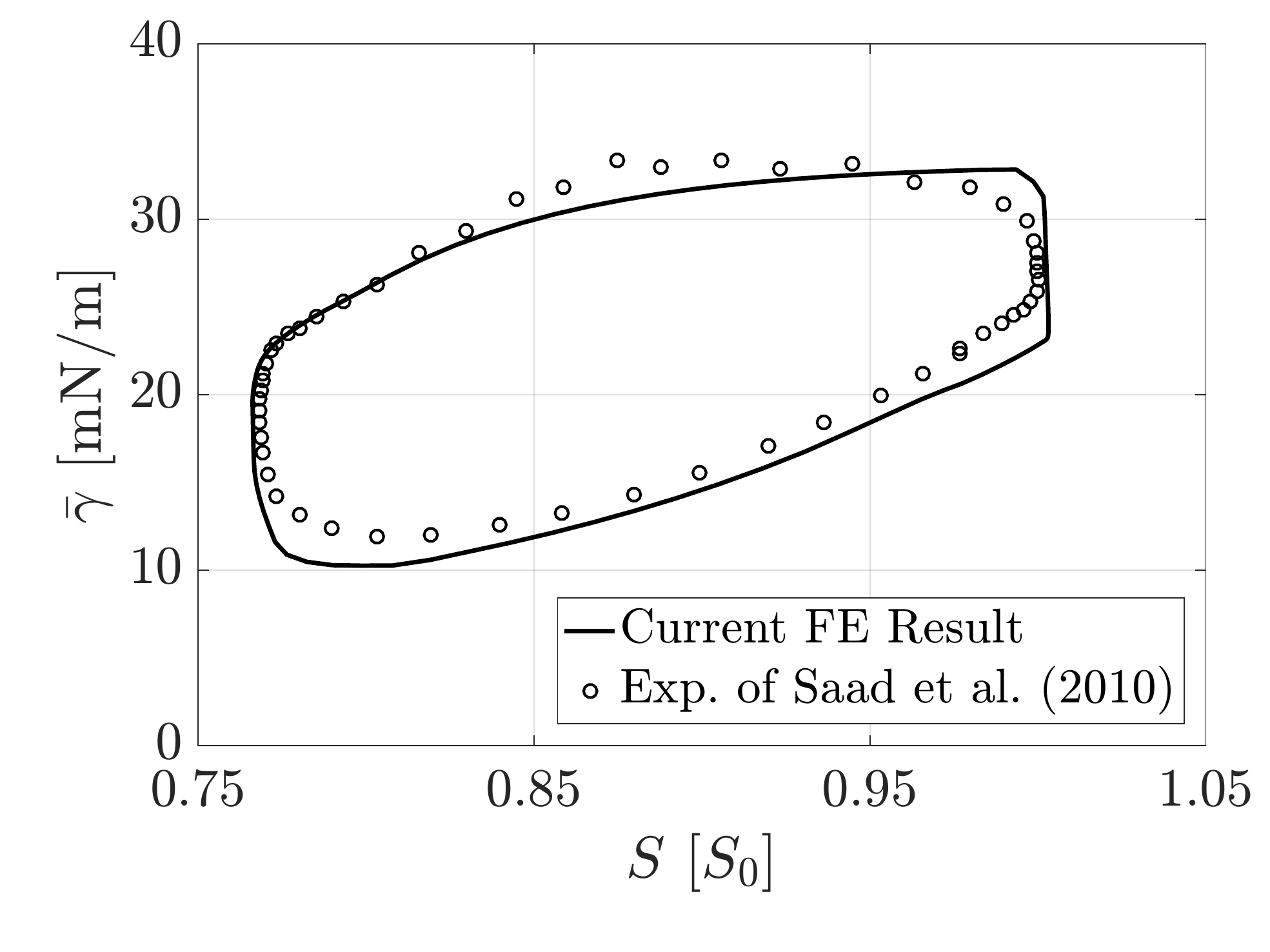}}
\put(-7.0,0.6){c)}
\put( 0.3,0.0){\includegraphics[height=45mm]{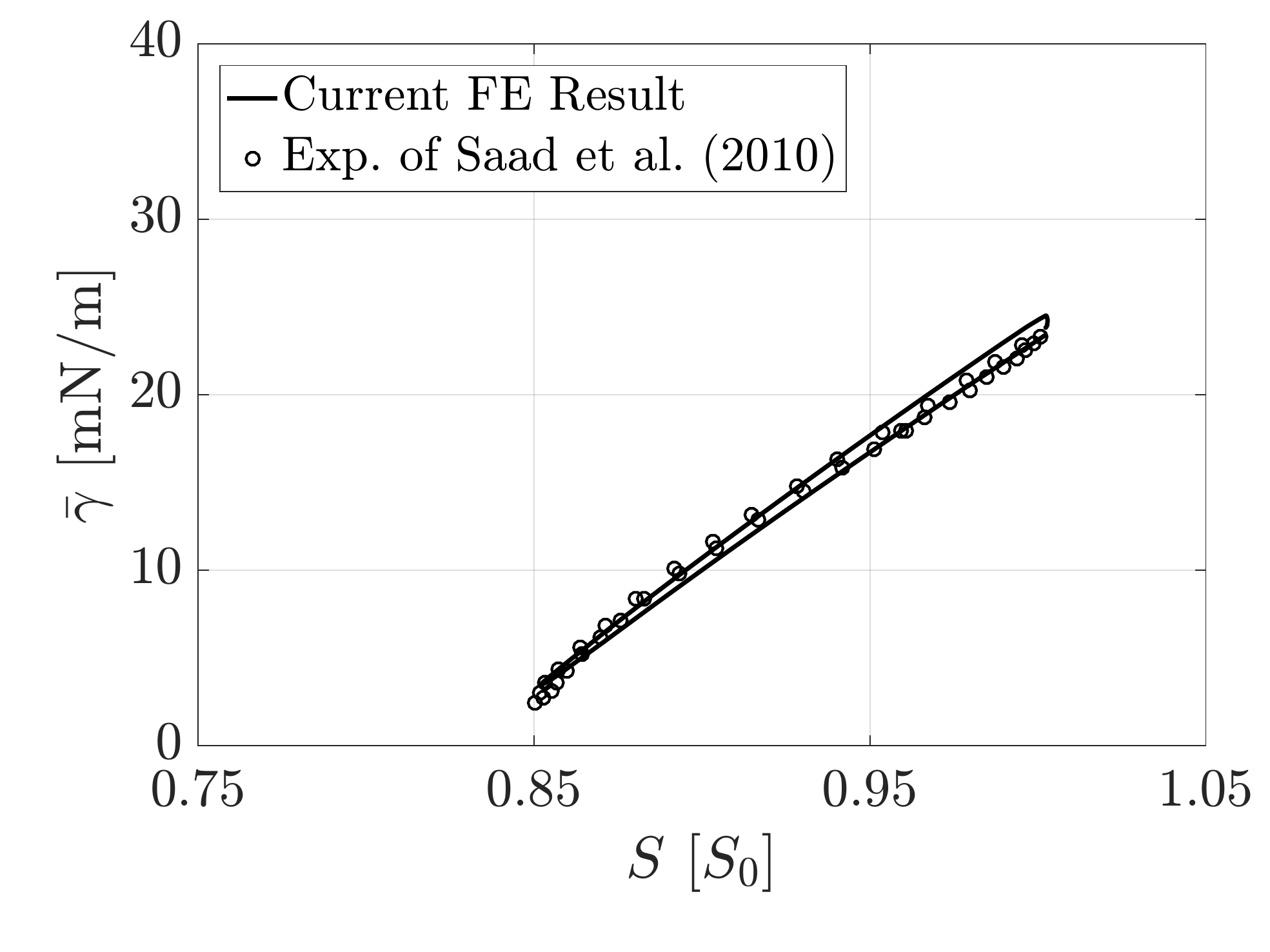}}
\put( 0.1,0.6){d)}
\end{picture}
\caption{The CSD test: FE results of the CR model showing the change of the average surface tension $\bar\gamma$ vs.~the relative surface area $S/S_0$, compared with the experimental results of \citet{saad10}. Figs. a-d correspond to the cases 1-4 in Tab.~\ref{t:CRM_par}, respectively. The loop orientation is clockwise.}
\label{f:csd_iso}
\end{center}
\end{figure}

In Fig.~\ref{f:csd_iso}, the corresponding isotherm is plotted for different materials. The simulation results are compared with the experimental results of \citet{saad10}. The qualitative agreement between our FE simulations and the experiments of \citet{saad10} is very good. However there are small differences. These differences can have different reasons: For example, all the required parameters to simulate the test exactly as in the experiment are not provided in the literature. We have estimated some variables such as $\gamma_\mathrm{min}$ and $\gamma_\mathrm{eq}$. Further, the loading pattern used here to control the droplet volume can be slightly different from the loading pattern used in the experiment of \citet{saad10}. 

\FloatBarrier
\subsection{Liquid film expansion/compression}\label{s:film_com}
\begin{figure}[ht!]
\begin{center} \unitlength1cm
\unitlength1cm
\begin{picture}(0,4.5)
\put(-6.8,0.0){\includegraphics[height=45mm]{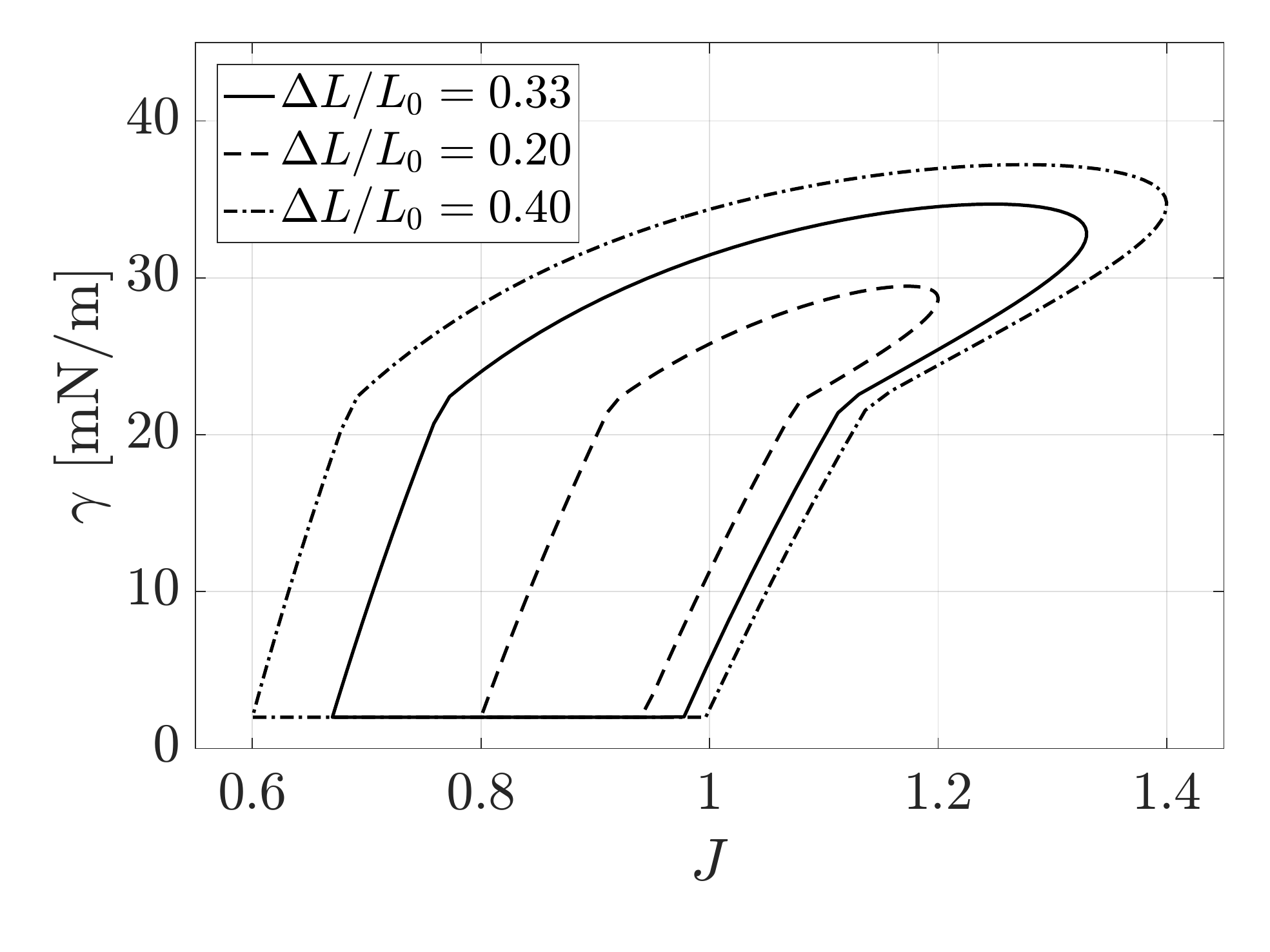}}
\put(-7.0,0.6){a)}
\put( 0.3,0.0){\includegraphics[height=45mm]{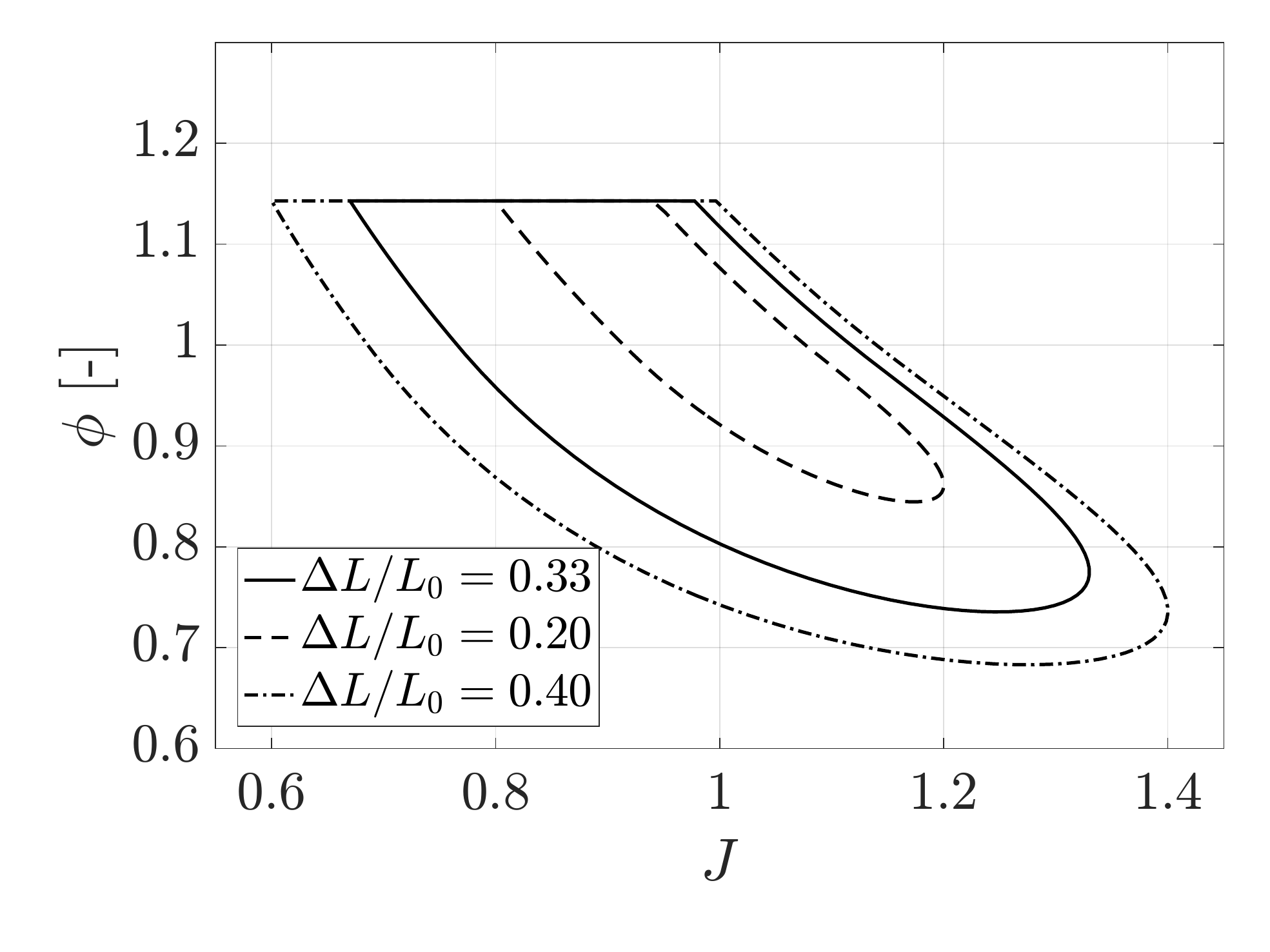}}
\put( 0.1,0.6){b)}
\end{picture}
\caption{Liquid film expansion/compression: Influence of $\Delta L/L_0$: a) Surface tension $\gamma$ vs.~\mbox{surface} area change~$J$ and b) normalized interfacial surfactant concentration $\phi$ vs.~surface area change~$J$. The loop orientation is clockwise.}
\label{f:flat_ALM_Aamp}
\end{center}
\end{figure}

In the \textit{Langmuir--Wilhelmy balance} \citep{butt06} a movable barrier controls the total area available for a thin film of interfacial molecules, and the surface tension is measured from the wetting force acting over the Wilhelmy plate. Similarly, here a thin liquid film is modeled by a $2\,L_0 \times L_0$ rectangular membrane, where $L_0$ is a length scale. The bulk effects are not considered and only the free surface is modeled. As shown in Fig.~\ref{f:comp_model}.b, three edges of the membrane are fixed as zero-displacement Dirichlet boundaries and the forth edge is controlled by a prescribed Dirichlet boundary condition $\bar\bu\,\hat{=}\,[\bar u_x,~0,~0]^\mrT$ that changes with time as
\eqb{l}
\bar u_x(t) = \Delta L\,\sin\big(t/T\big)~,
\eqe
where $T=3~\mrs$ and the loading amplitude is $\Delta L = 0.33\,L_0$ unless it is changed, e.g. in Fig.~\ref{f:flat_ALM_Aamp}.
In this example, the behavior of the surfactant monolayer is described by the AL model of Sec.~\ref{s:ALM}. The material constants are set  as listed in Tab.~\ref{t:ALM_par} following \citet{wiechert11thesis}.
\begin{table}[ht!]
\centering
{\renewcommand{\arraystretch}{1.2}
\begin{tabular}{ cccccc }
  \hline
  $K_1~[\mrs^{-1}]$ & $k_2~[\mrs^{-1}]$ & $m_1~[\mrm\mrN/\mrm]$ & $m_2~[\mrm\mrN/\mrm]$ & $\gamma_\mathrm{min}~[\mrm\mrN/\mrm]$ & $\gamma_0~[\mrm\mrN/\mrm]$ \\
  \hline
  1 & 0.016 & 48 & 140 & 10 & 70 \\
  \hline
\end{tabular}
\caption{Parameters of the {AL} model \citep{wiechert11thesis}}
\label{t:ALM_par}}
\end{table}

\begin{figure}[ht!]
\begin{center} \unitlength1cm
\unitlength1cm
\begin{picture}(0,4.5)
\put(-6.8,0.0){\includegraphics[height=45mm]{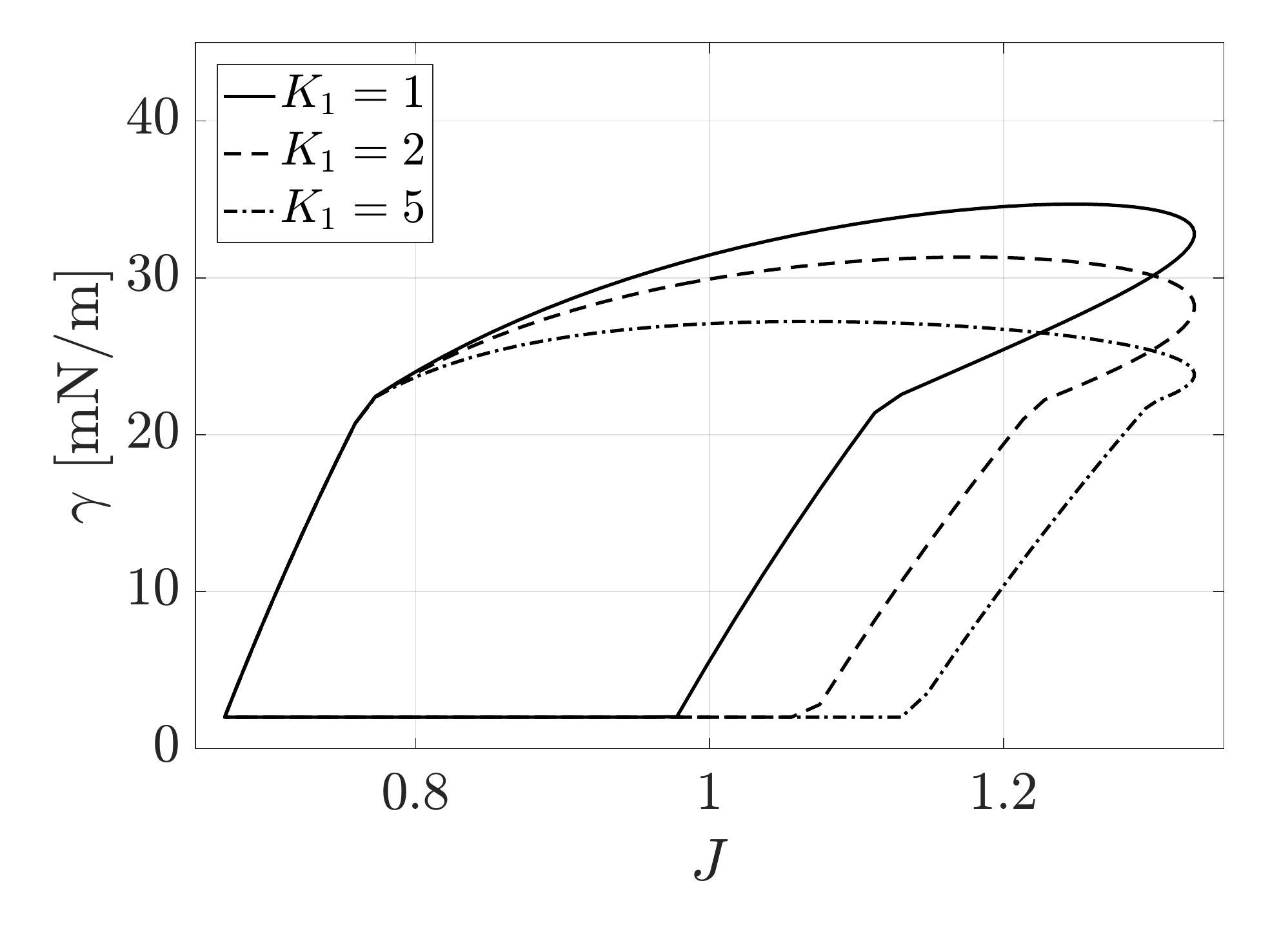}}
\put(-7.0,0.6){a)}
\put( 0.3,0.0){\includegraphics[height=45mm]{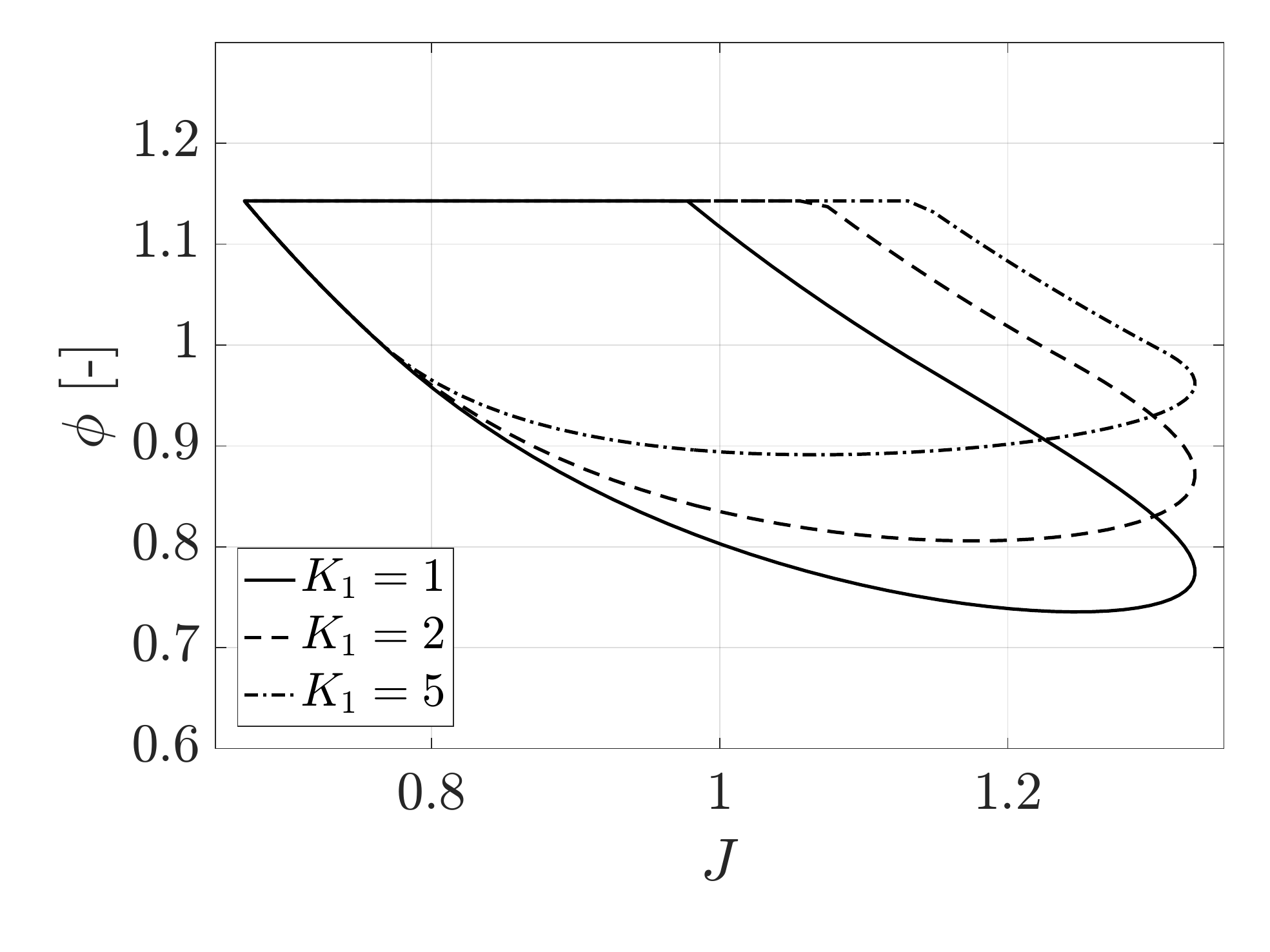}}
\put( 0.1,0.6){b)}
\end{picture}
\caption{Liquid film expansion/compression: Influence of $K_1$: a) Surface tension $\gamma$ vs.~\mbox{surface} area change~$J$ and b) normalized interfacial surfactant concentration $\phi$ vs.~surface area change~$J$. The unit of the legend is $\mrs^{-1}$ and the loop orientation is clockwise.}
\label{f:flat_ALM_K1}
\end{center}
\end{figure}

\begin{figure}[ht!]
\begin{center} \unitlength1cm
\unitlength1cm
\begin{picture}(0,4.5)
\put(-6.8,0.0){\includegraphics[height=45mm]{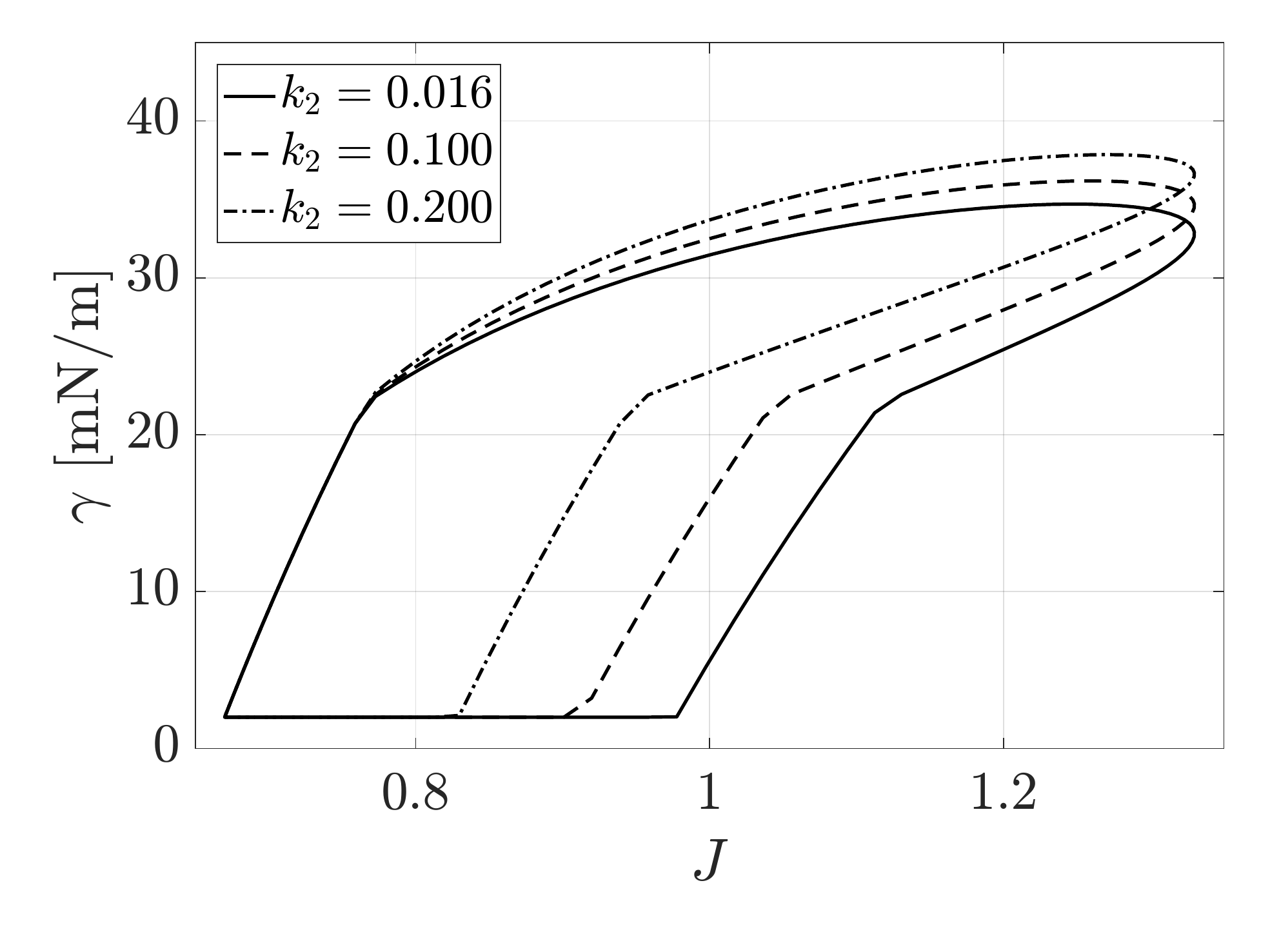}}
\put(-7.0,0.6){a)}
\put( 0.3,0.0){\includegraphics[height=45mm]{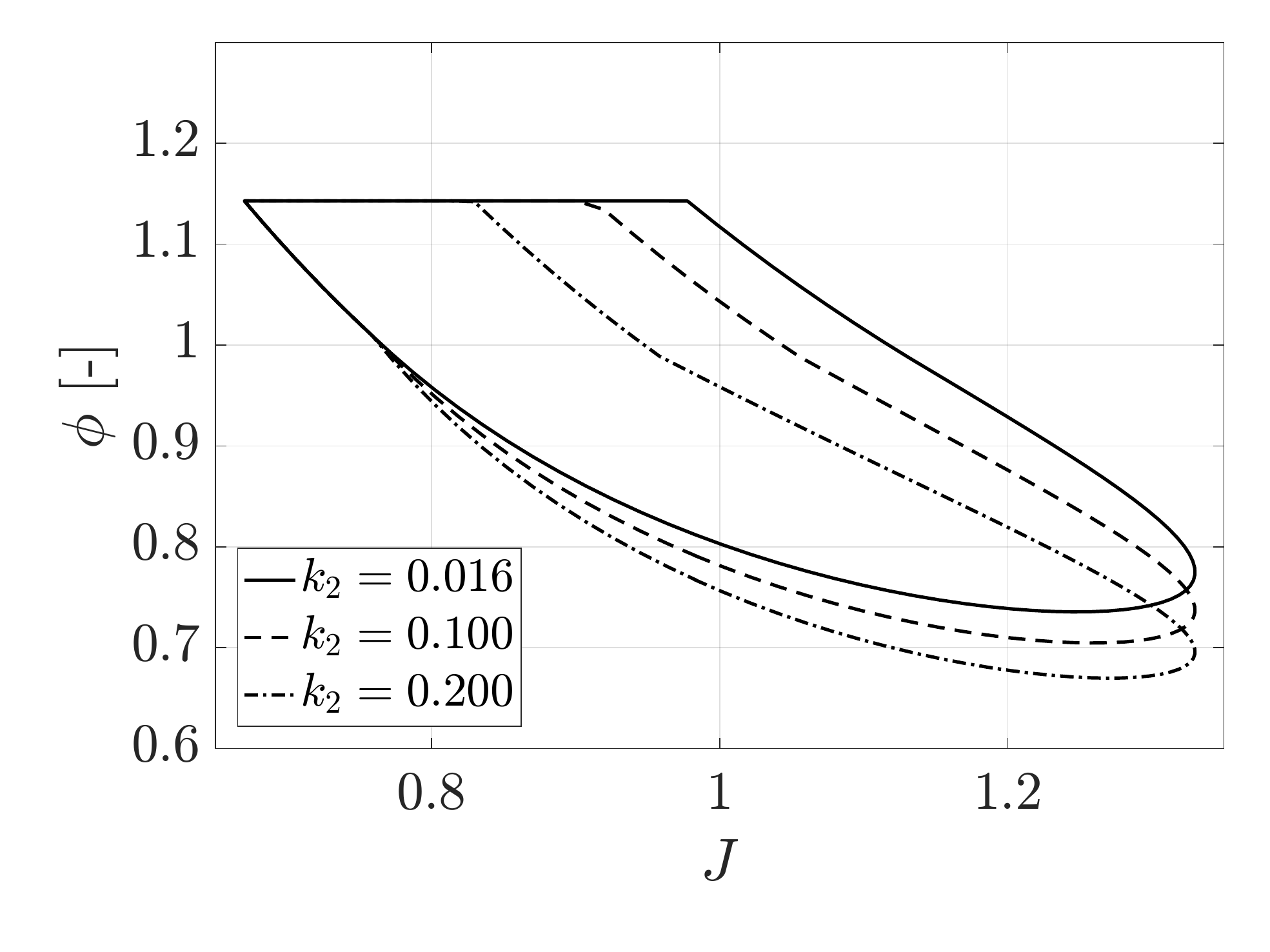}}
\put( 0.1,0.6){b)}
\end{picture}
\caption{Liquid film expansion/compression: Influence of $k_2$: a) Surface tension $\gamma$ vs.~\mbox{surface} area change~$J$ and b) normalized interfacial surfactant concentration $\phi$ vs.~surface area change~$J$. The unit of the legend is $\mrs^{-1}$ and the loop orientation is clockwise.}
\label{f:flat_ALM_k2}
\end{center}
\end{figure}

\begin{figure}[ht!]
\begin{center} \unitlength1cm
\unitlength1cm
\begin{picture}(0,4.5)
\put(-6.8,0.0){\includegraphics[height=45mm]{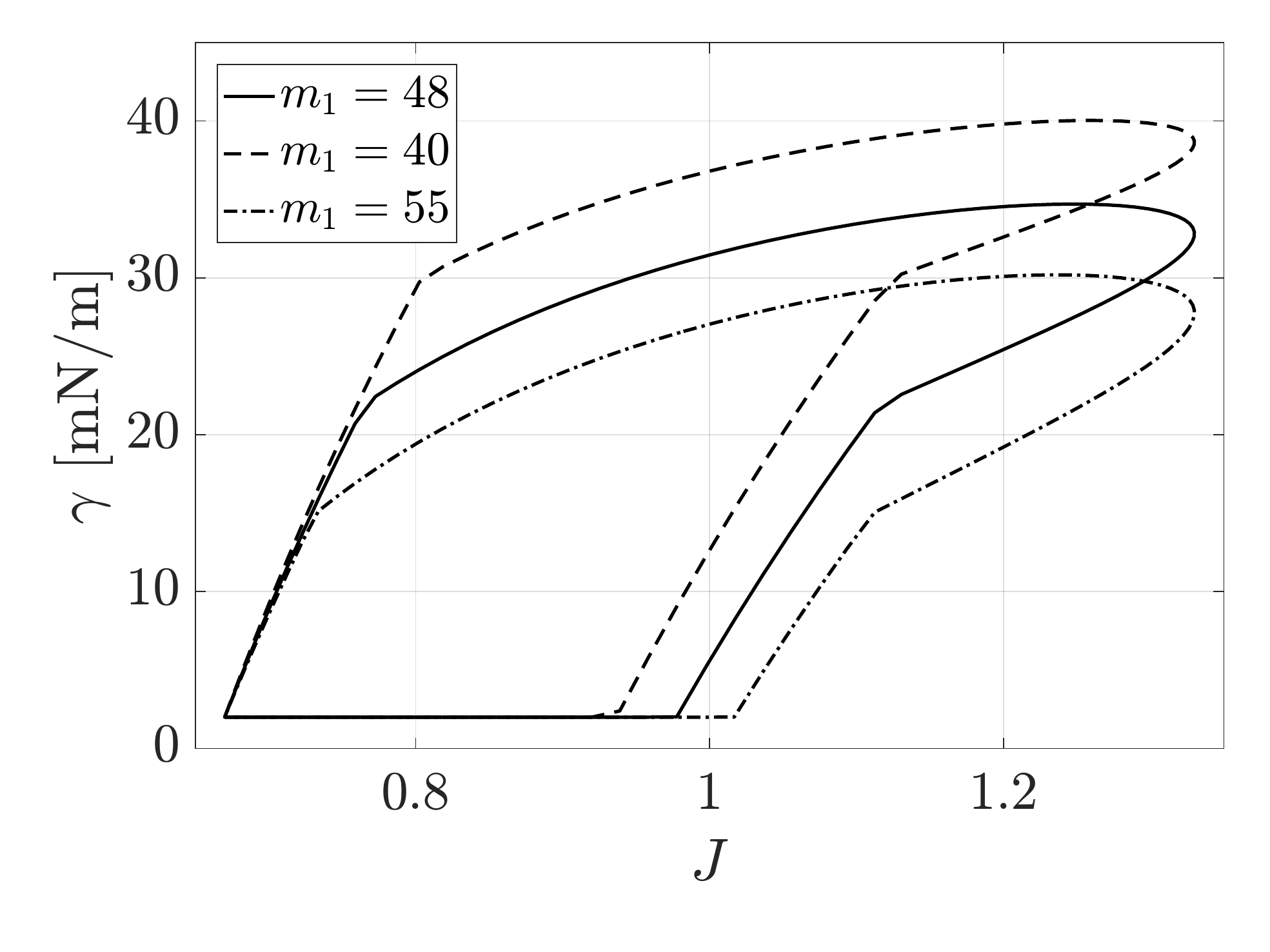}}
\put(-7.0,0.6){a)}
\put( 0.3,0.0){\includegraphics[height=45mm]{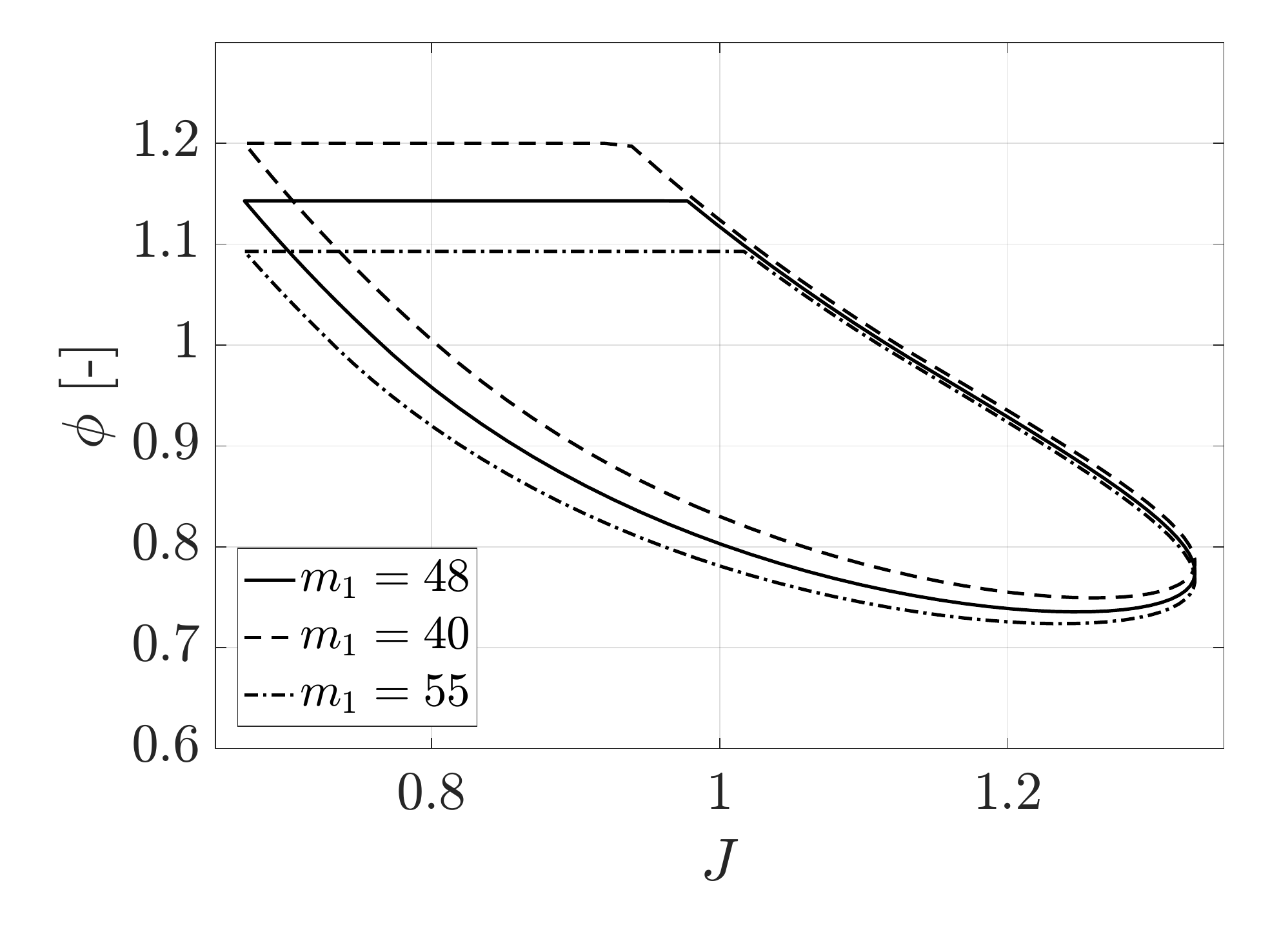}}
\put( 0.1,0.6){b)}
\end{picture}
\caption{Liquid film expansion/compression: Influence of $m_1$: a) Surface tension $\gamma$ vs.~\mbox{surface} area change~$J$ and b) normalized interfacial surfactant concentration $\phi$ vs.~surface area change~$J$. The unit of the legend is mN/m and the loop orientation is clockwise.}
\label{f:flat_ALM_m1}
\end{center}
\end{figure}

\begin{figure}[ht!]
\begin{center} \unitlength1cm
\unitlength1cm
\begin{picture}(0,4.5)
\put(-6.8,0.0){\includegraphics[height=45mm]{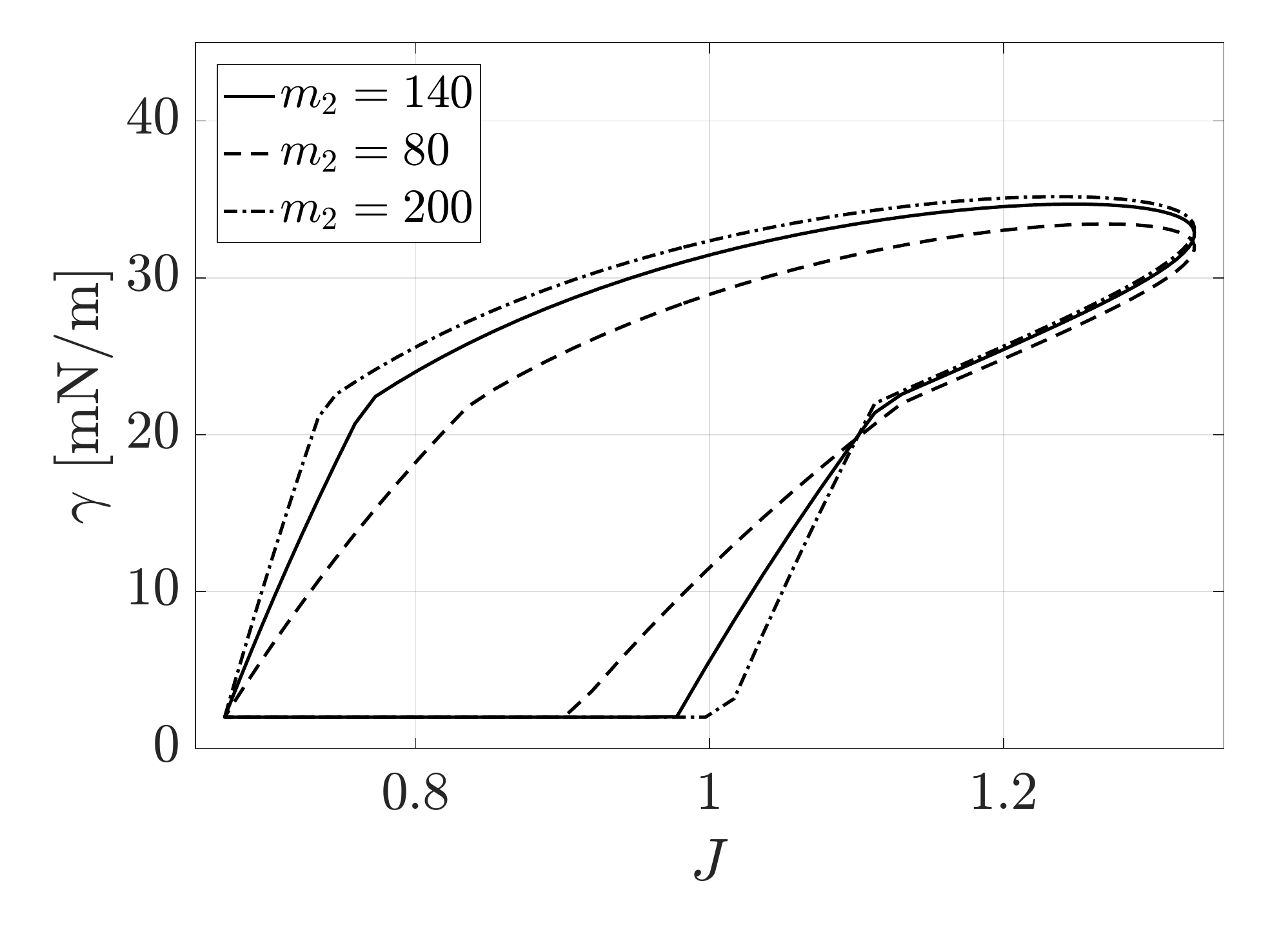}}
\put(-7.0,0.6){a)}
\put( 0.3,0.0){\includegraphics[height=45mm]{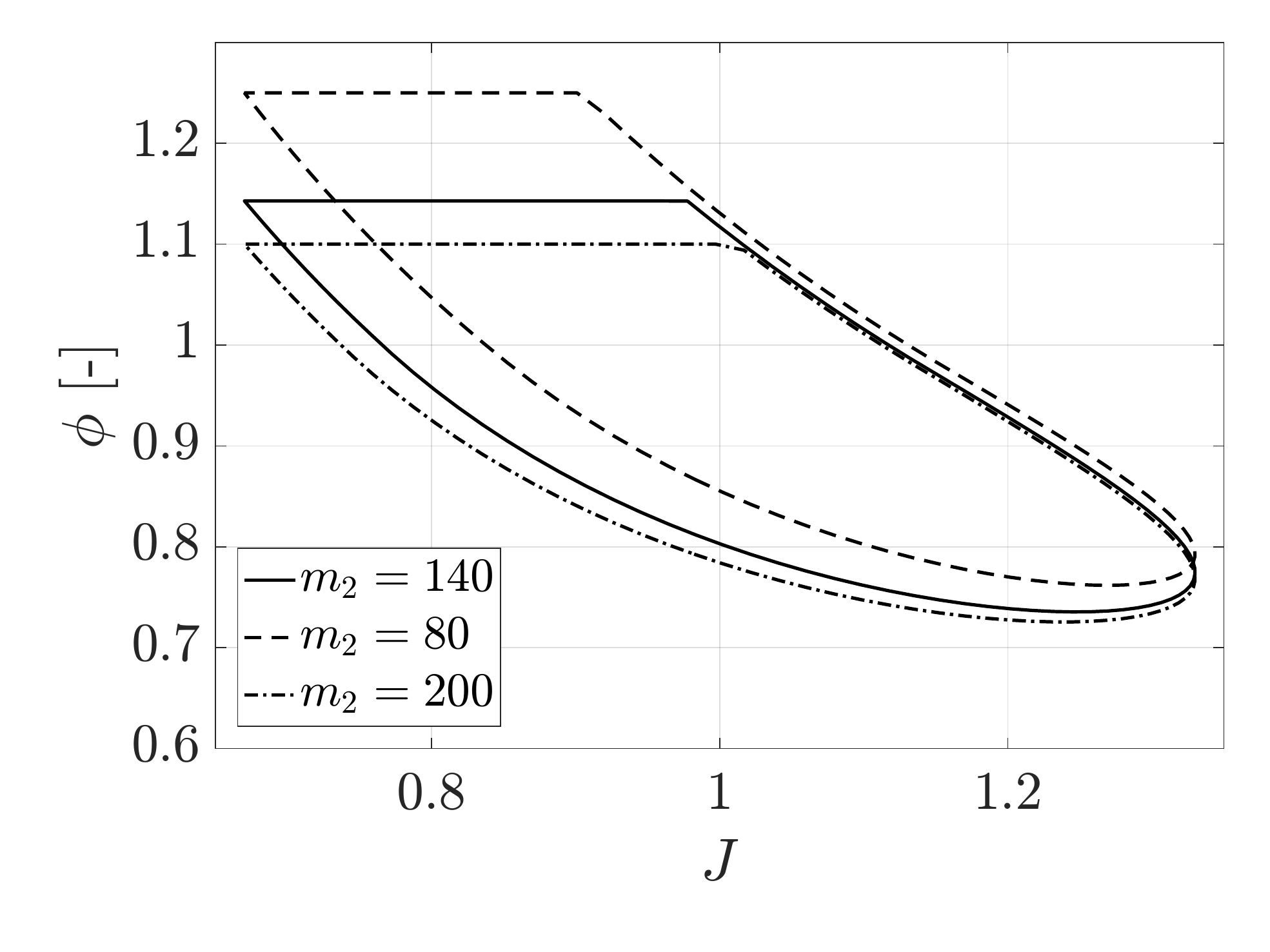}}
\put( 0.1,0.6){b)}
\end{picture}
\caption{Liquid film expansion/compression: Influence of $m_2$: a) Surface tension $\gamma$ vs.~\mbox{surface} area change~$J$ and b) normalized interfacial surfactant concentration $\phi$ vs.~surface area change~$J$. The unit of the legend is mN/m and the loop orientation is clockwise.}
\label{f:flat_ALM_m2}
\end{center}
\end{figure}

\begin{figure}[ht!]
\begin{center} \unitlength1cm
\unitlength1cm
\begin{picture}(0,4.5)
\put(-6.8,0.0){\includegraphics[height=45mm]{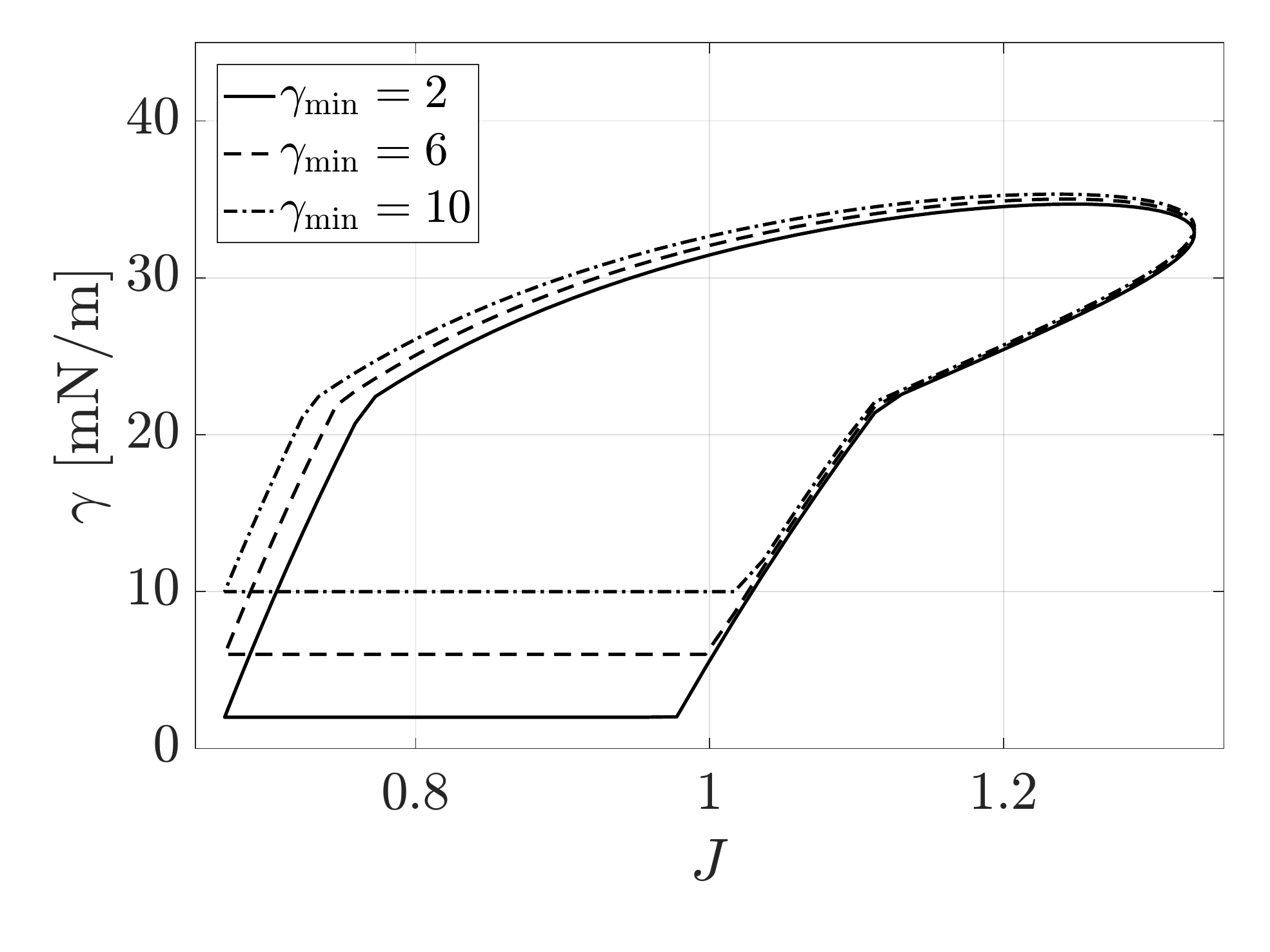}}
\put(-7.0,0.6){a)}
\put( 0.3,0.0){\includegraphics[height=45mm]{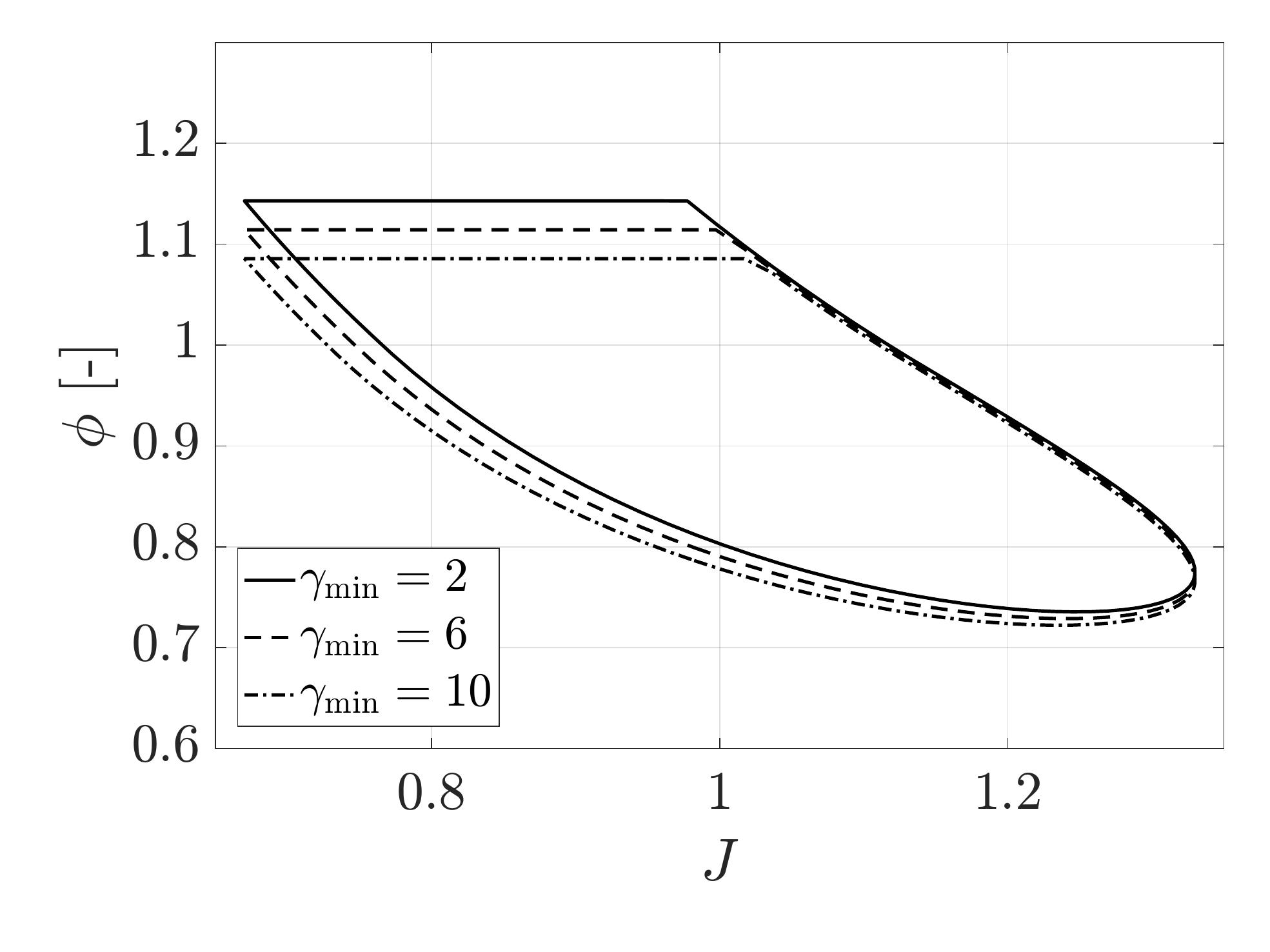}}
\put( 0.1,0.6){b)}
\end{picture}
\caption{Liquid film expansion/compression: Influence of $\gamma_\mathrm{min}$: a) Surface tension $\gamma$ vs.~\mbox{surface} area change~$J$ and b) normalized interfacial surfactant concentration $\phi$ vs.~surface area change~$J$. The unit of the legend is mN/m and the loop orientation is clockwise.}
\label{f:flat_ALM_gmin}
\end{center}
\end{figure}

In Figs.~\ref{f:flat_ALM_K1} to \ref{f:flat_ALM_Aamp}, the changes of surface tension $\gamma$ and the normalized interfacial surfactant concentration $\phi$ are plotted against the surface area change $J$ with different values for the model parameters. As is expected, our finite element results are in agreement with the finite element results of \citet{wiechert11thesis}.  

\FloatBarrier
\subsection{Relaxation of a liquid film}\label{s:film_rel}
\begin{figure}[ht]
\begin{center} \unitlength1cm
\unitlength1cm
\begin{picture}(0,9.0)
\put(-6.8,4.5){\includegraphics[height=45mm]{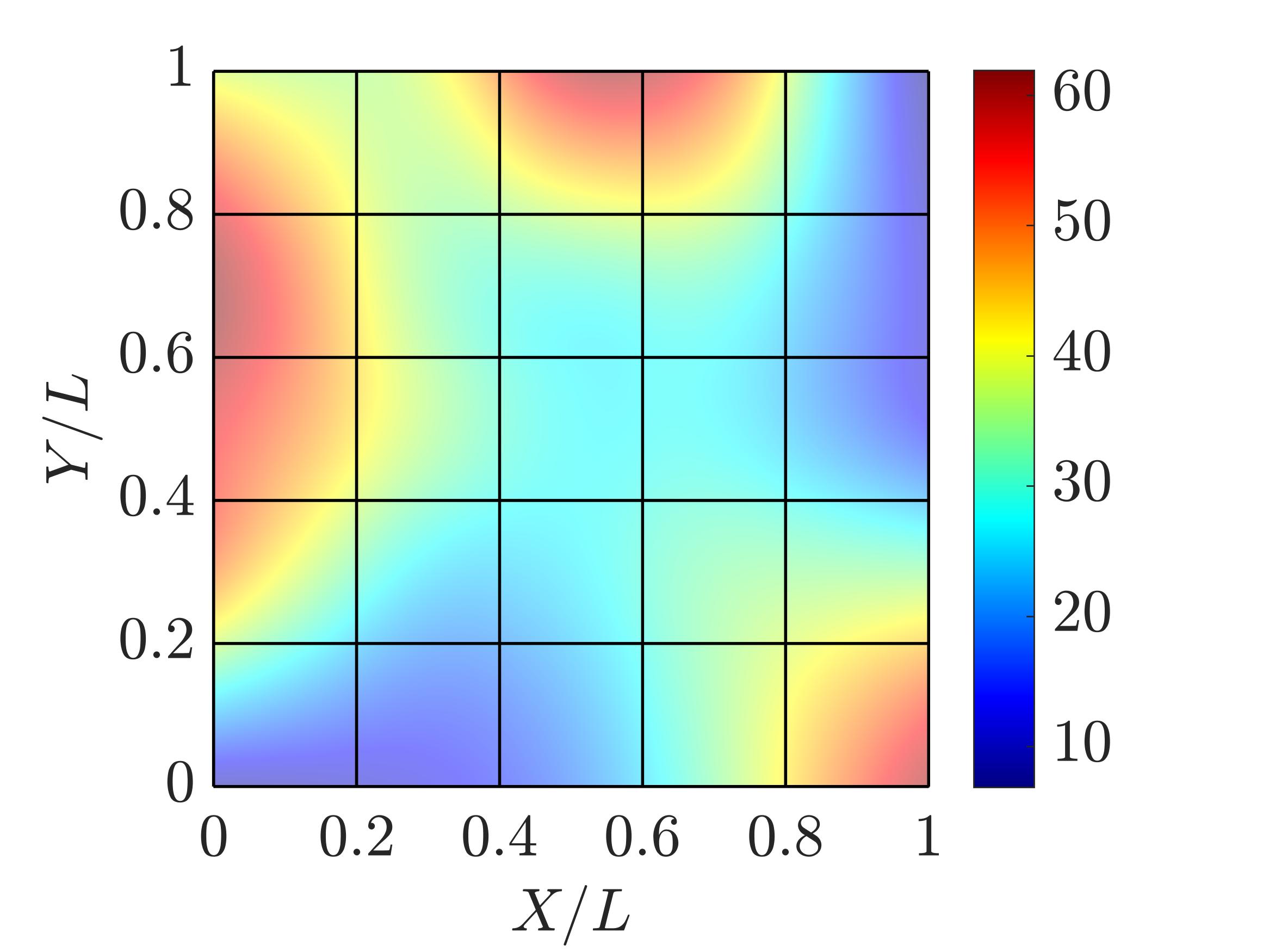}}
\put(-7.0,5.1){a)}
\put( 0.3,4.5){\includegraphics[height=45mm]{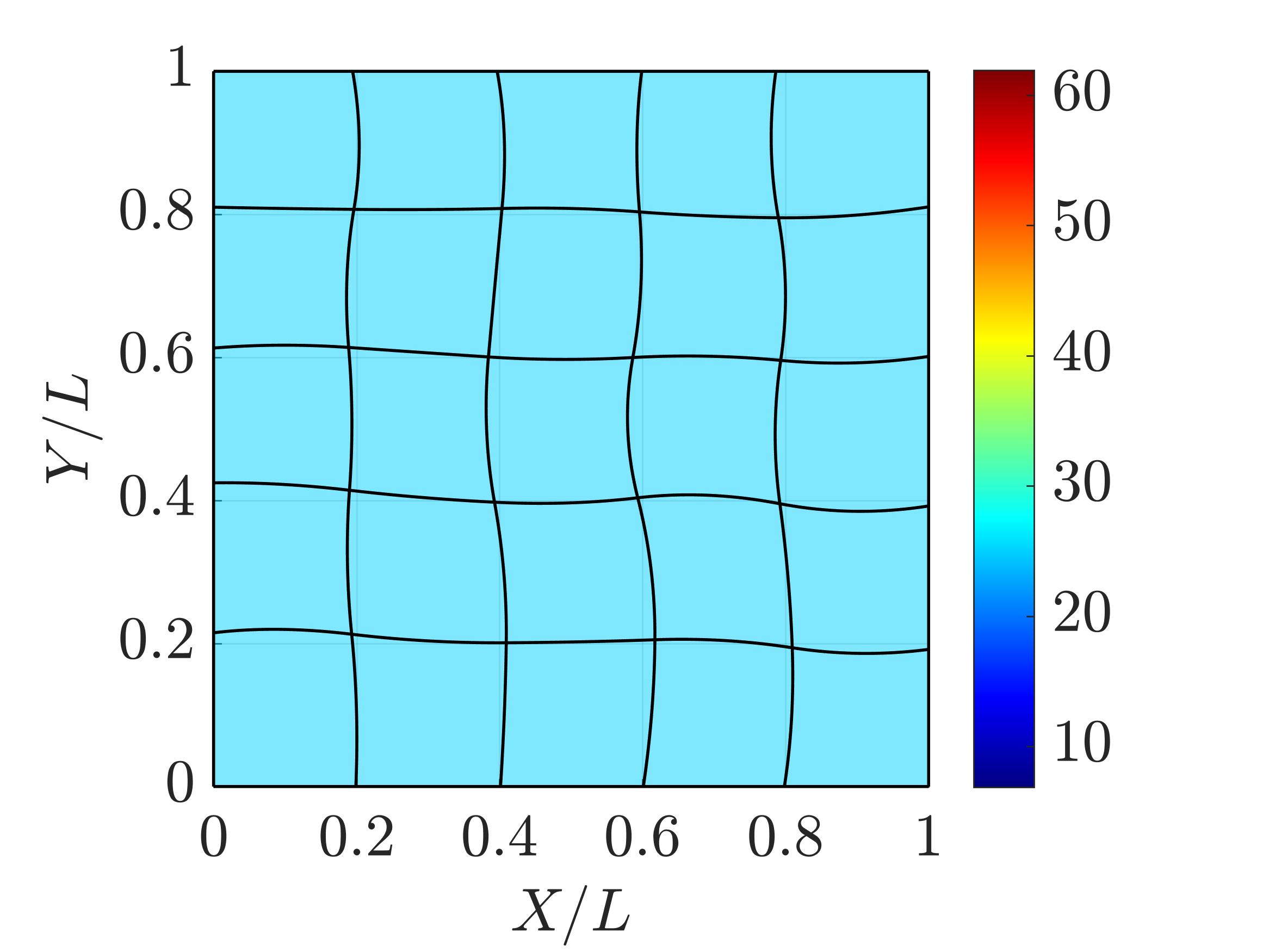}}
\put( 0.1,5.1){b)}
\put(-6.8,0.0){\includegraphics[height=45mm]{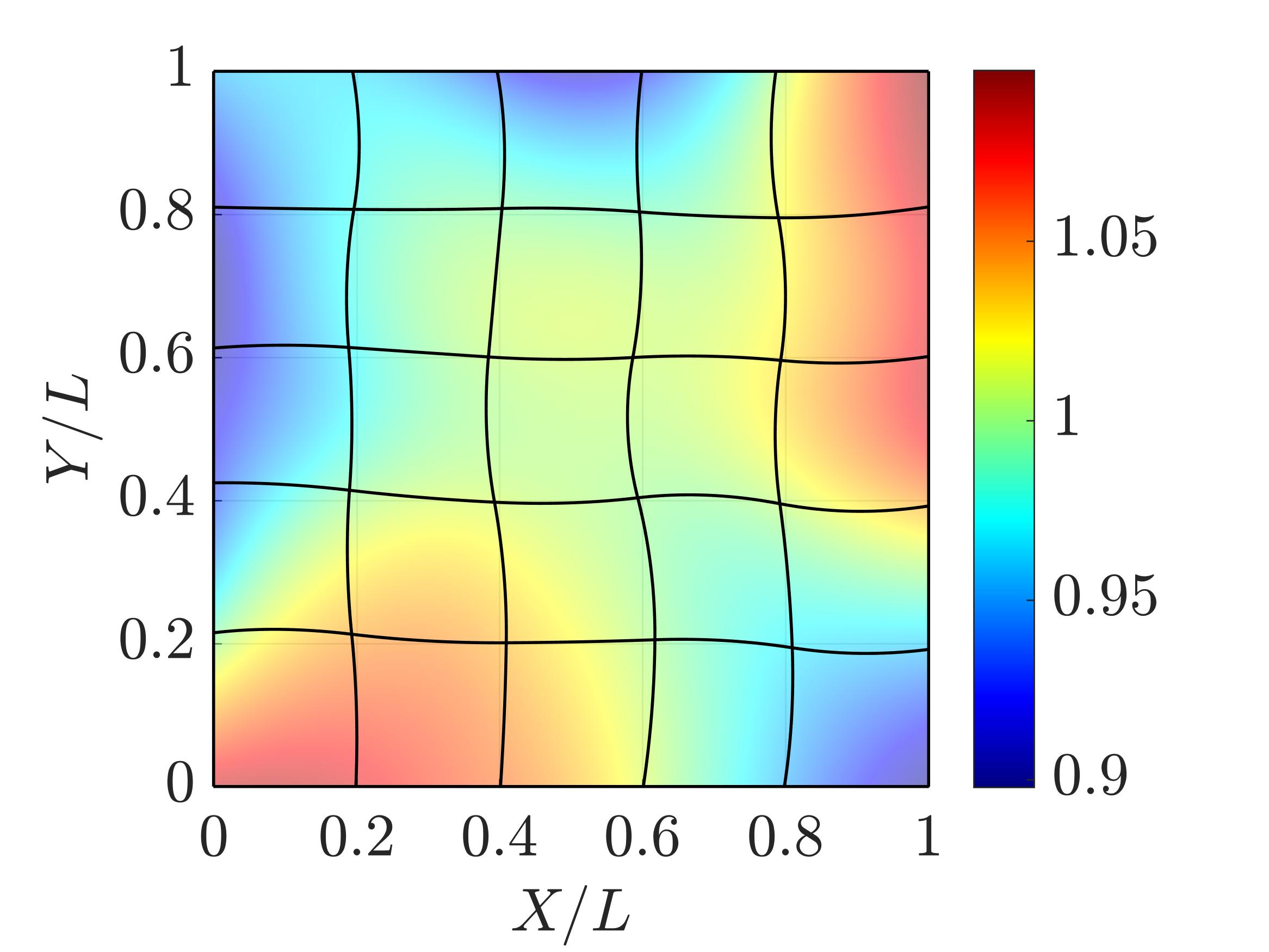}}
\put(-7.0,0.6){c)}
\put( 0.3,0.0){\includegraphics[height=45mm]{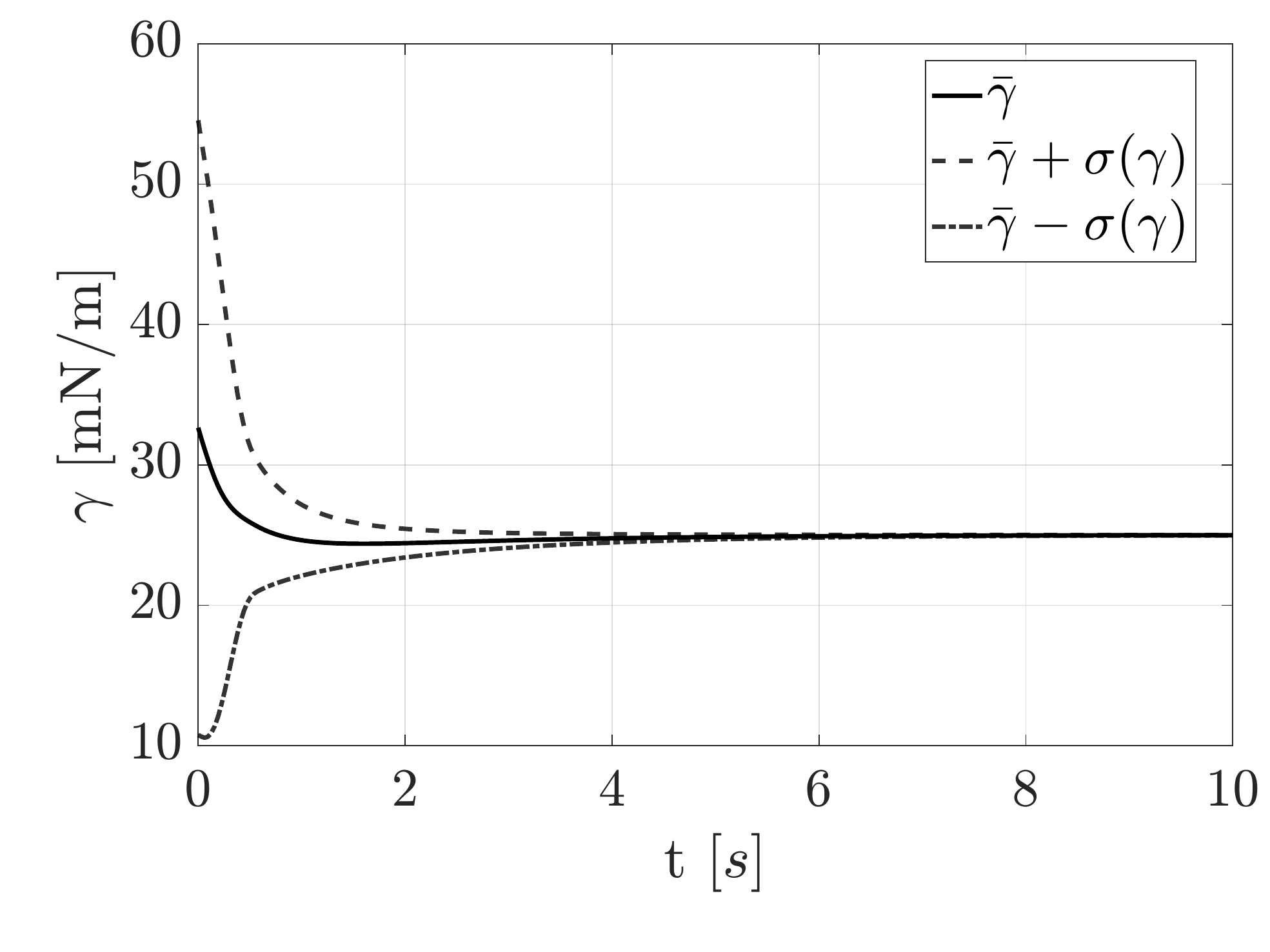}}
\put( 0.1,0.6){d)}
\end{picture}
\caption{Relaxation of a liquid film: a) Initial configuration with randomly distributed surface tension, b) equilibrated configuration with homogeneous surface tension, c) stretch distribution in the relaxed state and d) change of surface tension with time.}
\label{f:relsq}
\end{center}
\end{figure}
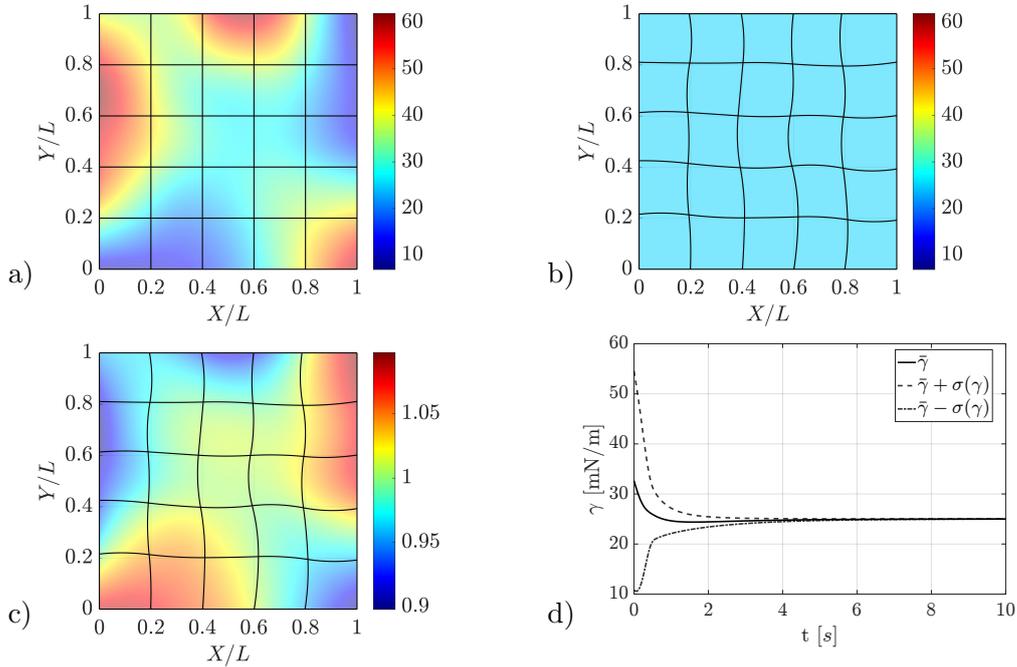

As already discussed in Remark~\ref{r:marangoni}, the point-wise approach to model the surface tension and interfacial concentration can result in stress gradients that in turn cause interfacial flows. Even though, the presented finite element formulation is Lagrangian and not Eulerian, it can represent such flows as long as they are not too large, which is demonstrated next. We therefore consider a $L\times L$ square liquid film, the edges of which are completely fixed. It is assumed that the film consists of BLES 0.5 mg/ml in humid conditions (see Tab.~\ref{t:CRM_par}) and it is modeled by the CR model. Further, it is supposed that the surfactant concentration is not homogeneous, which causes the initial surface tension distribution shown in Fig.~\ref{f:relsq}.a. As it can be seen in Fig.~\ref{f:relsq}.b, if the system is allowed to relax, the surface deforms due to the local variation in the surface tension and finally reaches a new configuration, where the surface tension is not only homogeneous but also equal to the equilibrium surface tension $\gamma_\mathrm{eq} = 25~\mrm\mrN/\mrm$. \tred{Marangoni flows generally need inertia or a suitable body force to be in (dynamic) equilibrium. In the current example, an in-plane body force is applied to ensure the initial equilibrium and then the body force is decreased to zero within $T = 0.6~\mrs$.}
Fig.~\ref{f:relsq}.c shows the corresponding deformation map colored by the surface stretch $J$ in the final relaxed state and Fig.~\ref{f:relsq}.d shows how the mean surface tension $\bar\gamma$, defined by Eq.~\eqref{e:mean_gam}, and the corresponding standard deviation $\sigma(\gamma)$ evolve during the relaxation process. Hence, by employing the presented computational model, where the local stretch and the local surfactant concentration (and equivalently the local surface tension) are mutually related, the system will indeed find a new equilibrated configuration with a homogeneous surface tension.

\FloatBarrier
\subsection{Pendant drop (PD) test}\label{s:pd}
\begin{figure}[ht!]
\begin{center} \unitlength1cm
\unitlength1cm
\begin{picture}(0,4.5)
\put(-6.8,0.0){\includegraphics[height=45mm]{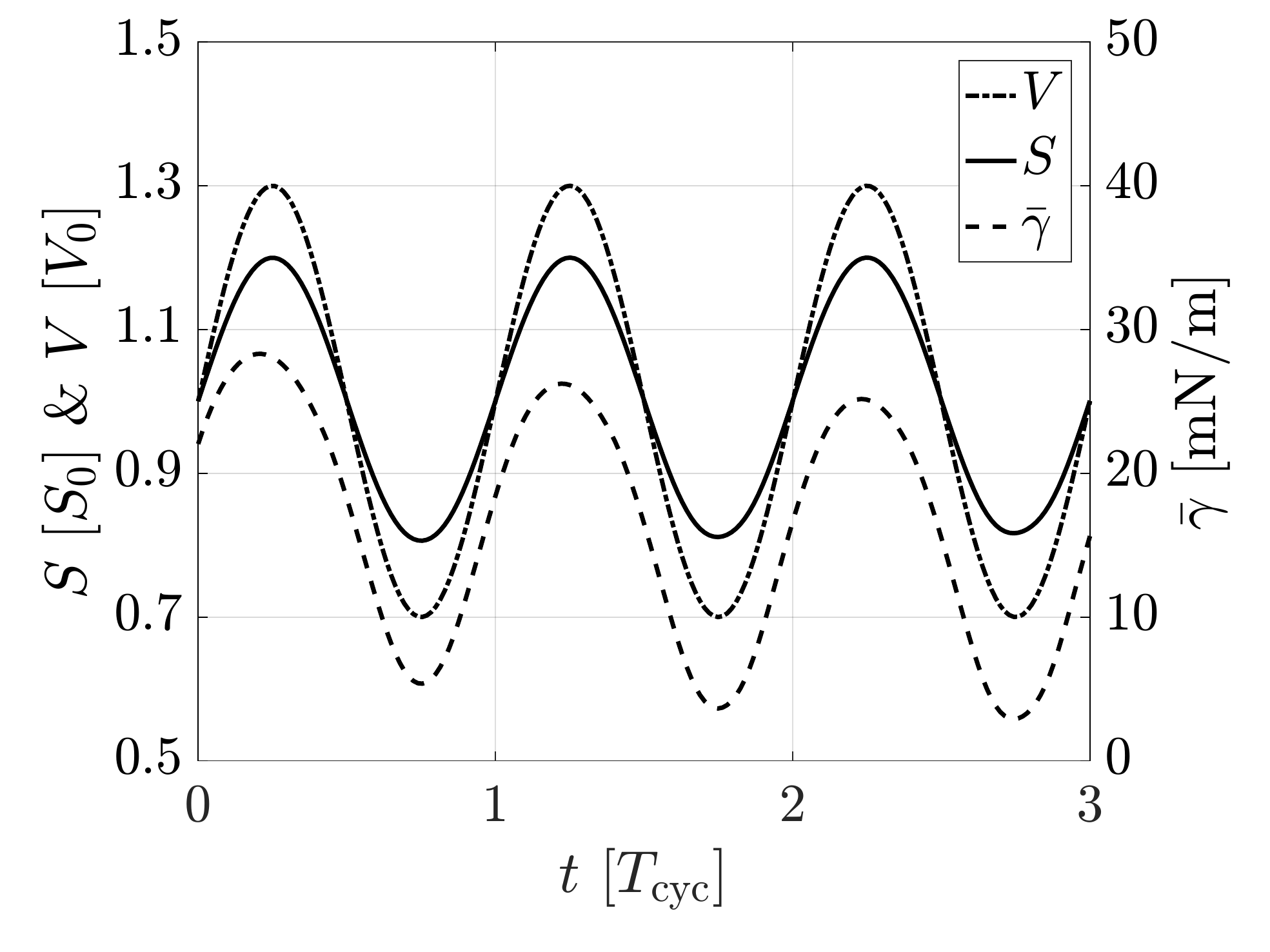}}
\put(-7.0,0.6){a)}
\put( 0.3,0.0){\includegraphics[height=45mm]{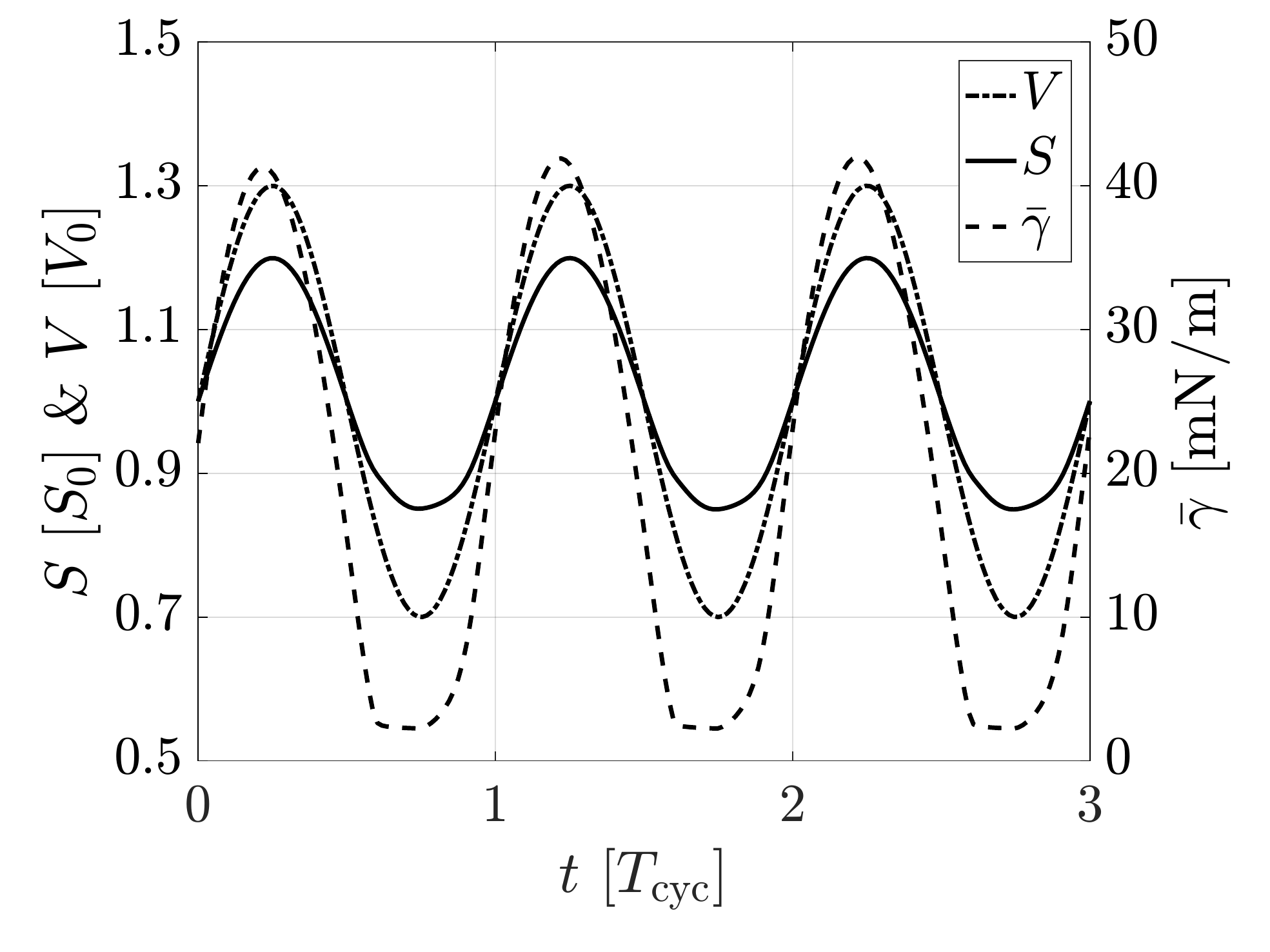}}
\put( 0.1,0.6){b)}
\end{picture}
\caption{The PD test: Cyclic alteration of volume $V$, surface area $S$ and the mean surface tension $\bar\gamma$ for a) the AL and b) the CR models.}
\label{f:pd_curves}
\end{center}
\end{figure}

In this section, the pendant drop test, introduced earlier, is simulated. In the reference configuration, the drop is modeled as 1/8 of a sphere with boundary conditions similar to the constrained droplet (see Fig.~\ref{f:comp_model}.a) with the difference that the drop is held downward in the direction of the gravity force. The initial radius of the drop is $1.0~\mrm\mrm$. Here, the volume of the droplet is controlled following the volume constraint \eqref{e:gv}. The corresponding profiles of the drop at different volumes are plotted in Figs.~\ref{f:pd_alm_drops}~and~\ref{f:pd_crm_drops} for the AL and CM material models, respectively. For the AL model, all the parameters are the same as Tab.~\ref{t:ALM_par} except $m_2 = 50~\mrm\mrN/\mrm$. For the CR model, all the parameters are set according to the case \tred{4} in Tab.~\ref{t:CRM_par}. Figs.~\ref{f:pd_alm_drops}~and~\ref{f:pd_crm_drops} show the finite element results for the cases where the volume is controlled in a monotonic way. As a second example, the drop volume is altered cyclically as
\eqb{l}
V(t) = V_0 + \Delta V_\mathrm{max}\,\sin\big(t/T\big)~, 
\label{e:DV}\eqe
where $T=3~\mrs$ and $\Delta V_\mathrm{max} = 0.3\,V_0$ here. Fig.~\ref{f:pd_curves} shows the dynamic changes of the mean surface tension $\gamma$, given by Eq.~\eqref{e:mean_gam}, and the surface area with cyclic alteration of the drop volume.
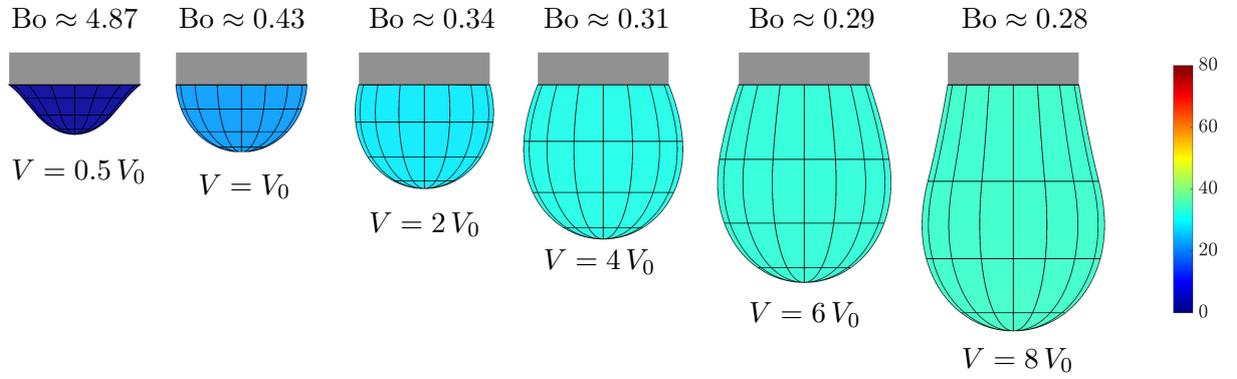
\begin{figure}[ht]
\begin{center} \unitlength1cm
\unitlength1cm
\begin{picture}(0,4.5)
\put(-7.9,4.2){$\mathrm{Bo}\approx4.87$}
\put(-8.0,0.0){\includegraphics[height=40mm,trim={450px 0 650px 0},clip]{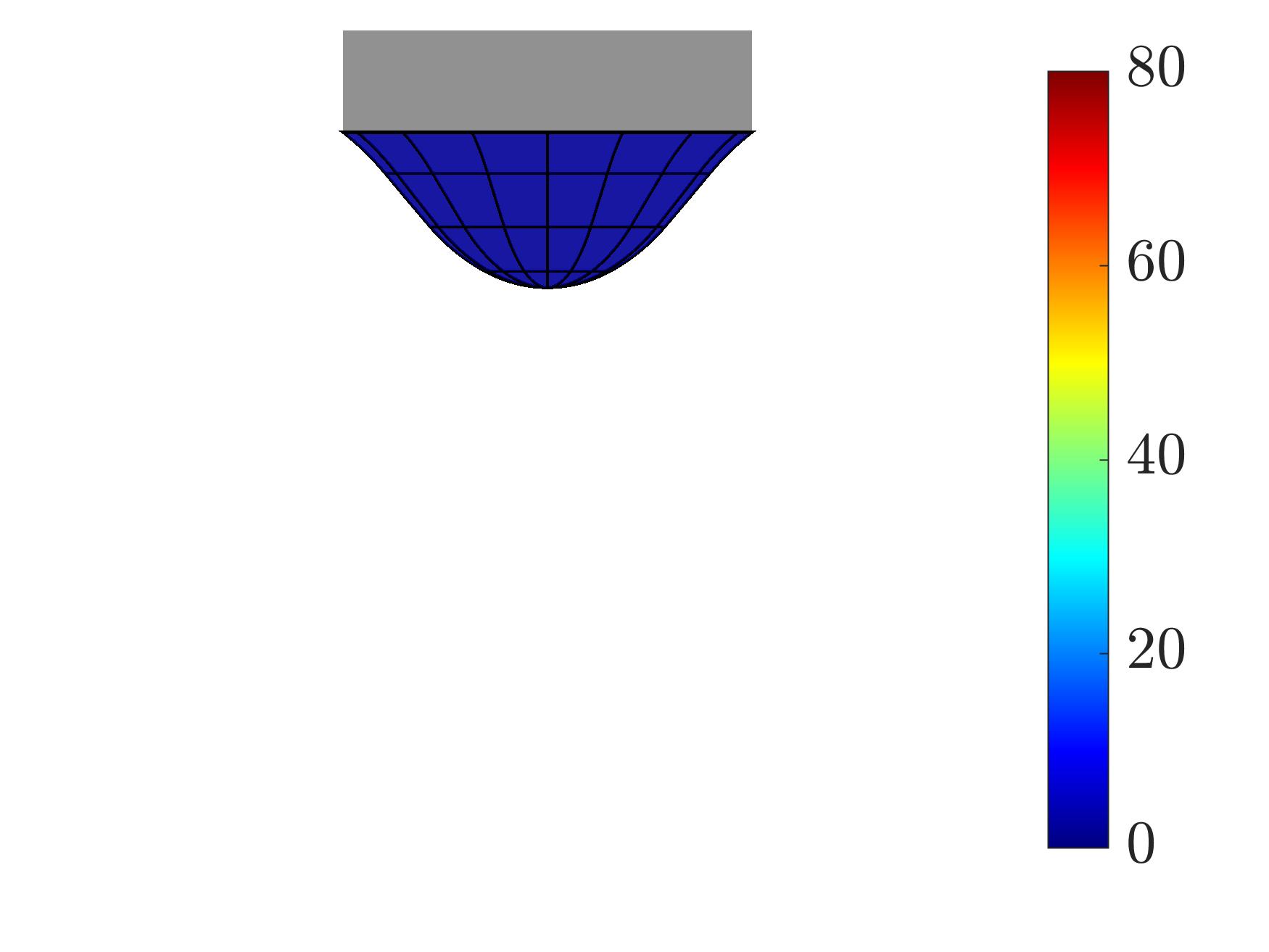}}
\put(-7.9,2.2){$V=0.5\,V_0$}
\put(-5.7,4.2){$\mathrm{Bo}\approx0.43$}
\put(-5.8,0.0){\includegraphics[height=40mm,trim={450px 0 650px 0},clip]{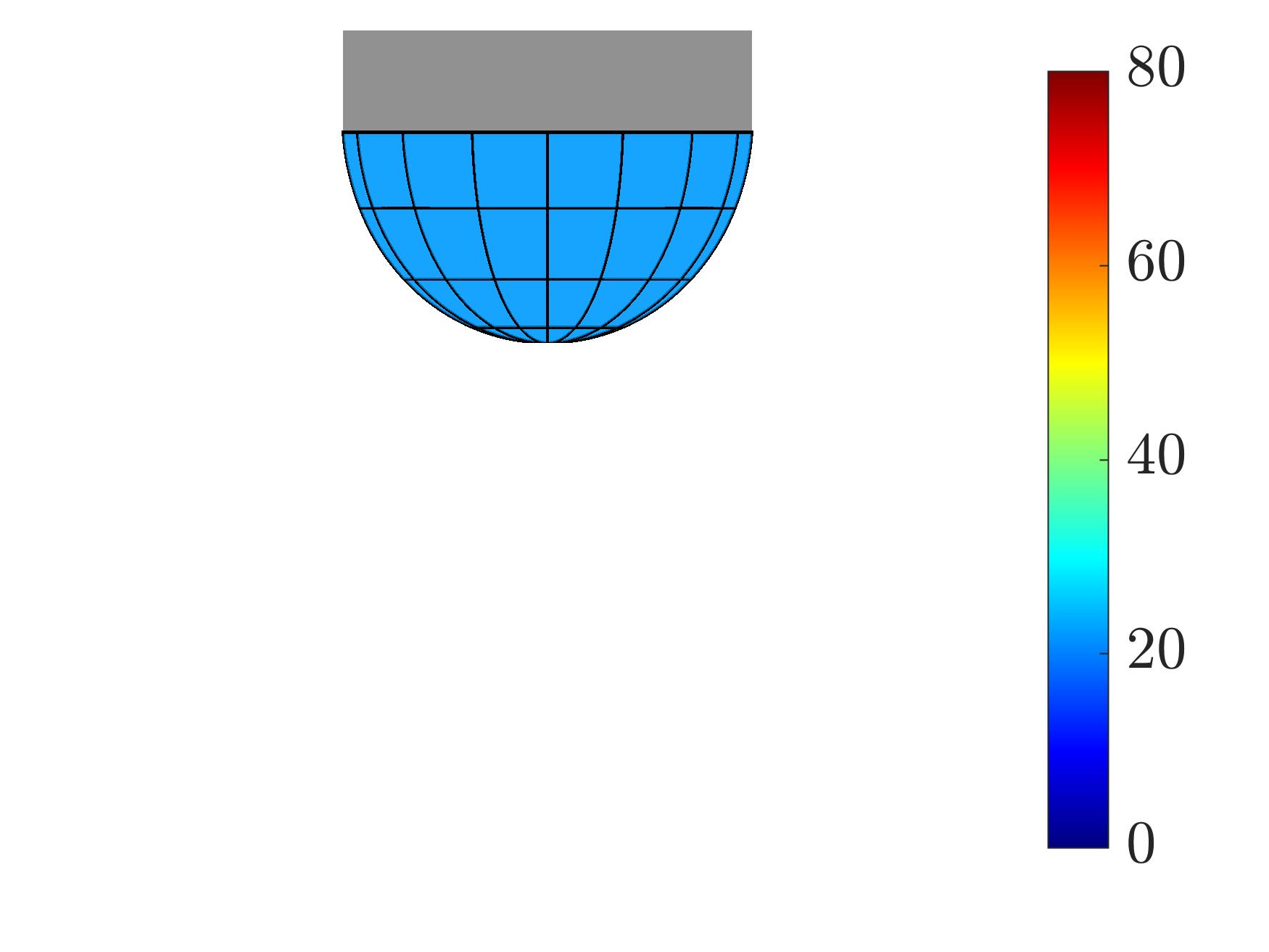}}
\put(-5.7,2.0){$~~V=V_0$}
\put(-3.2,4.2){$\mathrm{Bo}\approx0.34$}
\put(-3.4,0.0){\includegraphics[height=40mm,trim={450px 0 650px 0},clip]{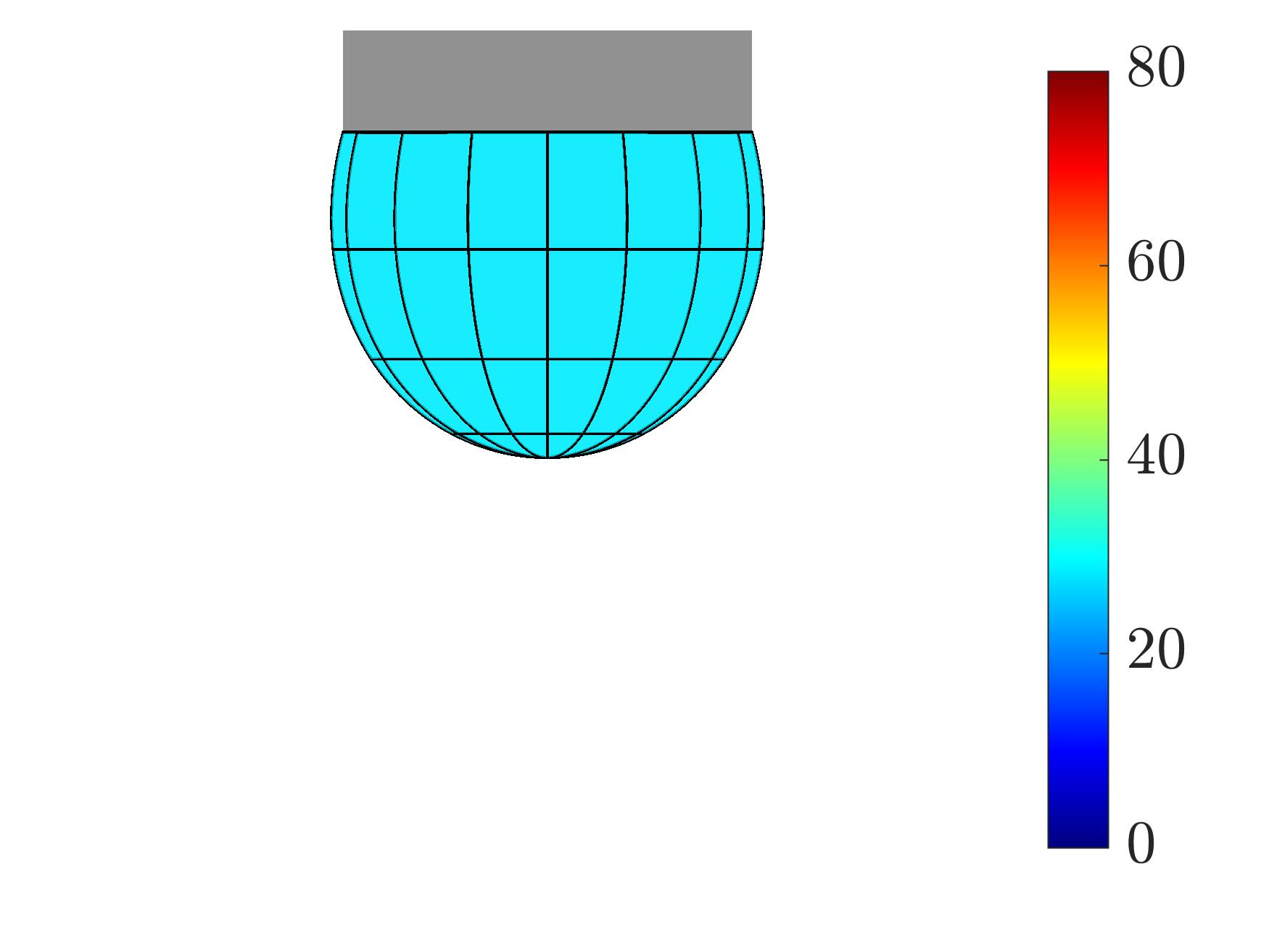}}
\put(-3.2,1.5){$V=2\,V_0$}
\put(-0.9,4.2){$\mathrm{Bo}\approx0.31$}
\put(-1.2,0.0){\includegraphics[height=40mm,trim={400px 0 620px 0},clip]{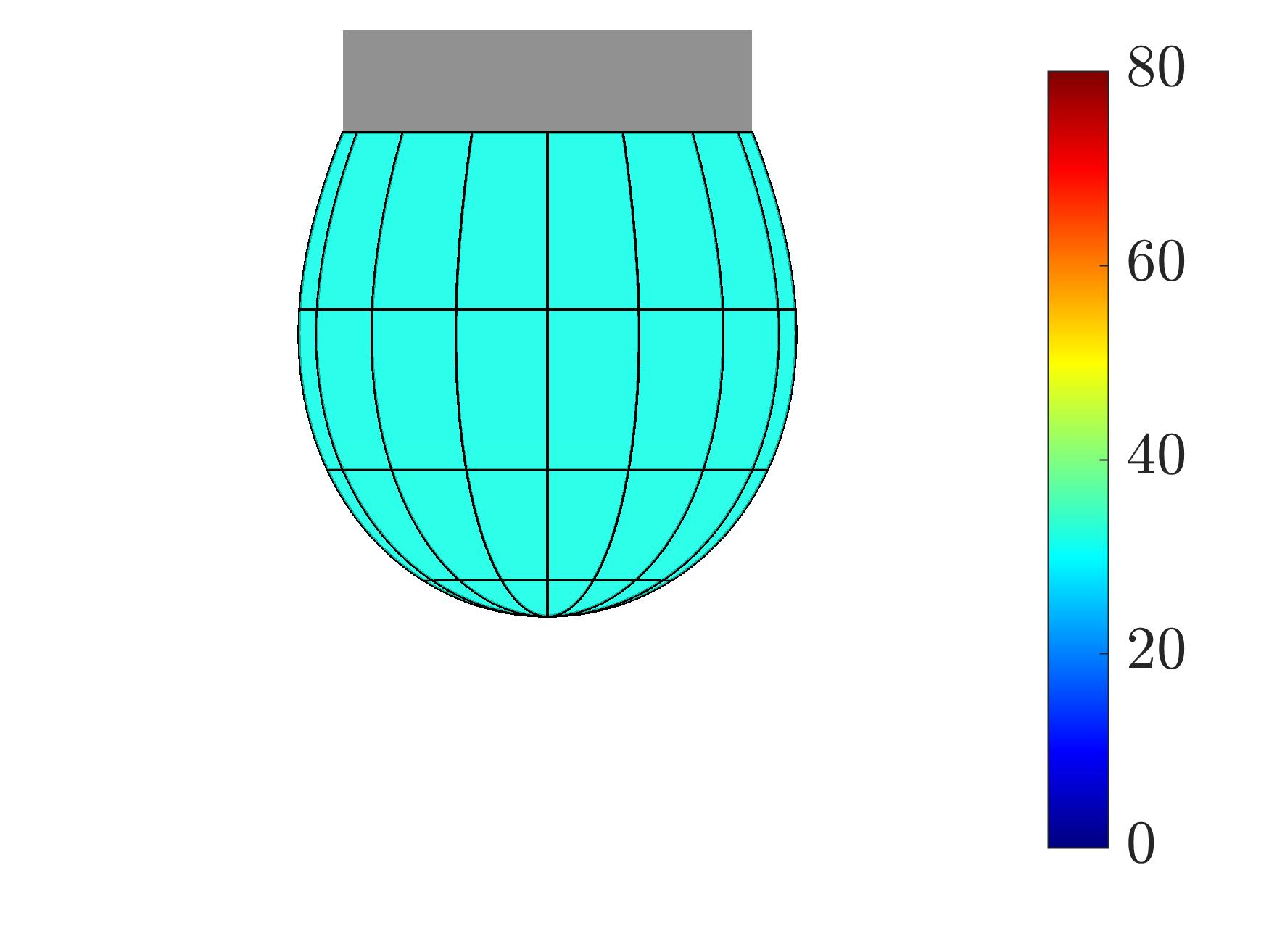}}
\put(-0.9,1.0){$V=4\,V_0$}
\put( 1.8,4.2){$\mathrm{Bo}\approx0.29$}
\put( 1.3,0.0){\includegraphics[height=40mm,trim={350px 0 600px 0},clip]{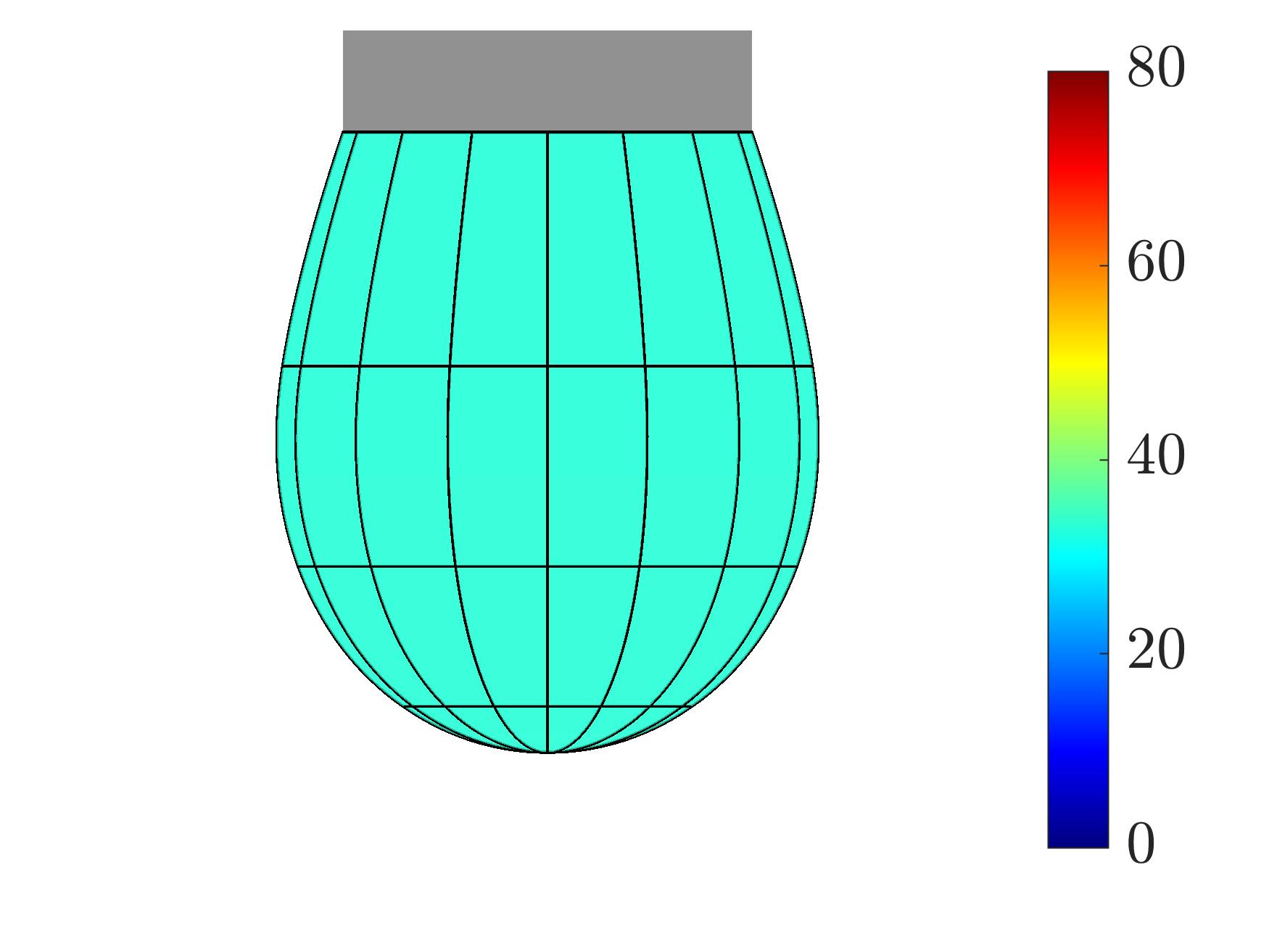}}
\put( 1.8,0.3){$V=6\,V_0$}
\put( 4.6,4.2){$\mathrm{Bo}\approx0.28$}
\put( 3.9,0.0){\includegraphics[height=40mm,trim={300px 0 100px 0},clip]{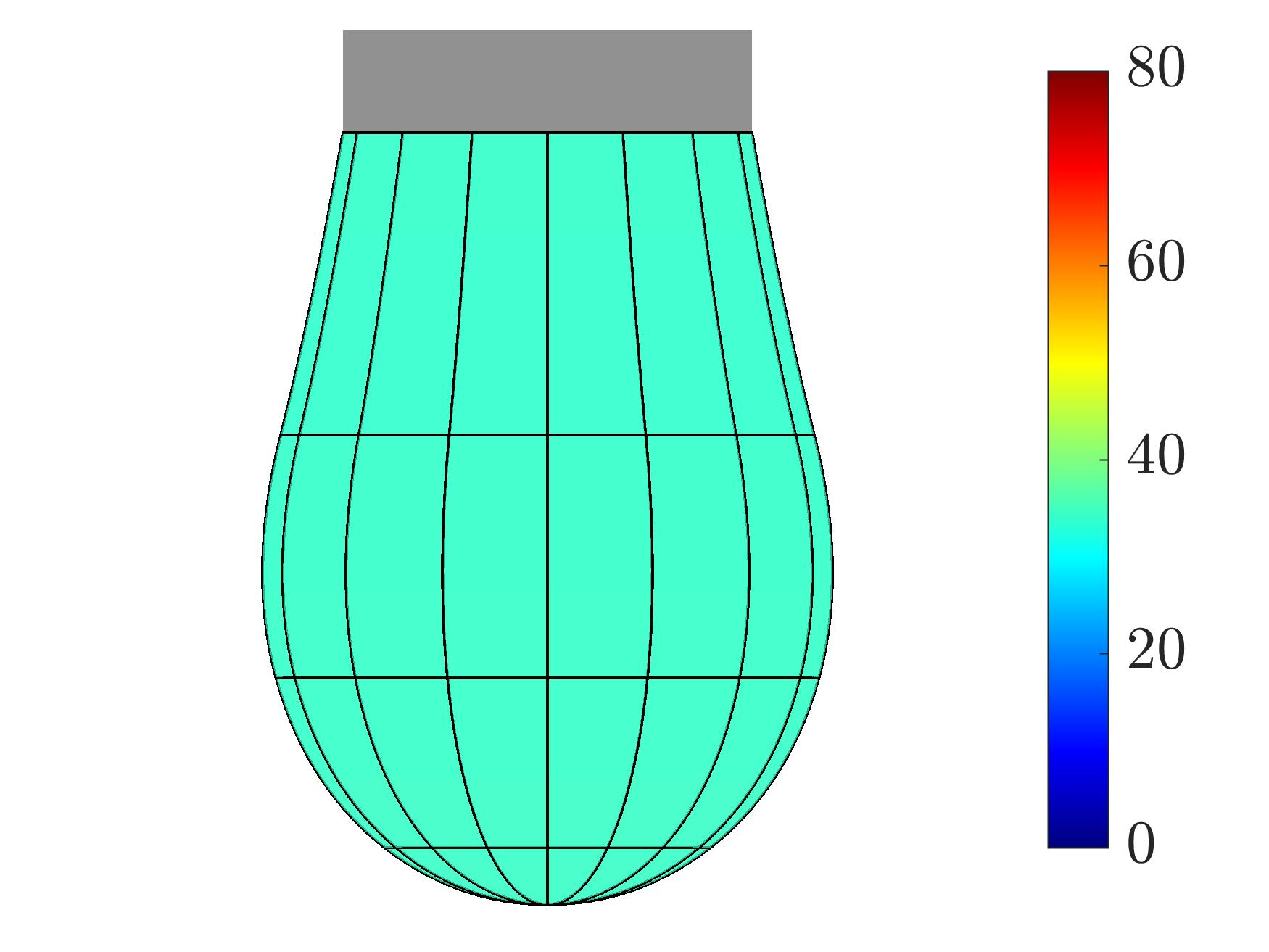}}
\put( 4.6,-0.3){$V=8\,V_0$}
\end{picture}
\caption{The PD test: Drop shapes modeled by the AL model and colored by $\gamma$ [mN/m].}
\label{f:pd_alm_drops}
\end{center}
\end{figure}
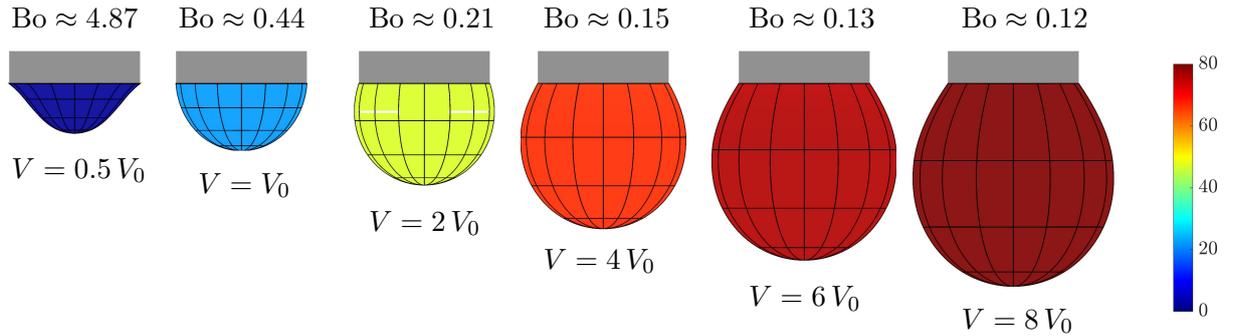
\begin{figure}[ht]
\begin{center} \unitlength1cm
\unitlength1cm
\begin{picture}(0,4.5)
\put(-7.9,4.2){$\mathrm{Bo}\approx4.87$}
\put(-8.0,0.0){\includegraphics[height=40mm,trim={450px 0 650px 0},clip]{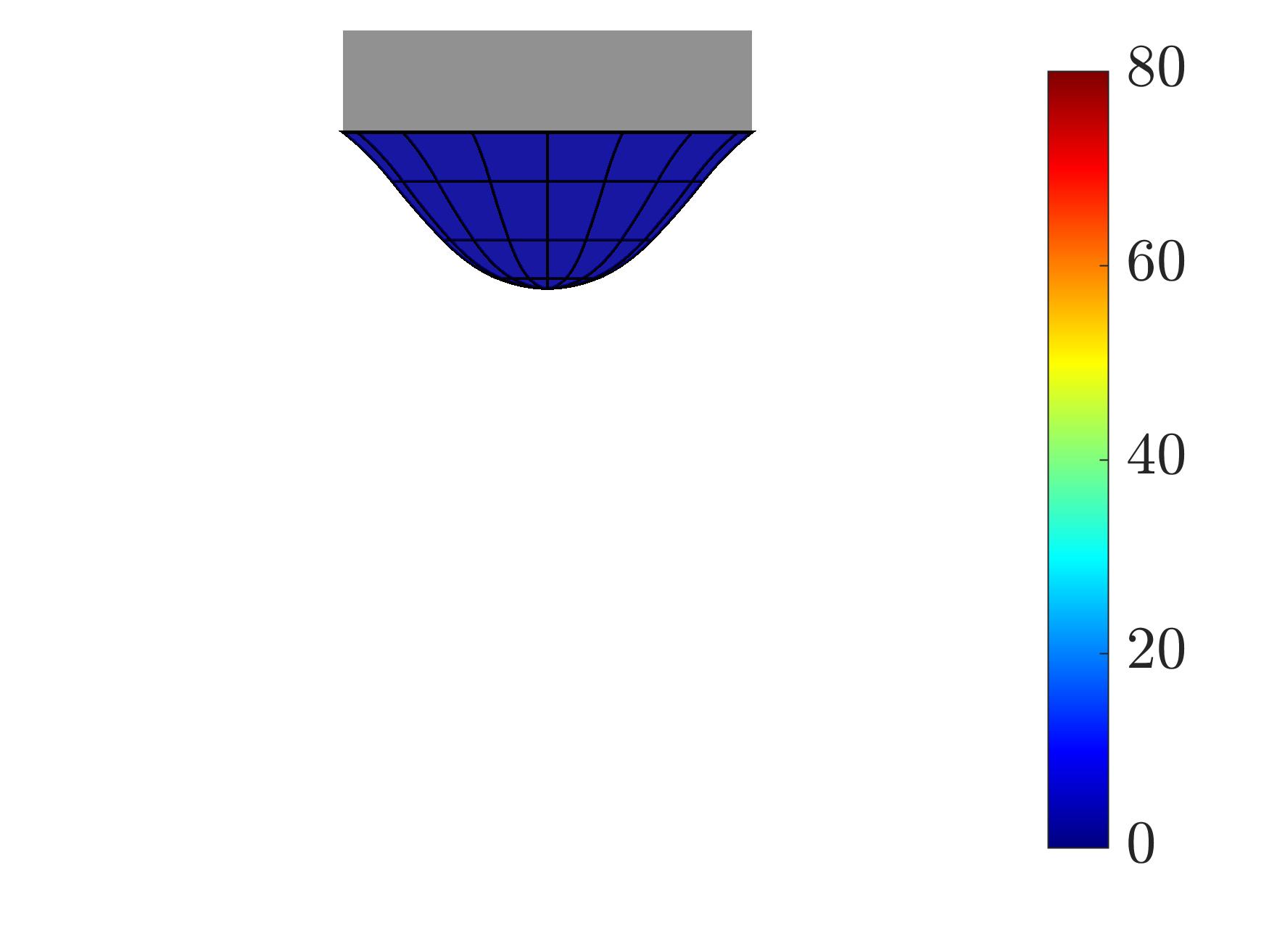}}
\put(-7.9,2.2){$V=0.5\,V_0$}
\put(-5.7,4.2){$\mathrm{Bo}\approx0.44$}
\put(-5.8,0.0){\includegraphics[height=40mm,trim={450px 0 650px 0},clip]{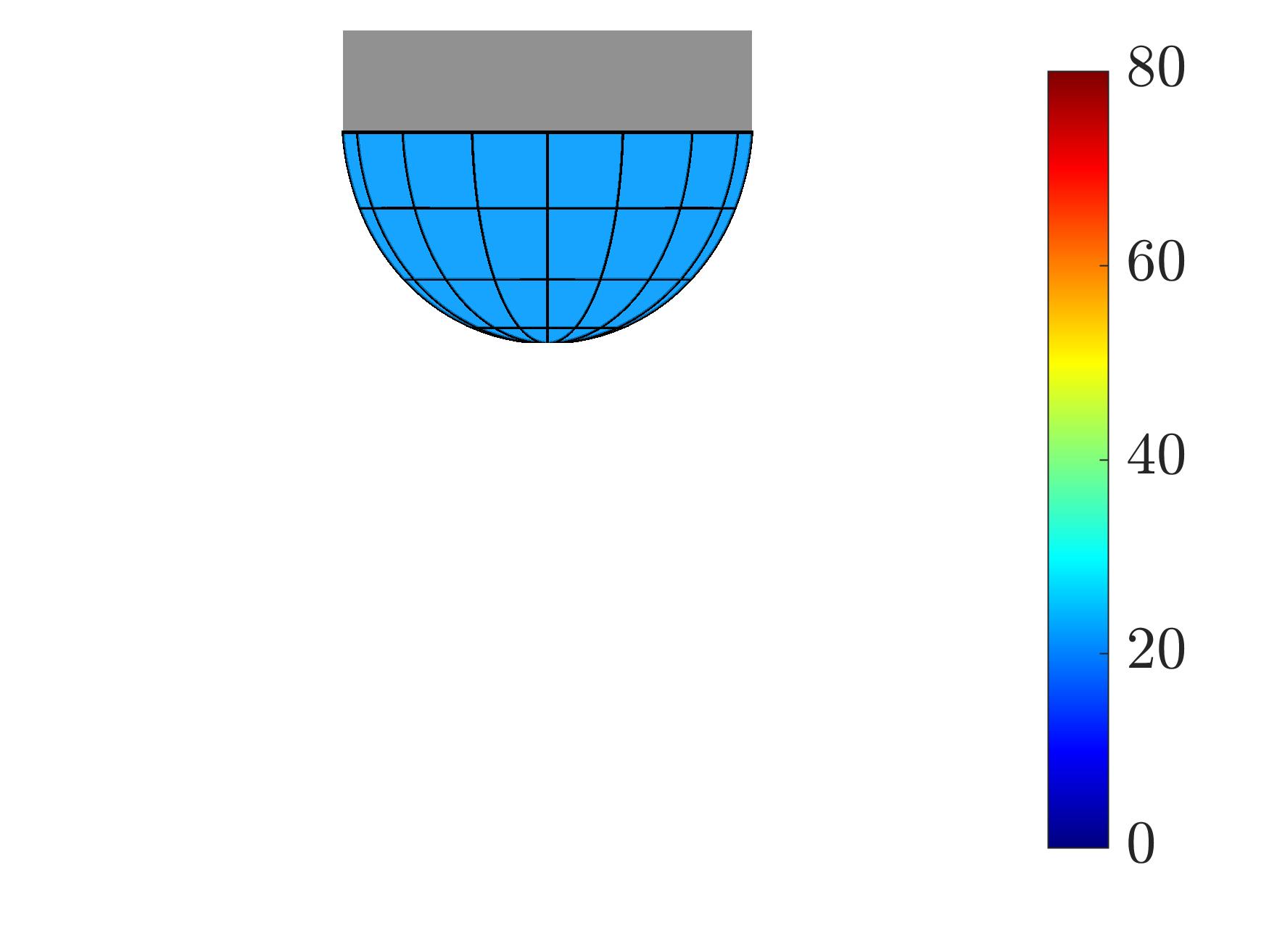}}
\put(-5.7,2.0){$~~V=V_0$}
\put(-3.2,4.2){$\mathrm{Bo}\approx0.21$}
\put(-3.4,0.0){\includegraphics[height=40mm,trim={450px 0 650px 0},clip]{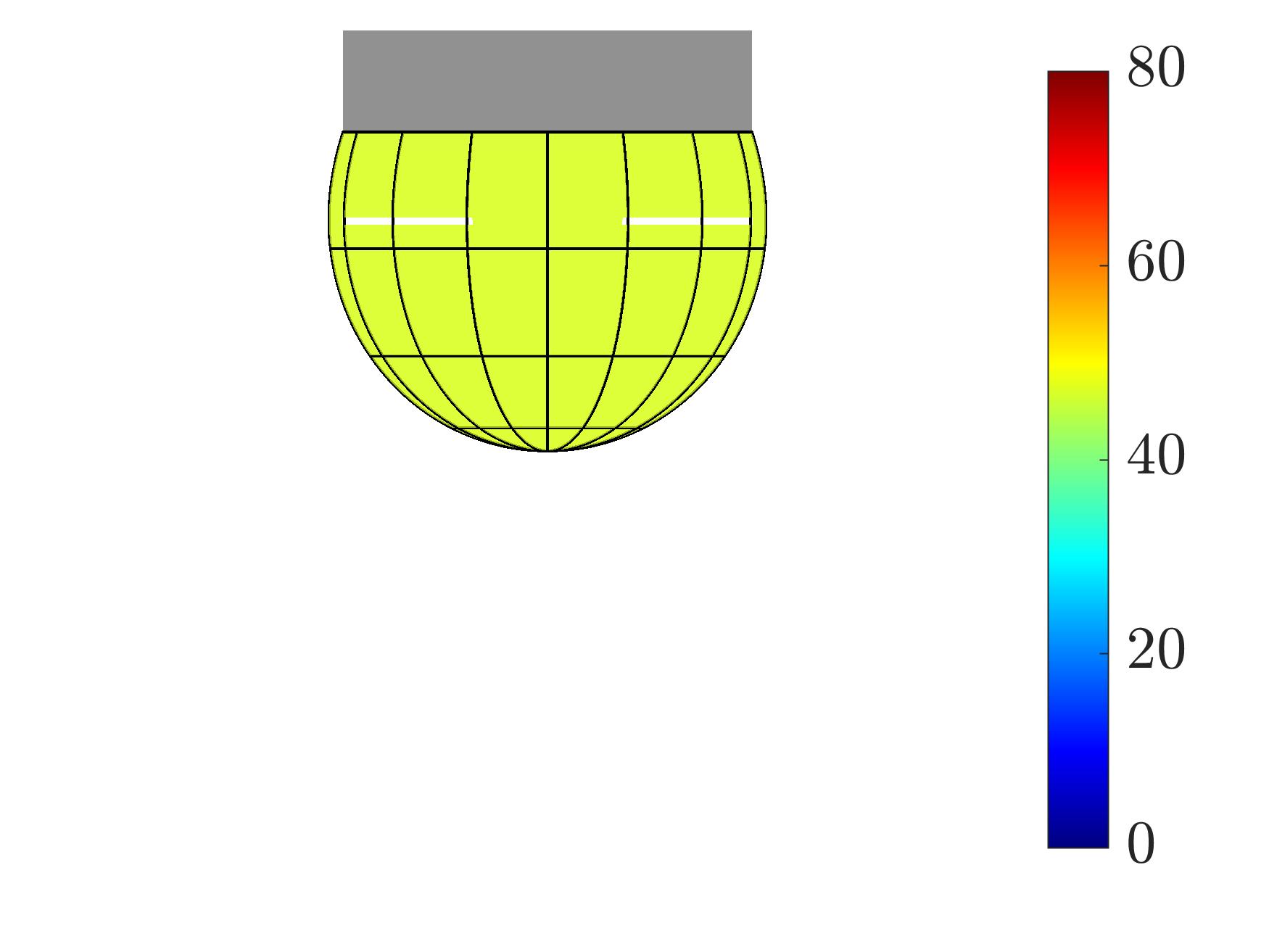}}
\put(-3.2,1.5){$V=2\,V_0$}
\put(-0.9,4.2){$\mathrm{Bo}\approx0.15$}
\put(-1.2,0.0){\includegraphics[height=40mm,trim={400px 0 620px 0},clip]{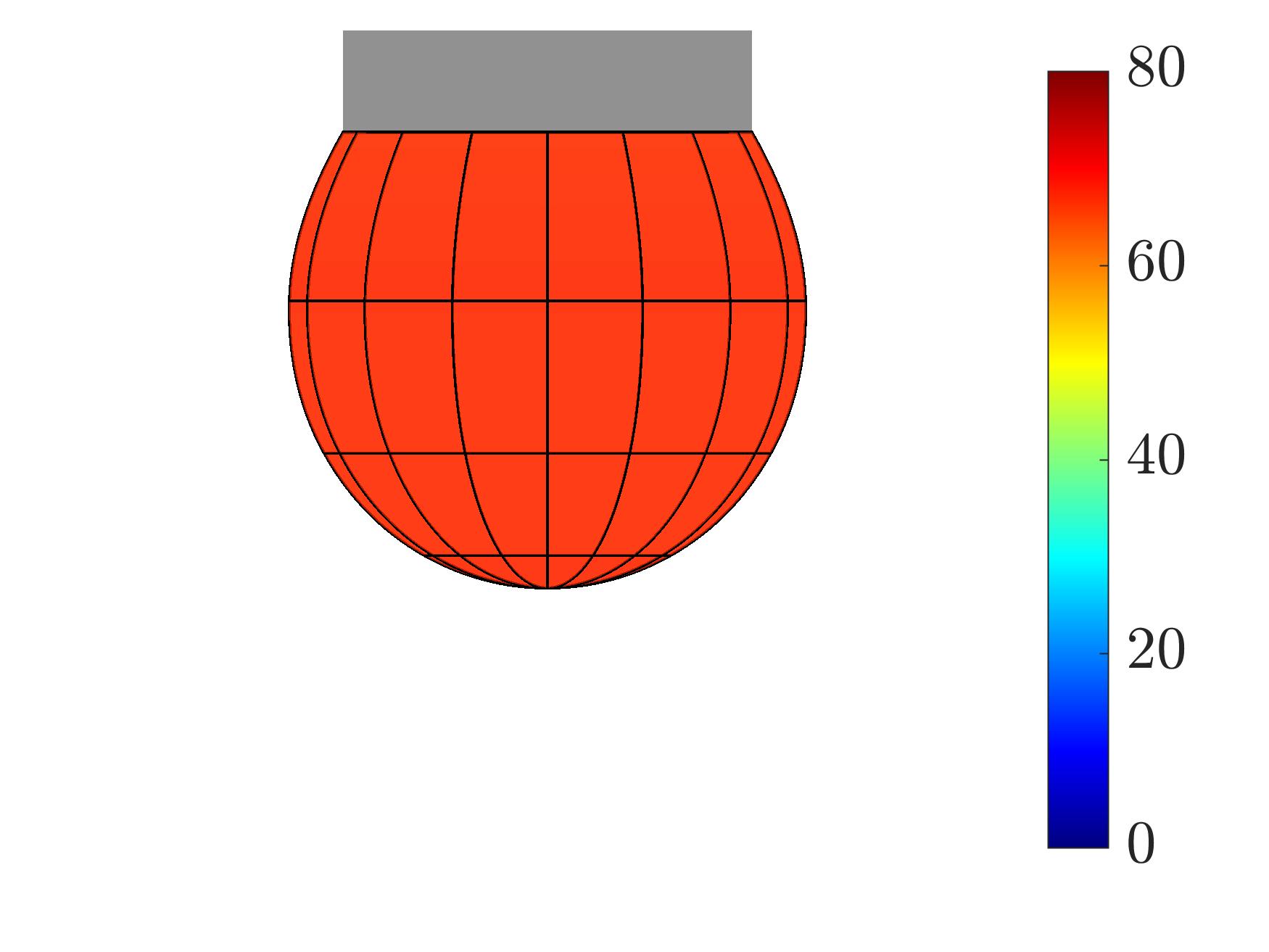}}
\put(-0.9,1.0){$V=4\,V_0$}
\put( 1.8,4.2){$\mathrm{Bo}\approx0.13$}
\put( 1.3,0.0){\includegraphics[height=40mm,trim={350px 0 600px 0},clip]{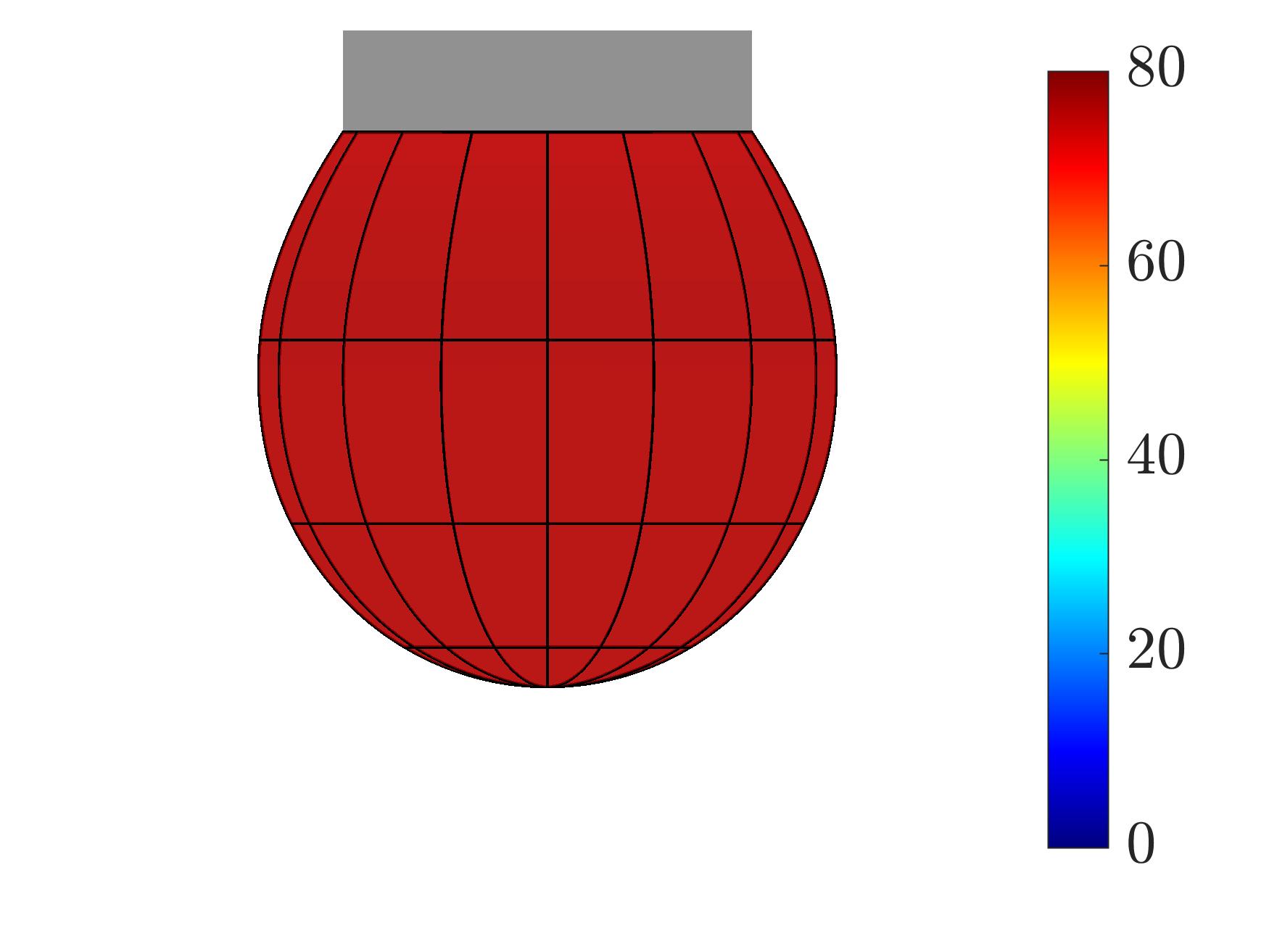}}
\put( 1.8,0.5){$V=6\,V_0$}
\put( 4.6,4.2){$\mathrm{Bo}\approx0.12$}
\put( 3.9,0.0){\includegraphics[height=40mm,trim={300px 0 100px 0},clip]{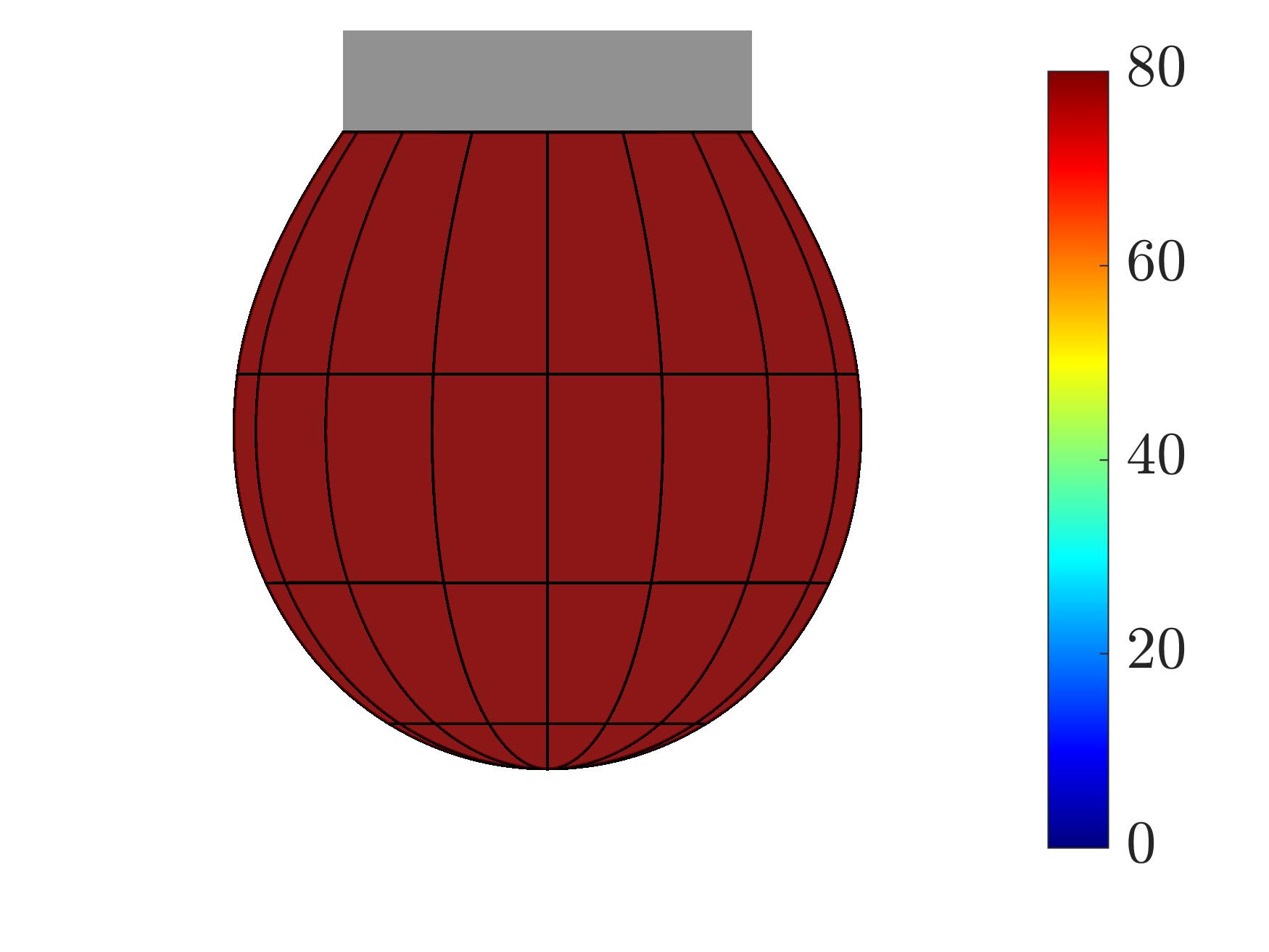}}
\put( 4.6,0.2){$V=8\,V_0$}
\end{picture}
\caption{The PD test: Drop shapes modeled by the CR model and colored by $\gamma$ [mN/m].}
\label{f:pd_crm_drops}
\end{center}
\end{figure}

As can be observed, for small changes in the volume, both models predict similar surface tension; however, as the drop volume increases, the models predict different surface tension and accordingly difference Bond numbers for an identical given volume. It should be noted that a detailed comparison of the two models is only possible when the parameters of each model is identified for the same experiment with the same type of surfactants. This is left for future work. Here, the parameters are adopted from two different resources. Nevertheless, the difference in the behavior of these two models is well discussed in \citetlist{saad10,saad11thesis}.

\subsection{Unconstrained sessile drop (SD) test}\label{s:sd}
\begin{figure}[ht!]
\begin{center} \unitlength1cm
\unitlength1cm
\begin{picture}(0,4.5)
\put(-8.0,0.0){\includegraphics[height=45mm,trim={0 150px 0 300px},clip]{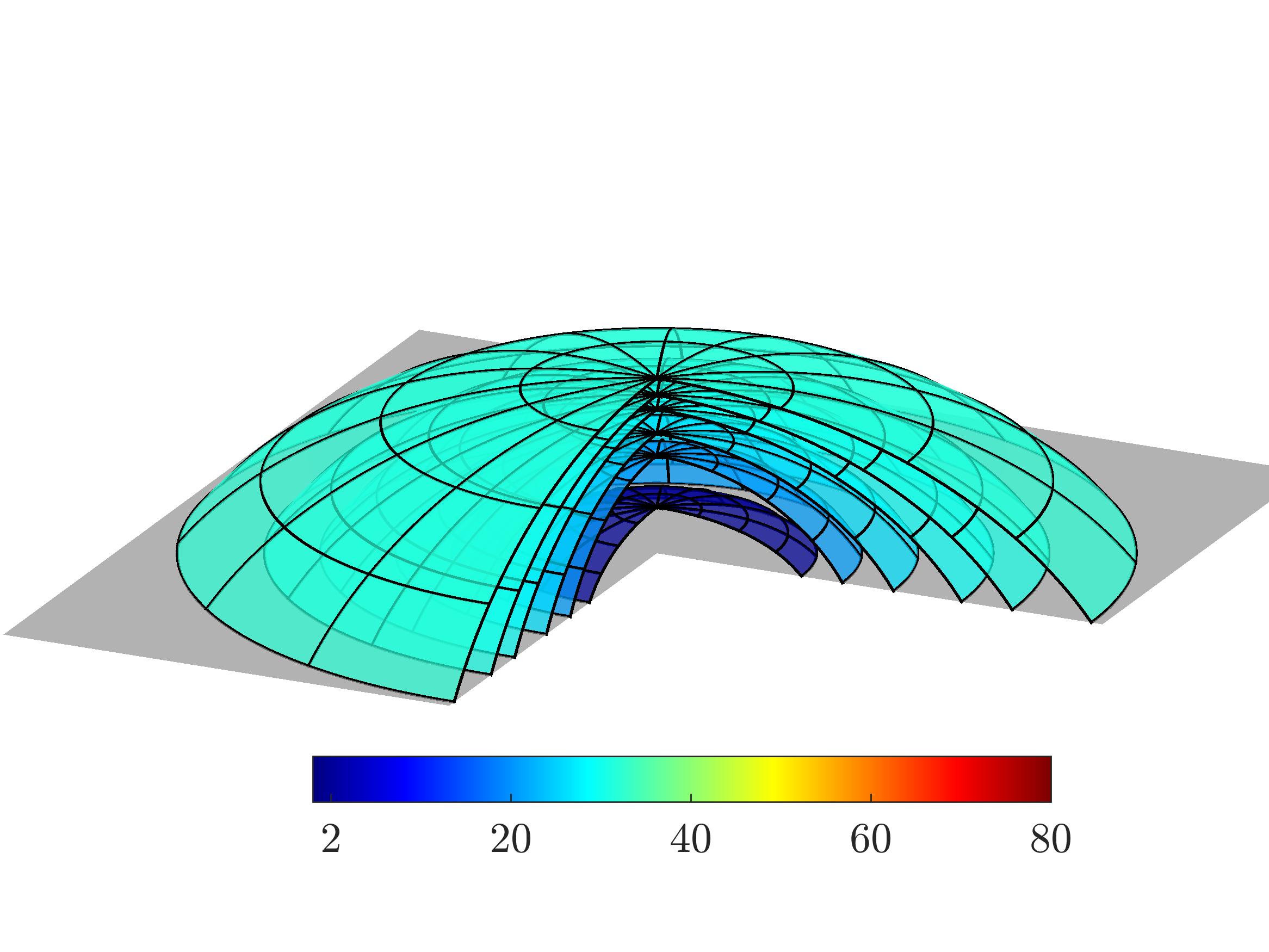}}
\put(-8.0,0.5){a)}
\put( 0.0,0.0){\includegraphics[height=45mm,trim={0 150px 0 300px},clip]{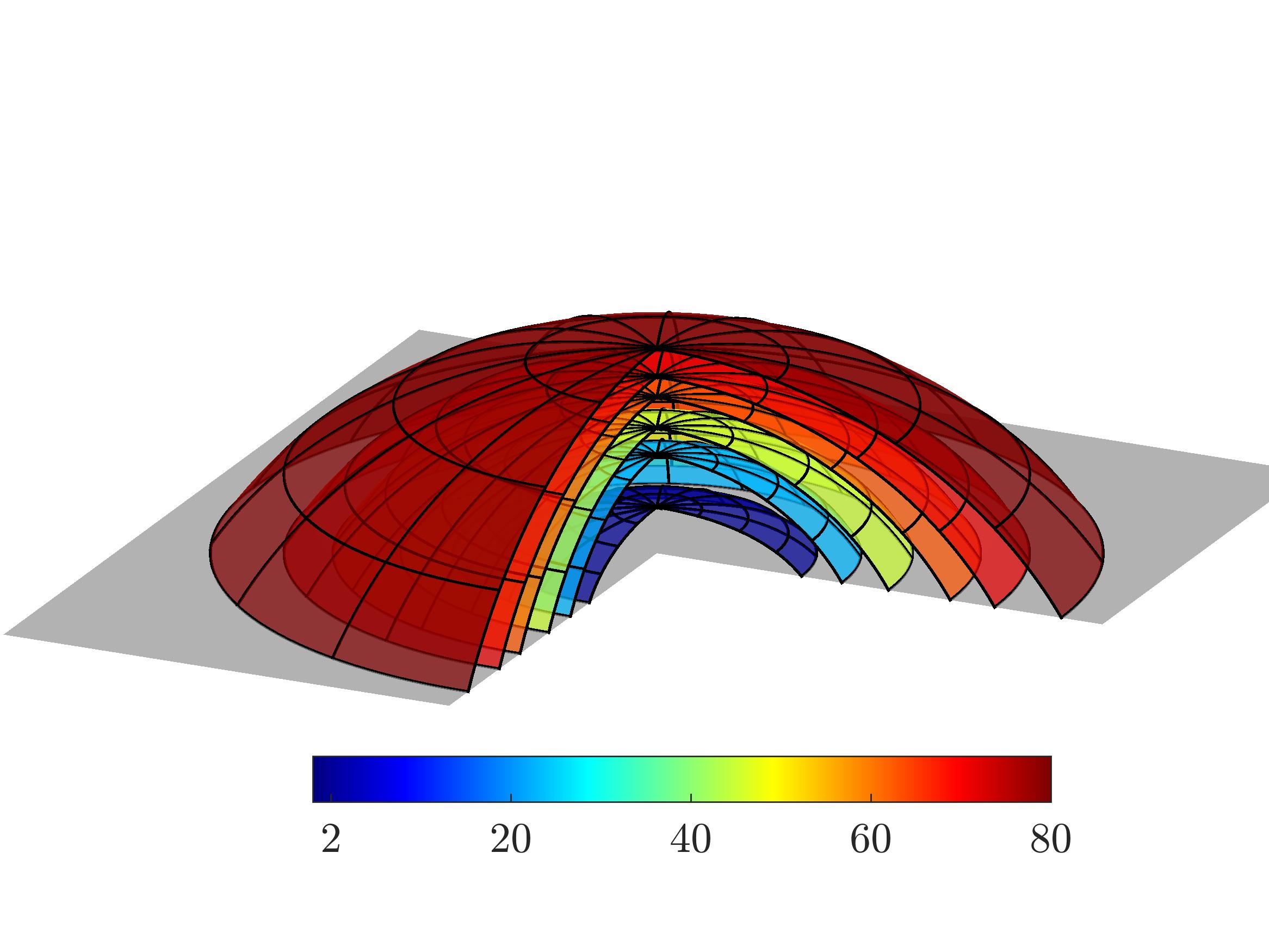}}
\put( 0.1,0.5){b)}
\end{picture}
\caption{The SD test ($\theta_\mrc = 60^\circ$): Drop shapes modeled by a) the AL and b) the CR models and colored by $\gamma$ at $V = 0.3,~1,~2,~4,~6~\&~10V_0 $.}
\label{f:sd_drops_thc60}
\end{center}
\end{figure}

\begin{figure}[ht!]
\begin{center} \unitlength1cm
\unitlength1cm
\begin{picture}(0,4.5)
\put(-8.0,0.0){\includegraphics[height=45mm,trim={0 150px 0 300px},clip]{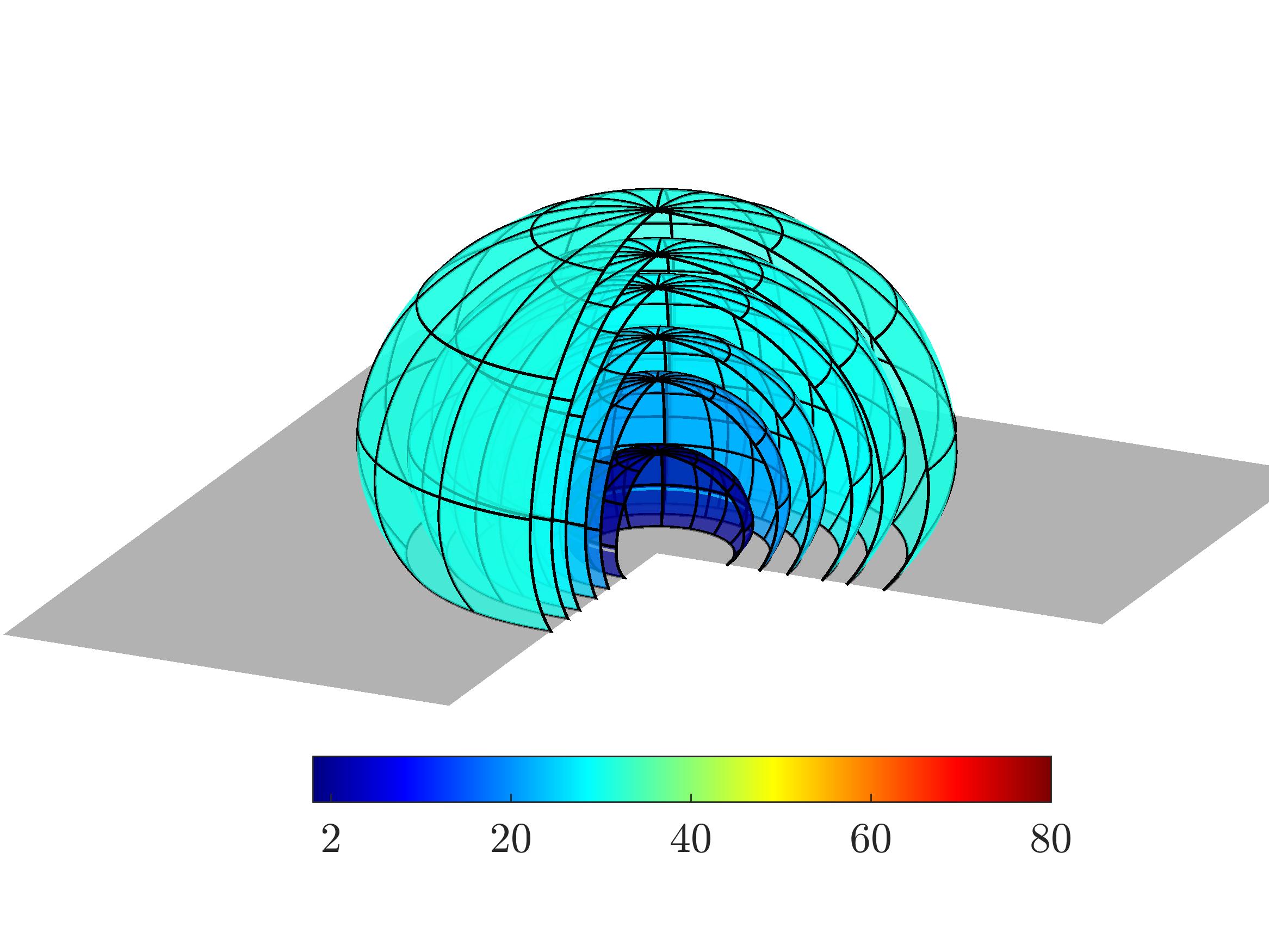}}
\put(-8.0,0.5){a)}
\put( 0.0,0.0){\includegraphics[height=45mm,trim={0 150px 0 300px},clip]{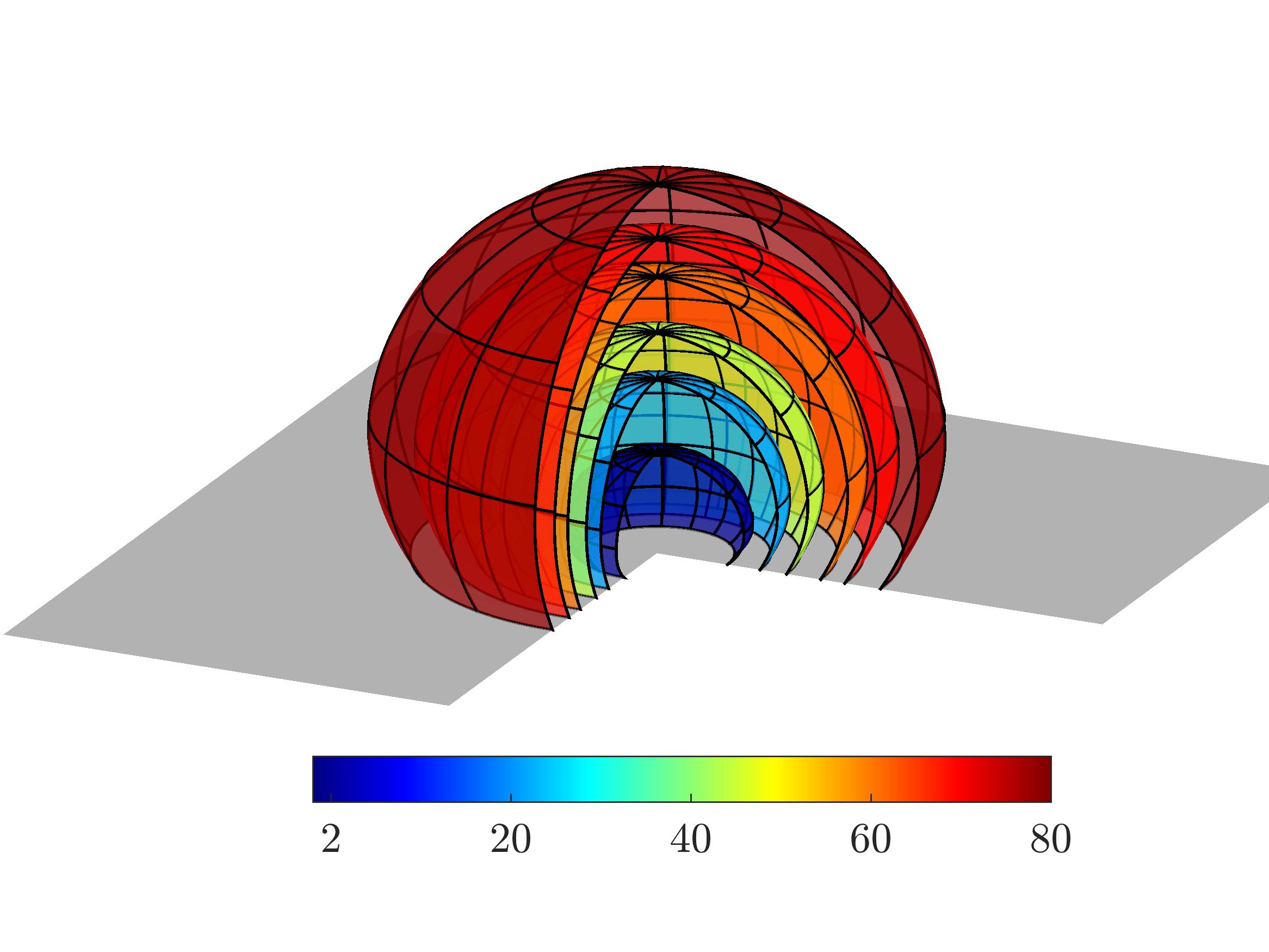}}
\put( 0.1,0.5){b)}
\end{picture}
\caption{The SD test ($\theta_\mrc = 120^\circ$): Drop shapes modeled by a) the AL and b) the CR models and colored by $\gamma$ at $V = 0.3,~1,~2,~4,~6~\&~10V_0 $.}
\label{f:sd_drops_thc120}
\end{center}
\end{figure}

\begin{figure}[ht!]
\begin{center} \unitlength1cm
\unitlength1cm
\begin{picture}(0,4.5)
\put(-6.8,0.0){\includegraphics[height=45mm]{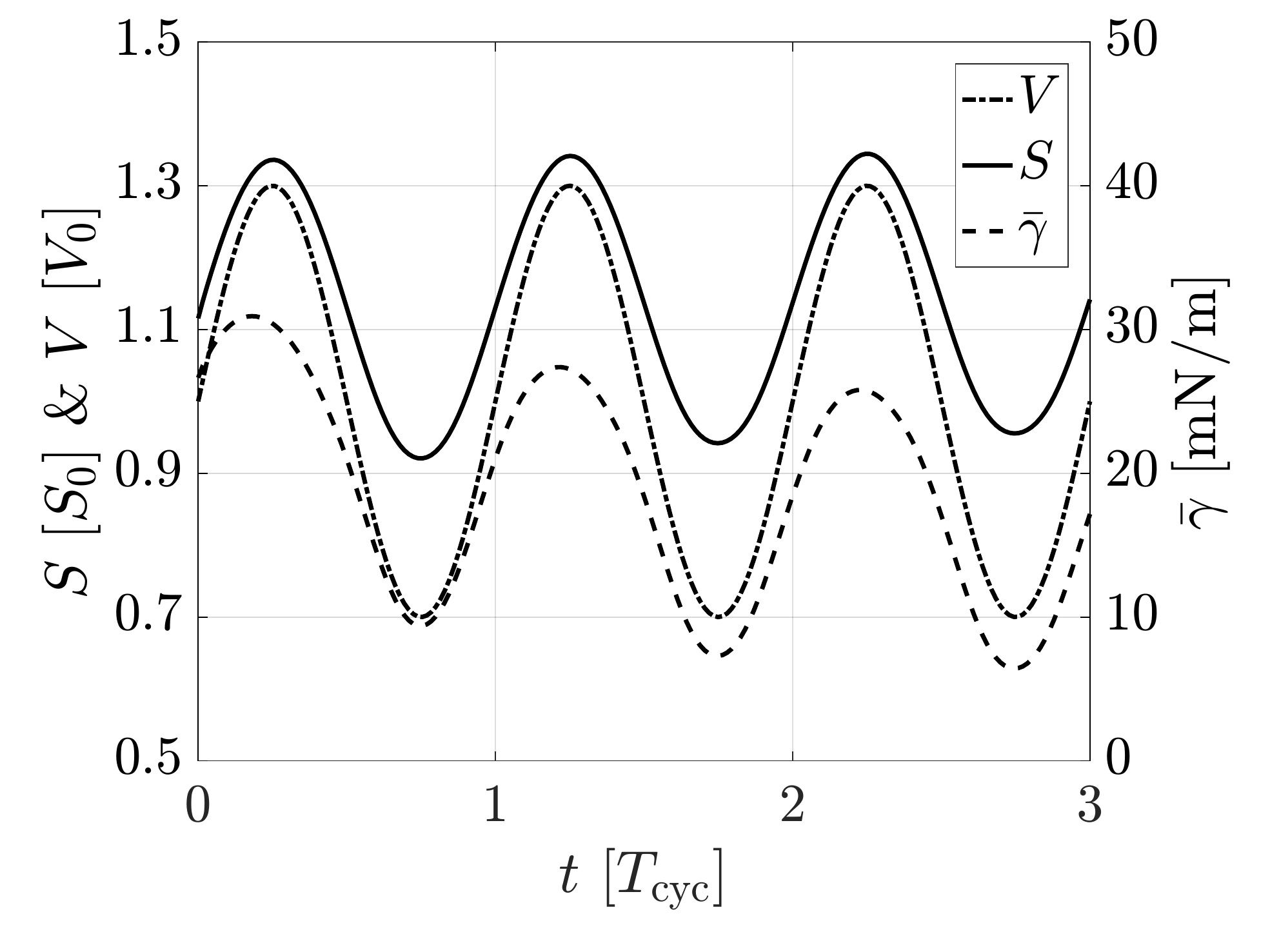}}
\put(-7.0,0.6){a)}
\put( 0.3,0.0){\includegraphics[height=45mm]{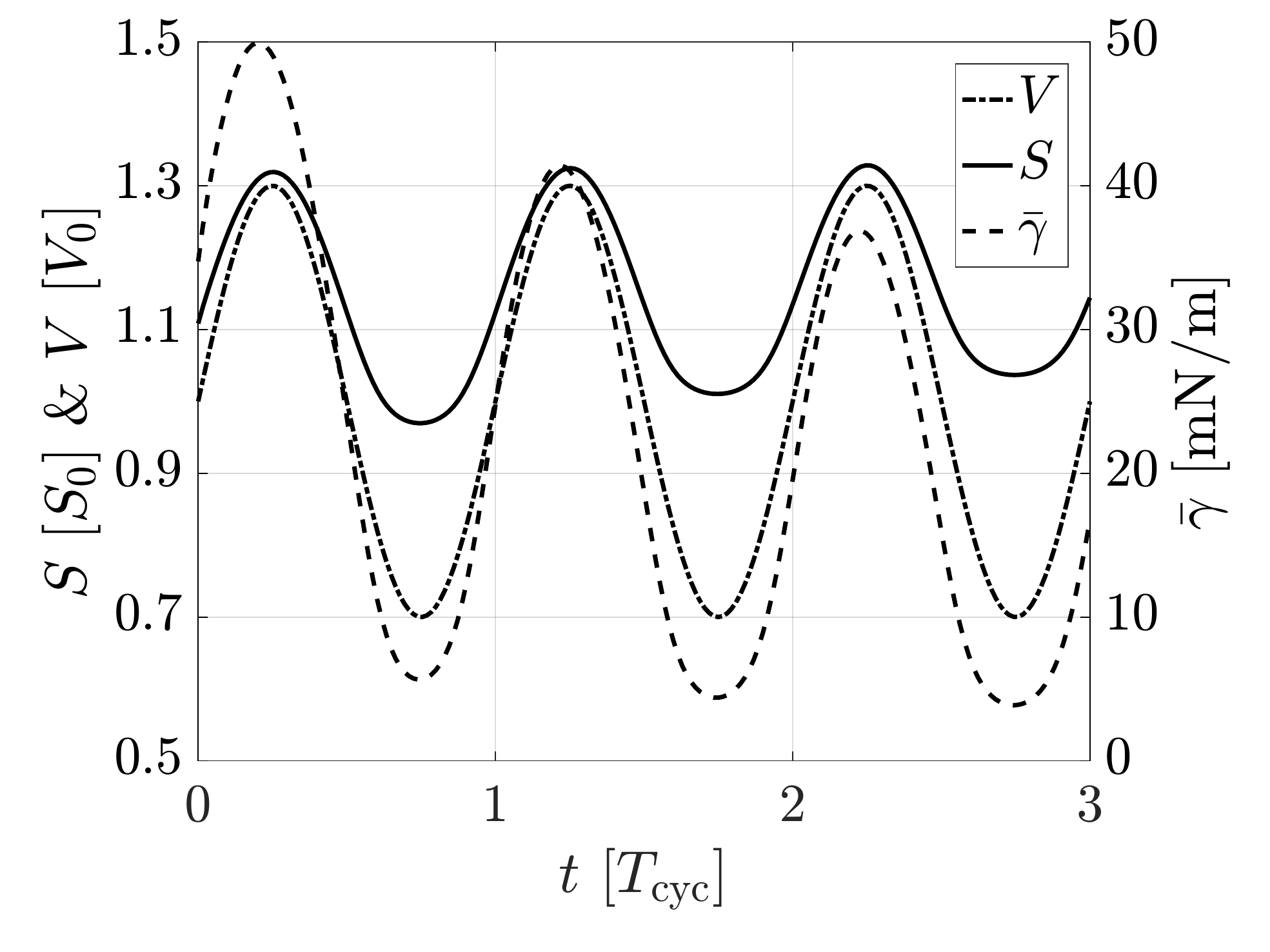}}
\put( 0.1,0.6){b)}
\end{picture}
\caption{The SD test ($\theta_\mrc = 60^\circ$): Cyclic alteration of volume $V$, surface area $S$ and the mean surface tension $\bar\gamma$ for a) the AL and b) the CR models.}
\label{f:sd_curves_thc60}
\end{center}
\end{figure}

\begin{figure}[ht!]
\begin{center} \unitlength1cm
\unitlength1cm
\begin{picture}(0,4.5)
\put(-6.8,0.0){\includegraphics[height=45mm]{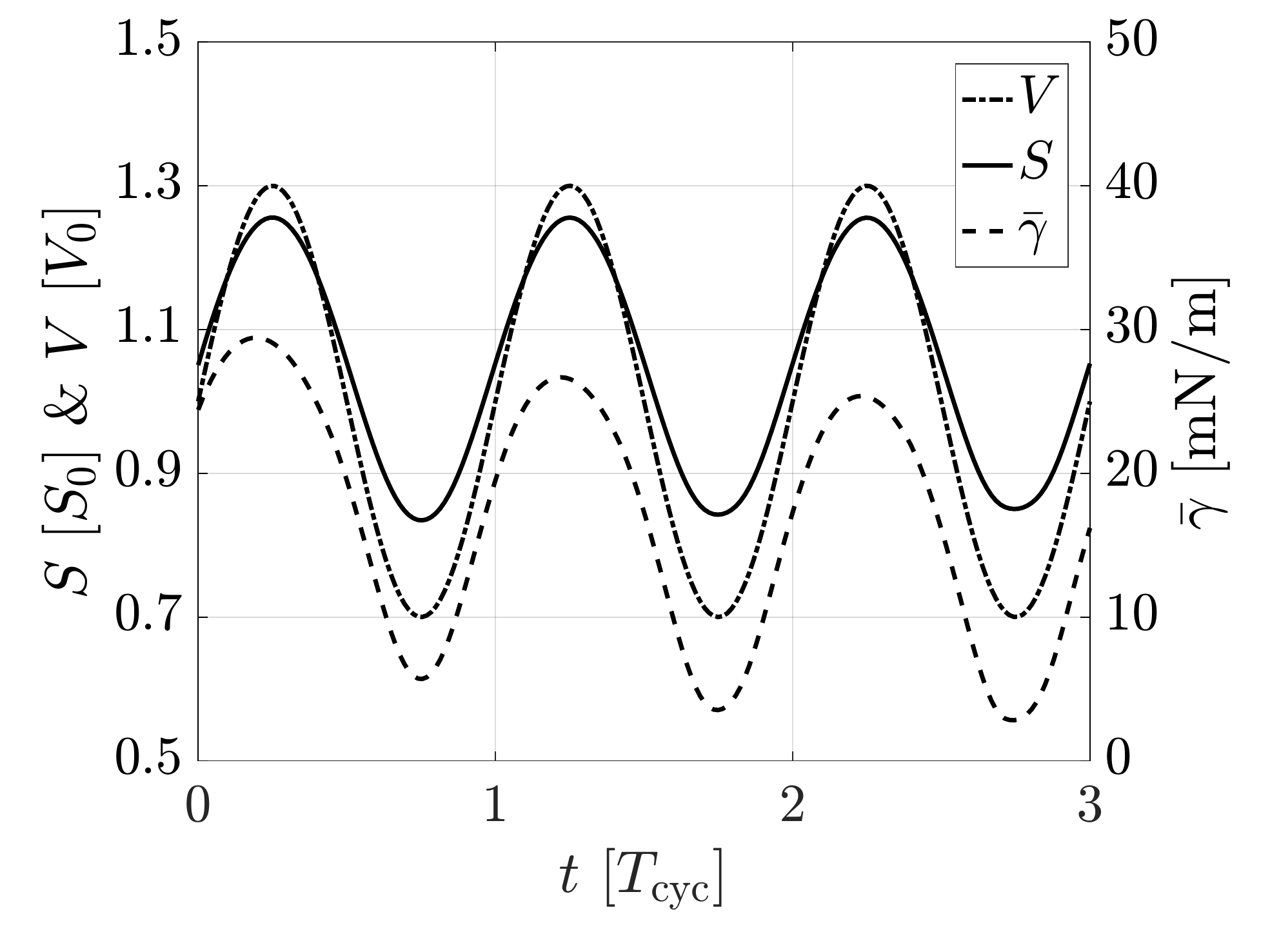}}
\put(-7.0,0.6){a)}
\put( 0.3,0.0){\includegraphics[height=45mm]{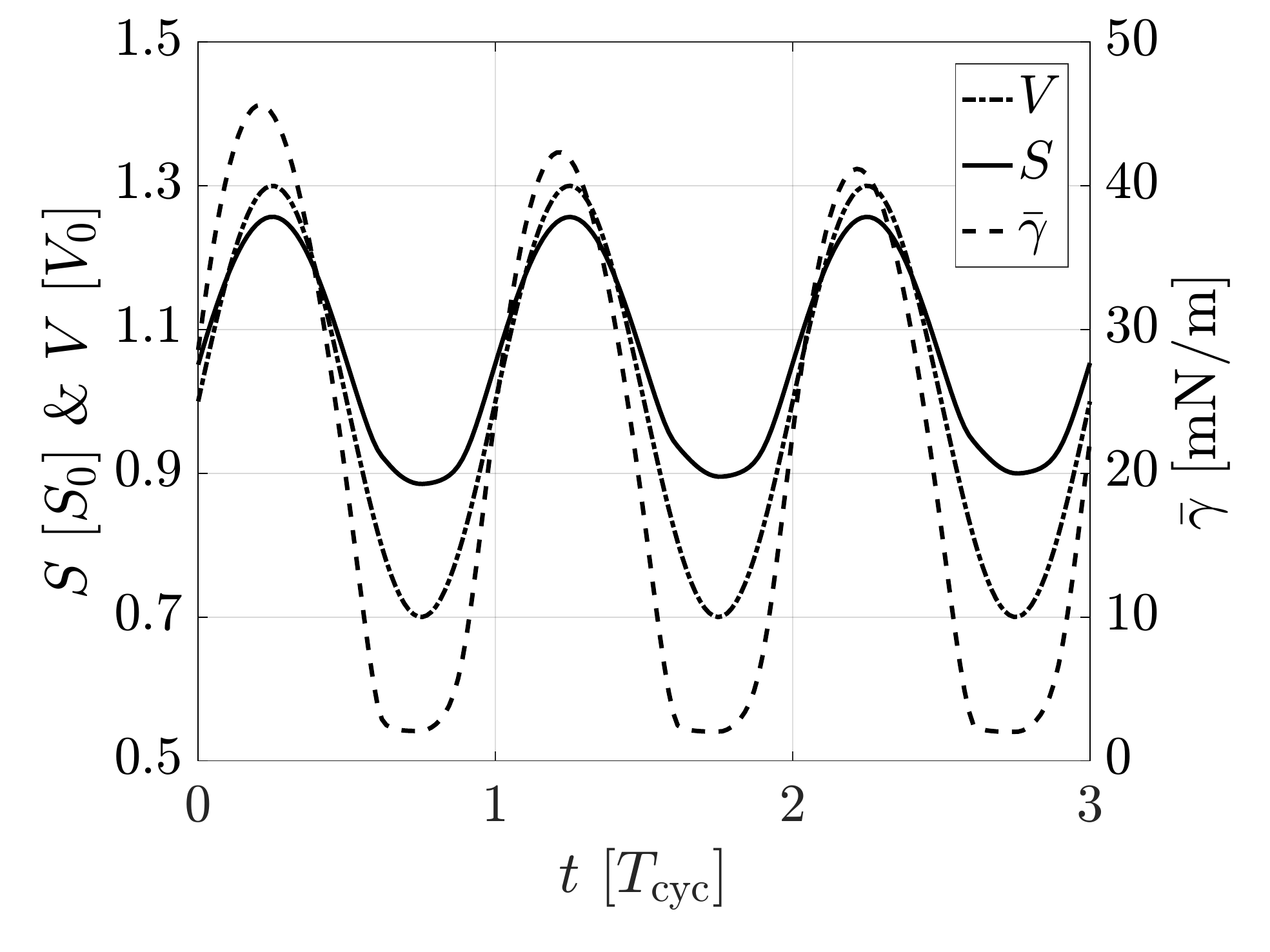}}
\put( 0.1,0.6){b)}
\end{picture}
\caption{The SD test ($\theta_\mrc = 120^\circ$): Cyclic alteration of volume $V$, surface area $S$ and the mean surface tension $\bar\gamma$ for a) the AL and b) the CR models.}
\label{f:sd_curves_thc120}
\end{center}
\end{figure}

In Sec.~\ref{s:csd}, a constrained sessile drop, that is pinned at the contact line, is studied. Here, an unconstrained sessile drop is simulated, where the contact line is free to move on the substrate and the contact angle is fixed. Particularly, it is assumed that the contact angle remains the same if the contact line is advancing or receding. One can use the proposed formulation of \citet{frictdroplet} to distinguish between the advancing and receding contact lines, which is not done here. The initial radius of the droplet is $R = 1.0~\mrm\mrm$ and the computational model with the corresponding boundary conditions are shown in Fig.~\ref{f:comp_model}.c. Here, the same material parameters as in the example of Sec.~\ref{s:pd} are used for both the AL and CR models. To enforce the contact line constraint, the formulation of Sec.~\ref{s:cont_drop} is used. The simulations are performed by controlling the drop volume following the volume constraint \eqref{e:gv}. Figs.~\ref{f:sd_drops_thc60}~and~\ref{f:sd_drops_thc120} show the drop shapes for the contact angles $\theta_\mrc = 60^\circ$ and $\theta_\mrc = 120^\circ$, respectively. In Figs.~\ref{f:sd_curves_thc60}~and~\ref{f:sd_curves_thc120}, the dynamic behavior of the models under cyclic change of the drop volume is shown. Similar to Sec.~\ref{s:pd}, the drop volume is controlled following Eq.~\eqref{e:DV} using $T=3~\mrs$ and $\Delta V_\mathrm{max} = 0.3\,V_0$.   


\subsection{Liquid bridge (LB) test}\label{s:lb}
\begin{figure}[ht!]
\begin{center} \unitlength1cm
\unitlength1cm
\begin{picture}(0,4.5)
\put(-7.0,0.0){\includegraphics[height=45mm]{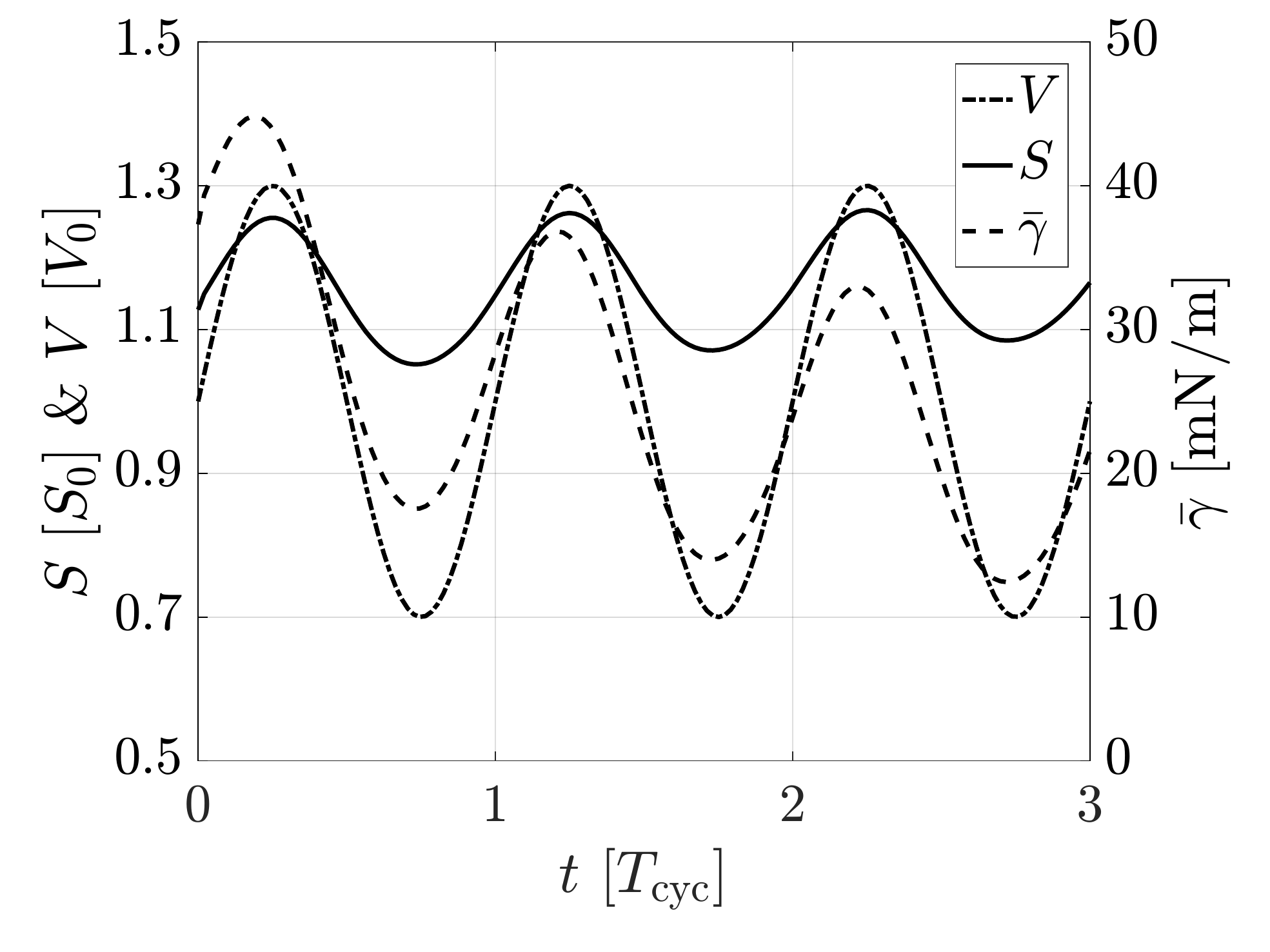}}
\put(-7.0,0.5){a)}
\put( 0.0,0.0){\includegraphics[height=45mm,trim={0 150px 0 300px},clip]{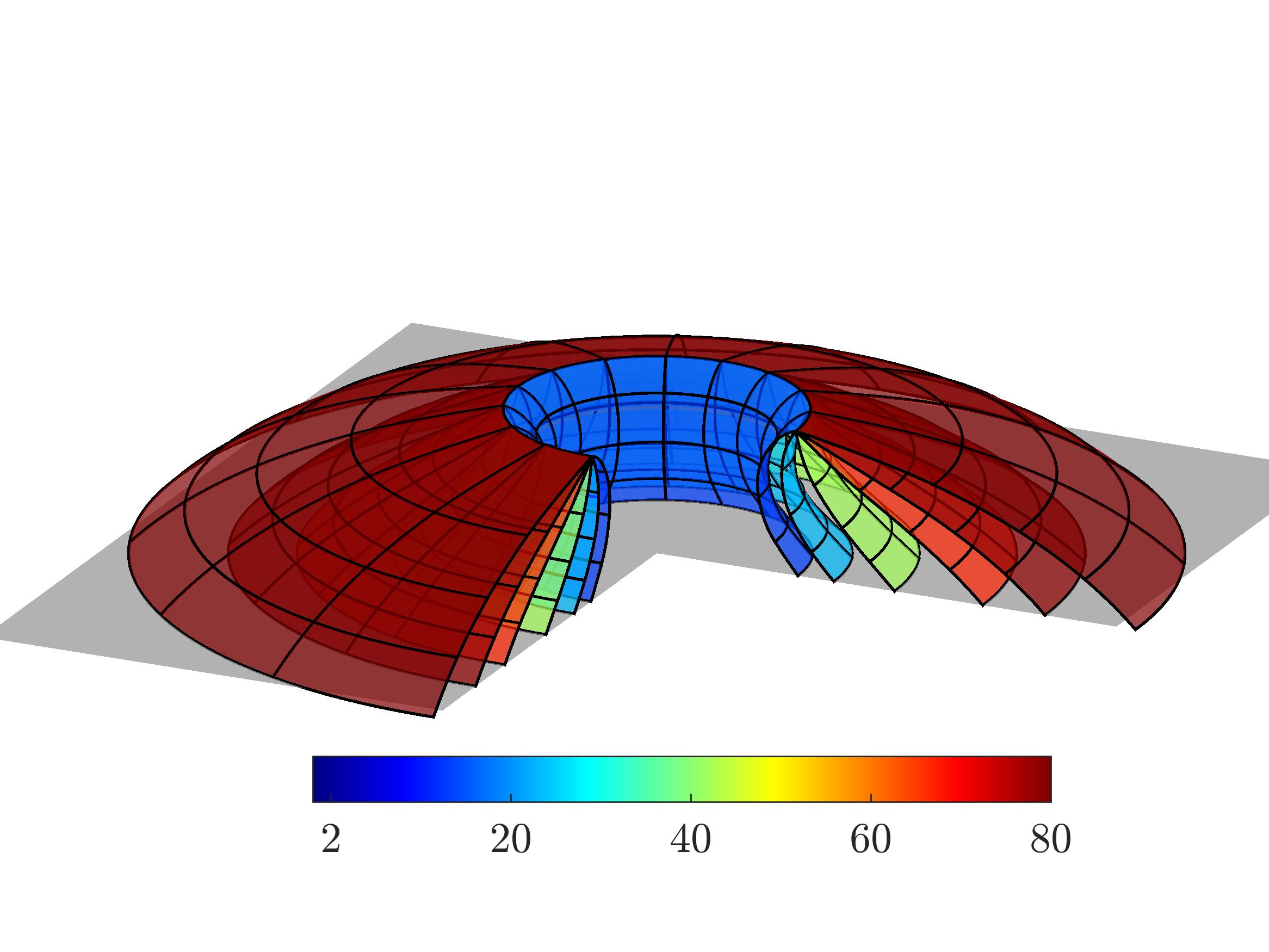}}
\put( 0.1,0.5){b)}
\end{picture}
\caption{The LB test ($\theta_\mrc = 45^\circ$): a) Cyclic alteration of volume $V$, surface area $S$ and the mean surface tension $\bar\gamma$ for and b) menisci shapes modeled by the CR models and colored by $\gamma$ at $V = 0.5,~1,~2,~4,~6~\&~10V_0 $.}
\label{f:lb_thc45}
\end{center}
\end{figure}

This section is devoted to show the capability of the presented computational model to simulate liquid bridges or menisci. As illustrated in Fig.~\ref{f:comp_model}.d, in the reference configuration, the liquid bridge is modeled as 1/4 of a cylinder to benefit from the symmetry of the problem. Accordingly, the symmetry boundary conditions are applied on the axial edges. On the top edge, the liquid bridge with length $L=2\,\mrm\mrm$ is pinned to the end of a needle or a small flat circular holder with radius $R=2\,\mrm\mrm$. On the bottom edge, the contact angle constraint is enforced following the approach of Sec.~\ref{s:cont_gen}. Here, only the results of the CR model are shown and the material parameters and the loading conditions are the same as in the example of Sec.~\ref{s:pd} \tred{and $\Delta t = 0.04~\mrs$}. As it can be seen in Fig.~\ref{f:lb_thc45}, the cyclic deformation of a liquid bridge can be efficiently predicted by the presented formulation.

\subsection{Sessile drop with line tension}\label{s:sd_lt}
\begin{figure}[H]
\begin{center} \unitlength1cm
\unitlength1cm
\begin{picture}(0,5)
\put(-8.0,0.0){\includegraphics[height=50mm]{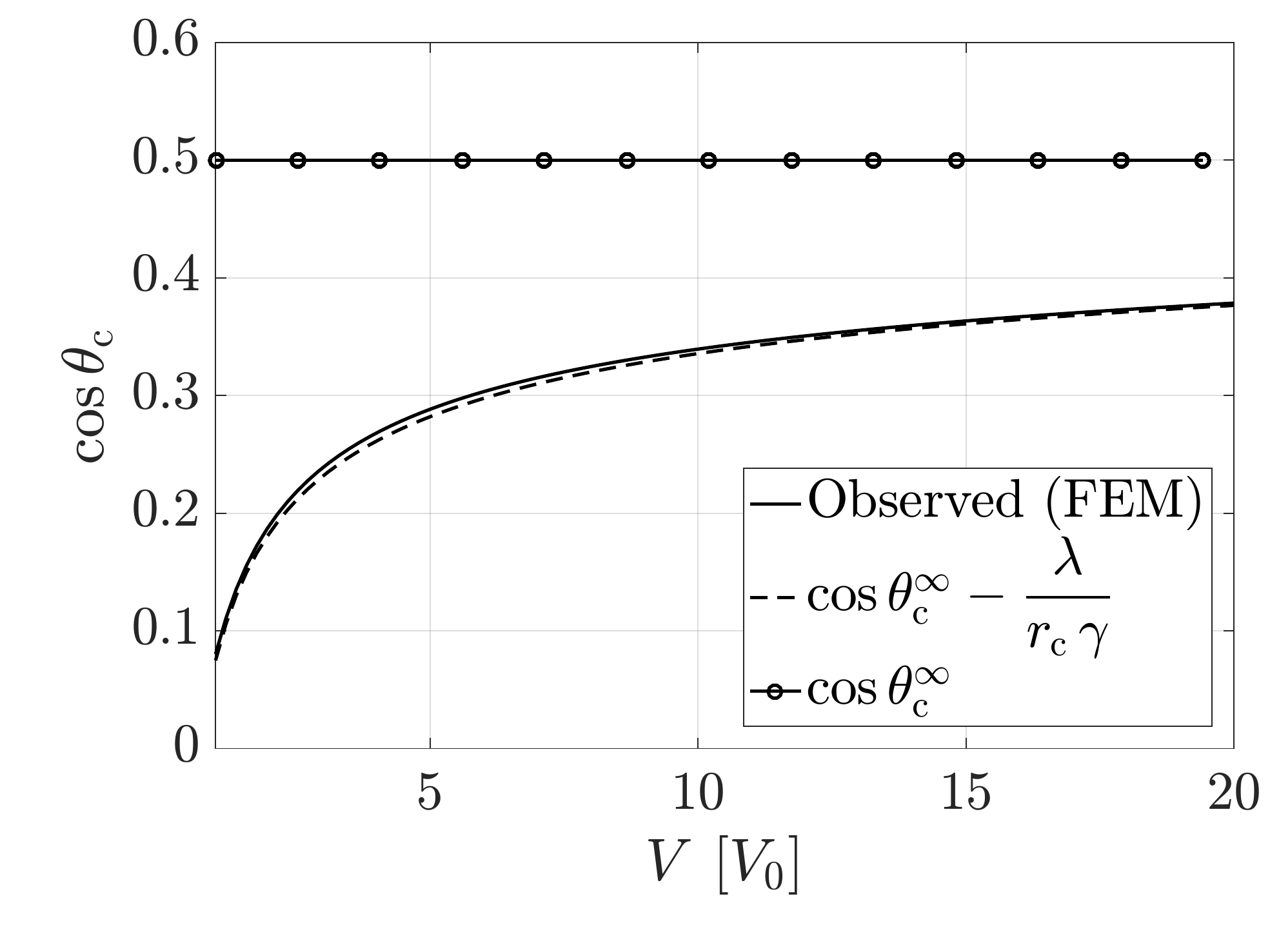}}
\put(-7.5,0.5){a)}
\put(-0.5,0.0){\includegraphics[height=50mm,trim={0 150px 0 200px},clip]{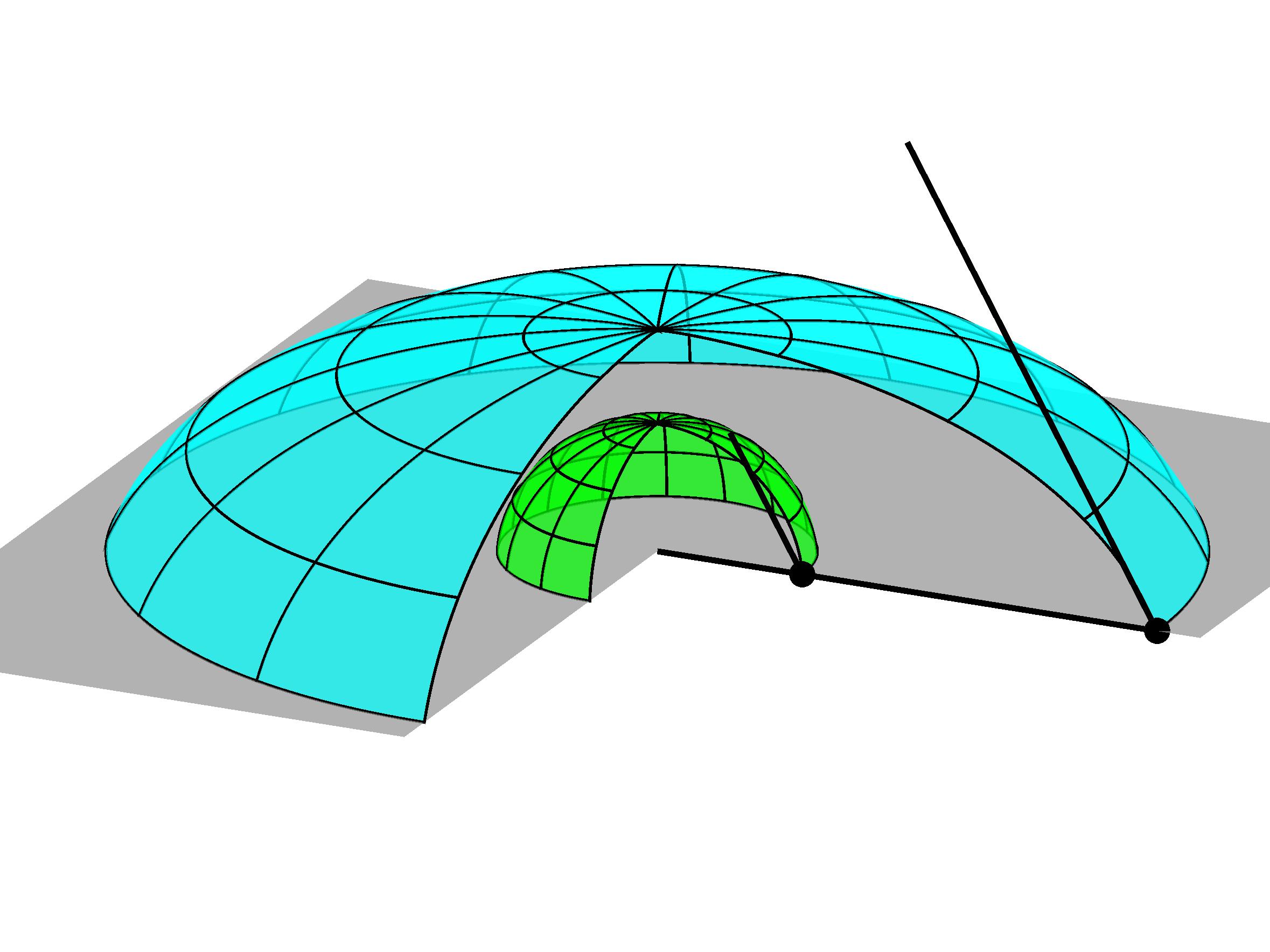}}
\put(-0.5,0.5){b)}
\end{picture}
\caption{The SD test with line tension ($\theta_\mrc^\infty = 60^\circ$): a) Changes of $\cos\theta_\mrc$ w.r.t.~the drop volume and b) two drop shapes at $V = V_0 $ and $V = 20\,V_0 $. The contact angle $\theta_\mrc = 60^\circ$ is shown by dark solid lines.}
\label{f:sd_lt}
\end{center}
\end{figure}

In the examples shown so far, the influence of line tension was neglected. Here, it is investigated how the line tension can affect the apparent contact angle. For this purpose, the unconstrained sessile drop of Sec.~\ref{s:sd} is re-examined. To study only the influence of the line tension, here the surface tension is assumed to be constant, namely $\gamma = 22~\mrm\mrN/\mrm$ and the line tension is set as $\lambda = \gamma\,R/2$. The contact angle $\theta_\mrc^\infty = 60^\circ$ is imposed following the droplet contact line model of Sec.~\ref{s:cont_drop}. The volume of the droplet is increased from $V=V_0$ to $V=20\,V_0$ and gravity is taken into account similar to the previous drop examples. As shown in Fig.~\ref{f:sd_lt}.a, here the contact angle observed during the FE simulation is compared with the contact angle predicted by Eq.~\eqref{e:yl3}. As it is expected, by increasing the radius of the contact line, the contribution of the line tension decreases and $\theta_\mrc$ approaches $\theta_\mrc^\infty$. Here, the contact angle is measured at the same point along the contact line, denoted by a filled circle in Fig.~\ref{f:sd_lt}.b, where $\theta_\mrc^\infty$ is also shown by thick solid lines. The small difference between the observed contact angle and the value predicted by Eq.~\eqref{e:yl3} decreases as the mesh is refined. Therefore, depending on the values of the surface tension, line tension and contact line radius, the contribution of line tension can be efficiently incorporated into the model.  

\section{Conclusion}\label{s:conc}
A new computational framework is presented to model dynamic concentration-dependent surface tension of liquids. In particular, the new formulation is adapted to two material models, namely the adsorption-limited (AL) model of \citet{otis94} and the compression-relaxation (CR) model of \citet{saad10}, which have been developed to study dynamic surface tension of pulmonary surfactants. The models are adapted here to a general continuum membrane formulation in the framework of arbitrary large deformations. The presented model can also be used for other similar constitutive laws, where the surface tension depends on the local concentration of surfactant and the local interface stretch. As discussed, the dynamics of pulmonary surface tension plays an important role in the proper function of lungs and any deficiency or disorder in surfactant behavior can result in serious pulmonary diseases such as respiratory distress syndrome (RDS). Thus, the development of computational models, such as the presented framework, is of great importance for a better understanding of lung biomechanics and for developing new methods that can prevent pulmonary disorders or provide better treatments for them. 

As it is shown through multiple numerical examples, the presented formulation can be used to simulate liquid films, drops (such as constrained and unconstrained sessile drops and pendant drops) and liquid menisci (or liquid bridges). An efficient formulation is presented to model the contact line for liquids of arbitrary shapes that are in contact with rigid planar smooth solid substrates. In this regard, a general formulation and a specific one for droplets is presented. Further, the contribution of the line tension to the contact line is investigated. The presented approach for modeling dynamic surface tension can be extended to liquid films with bending resistance \citep{liquidshell} and can be coupled to fluid flow within the droplet \citep{sauer18mono}. The introduced liquid membrane formulation can be used through a parameter identification process to obtain the material parameters of the presented surfactant models. The application of such an inverse analysis is left for future work. \tred{Furthermore, in future, the multiplicative decomposition of the deformation into elastic, viscous and swelling contribution can be considered following \citet{sauer18multi}.} 

%
\vspace{8mm}
{\bf Acknowledgment}\\
Financial support from the German Research Foundation (DFG) through grant GSC 111, is gratefully acknowledged. The authors also wish to thank {Katharina Immel} for her help in surveying the literature.
%
%
\appendix 

\section{Contribution of line tension to the weak form}\label{s:wf_lt}
Along the contact line $\sC$, 
\eqb{l}
\dif s = \norm{\dif\bx_\mrc} = \norm{\ba_\mrc}\,\dif\xi~,
\label{e:ds}\eqe
where $\xi\in[-1,~1]$ is the parameterization of curve $\sC$ according to $\bx_\mrc = \bx_\mrc(\xi)$ such that 
\eqb{l}
\ba_\mrc := \ds\pa{\bx_\mrc}{\xi}
\eqe
and $s$ denotes the arc length. It should be noted that this parameterization of curve $\sC$ is associated to the parametrization of surface $\sS$ introduced in Sec.~\ref{s:kin}. Strictly, $\xi:=\xi^1$ with $\xi^2\pm1$ or $\xi:=\xi^2$ with $\xi^1\pm1$. As shown by \citet{droplet}, 
\eqb{l}
\delta\norm{\ba_\mrc} = \norm{\ba_\mrc}^{-1}\,\ba_\mrc\cdot\delta\ba_\mrc 
\eqe
and from Eq.~\eqref{e:ds}
\eqb{l}
\delta\dif s = \delta\norm{\ba_\mrc}\,\dif\xi ~.
\label{e:dds}\eqe

Besides, from Eq.~\eqref{e:Wl}, 
\eqb{l}
\delta W_\lambda := \ds\int_\sC \lambda\,\delta\dif s ~.
\label{e:dWl}\eqe

Plugging Eq.~\eqref{e:dds} into Eq.~\eqref{e:dWl}, the contribution of line tension to weak form \eqref{e:wf_2} is obtained as given by Eq.~\eqref{e:Gextt}.

\section{FE tangent matrices}\label{s:fe_tangent}
In this appendix, the internal stiffness tangents that allow for the nonlinearities of surface tension $\gamma = \gamma(\bx,t)$ and the tangent matrix of the contact line force vector \eqref{e:fextc2} and \eqref{e:fextc1} are derived. The other tangent matrices can be found in \citetlist{membrane,droplet,frictdroplet}. 

\subsection{Internal stiffness tangent}\label{s:int_tang}
The spacial variation of the internal force vector $\mf^e_\mathrm{int}$, given in Eq.~(\ref{e:fint}), is
\eqb{l}
\Delta_\mrx\mf_\mathrm{int}^e 
= \ds\int_{\Omega^e_0}\mN^T_{,\alpha}\,\Delta_\mrx\tau^{\alpha\beta}\,\ba_\beta\,\dif A
+ \ds\int_{\Omega^e_0}\mN^T_{,\alpha}\,\tau^{\alpha\beta}\,\mN_{,\beta}\,\dif A\,\Delta\mx_e ~,
\label{e:Dfint}\eqe
where $\tau^{\alpha\beta} := J\,\sigma^{\alpha\beta}$. Since for all the models considered here, $\gamma$ depends on $\bx$ through $J$ only, the linearization of the stress tensor $\tau^{\alpha\beta}$ \tred{from Eq.~\eqref{e:sigab1}} yields
\eqb{lll}
\Delta_\mrx\tau^{\alpha\beta} \is \gamma\,J\,\Delta_\mrx a^{\alpha\beta} + a^{\alpha\beta}\,\Big(\gamma + J\,\dfrac{\partial\gamma}{\partial J}\Big)\Delta_\mrx J \\[3mm]
\mi \tred{\eta\,\Big(\dot{a}^{\alpha\beta}\,\Delta_\mrx J + J\,\Delta_\mrx\dot{a}^{\alpha\beta} + \Delta_\mrx \dot{J}\,a^{\alpha\beta} + \dot{J}\,\Delta_\mrx a^{\alpha\beta}\Big)}~,
\label{e:Dtau1}\eqe
where the two front terms are the variations due to the surface tension and the rear \tred{terms are the variations} due to the time-discretized viscous stress. Following \citet{shelltheo}
\eqb{l}
\Delta_\mrx a^{\alpha\beta} = a^{\alpha\beta\gamma\delta}\,\Delta_\mrx a_{\gamma\delta} ~,
\label{e:Daab}\eqe
with
\eqb{l}
a^{\alpha\beta\gamma\delta} := -\dfrac{1}{2}\big( a^{\alpha\gamma}\,a^{\beta\delta} + a^{\alpha\delta}\,a^{\beta\gamma} \big)~, 
\eqe
\eqb{l} 
\Delta_\mrx J = \dfrac{J}{2}a^{\alpha\beta}\,\Delta_\mrx a_{\alpha\beta}~,
\label{e:DJ}\eqe
and
\eqb{l}
\Delta_\mrx a_{\alpha\beta} = \big( \ba_\alpha\cdot\mN_{,\beta} + \ba_\beta\cdot\mN_{,\alpha} \big)\,\Delta\mx_e ~.
\eqe
\tred{Further, from Eqs.~\eqref{e:BaEu_aab}~and~\eqref{e:Daab},
\eqb{l}
\Delta_\mrx\dot{a}^{\alpha\beta} = \dfrac{1}{\Delta t}\,\Delta_\mrx a^{\alpha\beta} = \dfrac{1}{\Delta t}\,a^{\alpha\beta\gamma\delta}\,\Delta_\mrx a_{\gamma\delta} ~, 
\label{e:Ddotaab}\eqe
and
\eqb{lll}
\Delta_\mrx \dot{J} \is -\dfrac{1}{2}\,\big(\dot{a}^{\alpha\beta}\,a_{\alpha\beta}\,\Delta_\mrx J + J\,a_{\alpha\beta}\,\Delta_\mrx \dot{a}^{\alpha\beta} + J\,\dot{a}^{\alpha\beta}\,\Delta_\mrx a^{\alpha\beta}\big) \\[3mm]
\is \dfrac{1}{2}\,\Big[\Big(\dot{J} + \dfrac{1}{\Delta t}\,J\Big)\,a^{\alpha\beta} - J\,\dot{a}^{\alpha\beta} \Big]\,\Delta_\mrx a_{\alpha\beta} ~,
\eqe
according to Eqs.~\eqref{e:dotJ},~\eqref{e:DJ}~and~\eqref{e:Ddotaab}. Here, $\dot{J}$ and $\dot{a}^{\alpha\beta}$ are obtained following Eqs.~\eqref{e:dotJ}~and~\eqref{e:BaEu_aab}, respectively.} Thus, Eq.~\eqref{e:Dtau1} can be reformulated as
\eqb{l}
\Delta_\mrx\tau^{\alpha\beta} = c^{\alpha\beta\gamma\delta}\,\dfrac{1}{2}\Delta_\mrx a_{\alpha\beta} ~,
\label{e:Dtau2}\eqe
with
\eqb{lll}
c^{\alpha\beta\gamma\delta} := 2\,\dfrac{\partial \tau^{\alpha\beta}}{\partial a_{\gamma\delta}}
\is 2\,\gamma\,J\,a^{\alpha\beta\gamma\delta} +
\Big(\gamma + J\,\dfrac{\partial\gamma}{\partial J}\Big)\,J\,a^{\alpha\beta}\,a^{\gamma\delta} \\[3mm]
\mi \tred{\eta\,\Big[J\,\big(\dot{a}^{\alpha\beta}\,a^{\gamma\delta} - a^{\alpha\beta}\,\dot{a}^{\gamma\delta}\big) + \Big(\dot{J} + \dfrac{1}{\Delta t}\,J\Big)\big(a^{\alpha\beta}\,a^{\gamma\delta} + 2\,a^{\alpha\beta\gamma\delta}\big)\Big]}~,
\eqe
where $\partial\gamma/\partial J$ is derived in Secs.~\ref{s:tiALM} and \ref{s:tiCRM} for the {AL} and {CR} models, respectively. 
To determine the element stiffness tangents, Eq.~(\ref{e:Dfint}) can be arranged as \citep{membrane}
\eqb{l}
\Delta\mf^e_{\mathrm{int}}=\big(\mk^e_{\mathrm{mat}} + \mk^e_{\mathrm{geo}}\big)\,\Delta \mx_e~,
\eqe
where the material stiffness matrix
\eqb{l}
\mk^e_{\mathrm{mat}} = \ds\int_{\Omega^e_0}c^{\alpha\beta\gamma\delta}\,\mN^T_{,\alpha}\, (\ba_{\beta}\otimes\ba_{\gamma})\,\mN_{,\delta}\, \dif A
\eqe
and the geometric stiffness matrix
\eqb{l}
\mk^e_{\mathrm{geo}} = \ds\int_{\Omega^e_0}\mN_{,\alpha}^T\,\tau^{\alpha\beta}\,\mN_{,\beta}\,\dif A
\eqe
are introduced.

\subsection{Tangent matrix of the contact line force vector}\label{dcont}
According to Eq.~\eqref{e:ds}, Eq.~\eqref{e:fextc1} can be written as
\eqb{l}
\mf_\mathrm{extc}^e = \ds\frac{1}{2}\,\cot\theta_\mrc\ds\int_{-1}^{+1} \mN^\mrT_\mrt\,p_\mrc\,\bm_\mrc\,\norm{\br_\mrc}\,\norm{\ba_\mrc}\,\dif\xi ~.
\label{e:fcont}
\eqe
The linearization of Eq.~\eqref{e:fcont} results in
\eqb{l}
\Delta\mf_\mathrm{extc}^e = \mk_\mathrm{extc}^e\,\Delta \mx_e + \ml_\mathrm{extc}^e\,\Delta p ~.
\label{e:dfcont}
\eqe
For a rigid planar substrate, where $\Delta\bn_\mrc = 0$, \citep{droplet}
\eqb{l}
\Delta\bm_\mrc =-\big(\ba^\mrc \otimes \bm_\mrc\big)\Delta\ba_\mrc~,
\eqe
with $\ba^\mrc$ being the dual vector of $\ba_\mrc$, such that $\ba^\mrc\cdot\ba_\mrc=1$ as described in Sec.~\ref{s:kin}, and $\Delta\ba_\mrc = \mN_{\mrt,\xi}\Delta \mx_e $. Furthermore, $\Delta \norm{\br_\mrc} = \norm{\br_\mrc}^{-1}\,\br_\mrc\cdot\Delta\bx_\mrc$, where $\Delta\bx_\mrc = \mN_\mrt\,\Delta\mx_e$. Thus, 

\eqb{lll}
\mk_\mathrm{extc}^e =
\ds\frac{1}{2}\,\cot\theta_\mrc\,\ds\int^{1}_{-1}\mN^\mrT_\mrt\,p_\mrc\hspace*{-2ex}&\Big[\hspace*{3pt}
  \norm{\br_\mrc}^{-1}\,\norm{\ba_\mrc}\,\big(\bm_\mrc\otimes\bx\big)\,\mN_\mrt \\[3mm]
&+\norm{\br_\mrc}\,\norm{\ba_\mrc}^{-1}\,\big(\bm_\mrc\otimes\ba_\mrc\big)\,\mN_{\mrt,\xi} \\[3mm]
&-\norm{\br_\mrc}\,\norm{\ba_\mrc}\,\big(\ba^\mrc\otimes\bm_\mrc\big)\,\mN_{\mrt,\xi} 
\Big]\,\dif\xi
\label{e:Kcont}
\eqe
and
\eqb{l}
\ml_\mathrm{extc}^e = \ds\frac{1}{2}\,\cot \theta_\mrc\ds\int_{-1}^{+1} \mN^\mrT_\mrt\,\bm_\mrc\,\norm{\br_\mrc}\,\norm{\ba_\mrc}\,\dif\xi~.
\label{e:lcont}
\eqe

Alternatively, if Eq.~\eqref{e:fextc2} is used, the linearization yields
\eqb{lll}
\mk_\mathrm{extc}^e \is
\cos\theta_\mrc\,\ds\int^{1}_{-1}\mN^\mrT_\mrt\,\gamma(\bx_\mrc)\Big[\hspace*{3pt}
\norm{\ba_\mrc}^{-1}\,\big(\bm_\mrc\otimes\ba_\mrc\big)\,\mN_{\mrt,\xi}
-\norm{\ba_\mrc}\,\big(\ba^\mrc\otimes\bm_\mrc\big)\,\mN_{\mrt,\xi} 
\Big]\,\dif\xi  \\[3mm]
\plus \cos\theta_\mrc\,\ds\int^{1}_{-1}\mN^\mrT_\mrt\,\dfrac{\partial\gamma(J)}{\partial J}\,J\,\big(\bm_\mrc\otimes\ba^\alpha\big)\,\mN_{,\alpha}\,\dif\xi ~,
\label{e:Kcont2}
\eqe
where the rear term explicitly depends on the nonlinearities of the surface tension $\gamma = \gamma(J) = \gamma(\bx_\mrc)$ in general and vanishes if the surface tension is constant as for typical pure liquids. 

If the effect of line tension is considered, following Eq.~\eqref{e:fintt}, the tangent matrix 
\eqb{l}
\mk_{\mathrm{int}\lambda}^e := \ds\int_{-1}^{+1} \mN_{\mrt,\xi}^\mrT\,\lambda\,\norm{\ba_\mrc}^{-1}\,\mN_{\mrt,\xi}\,\dif\xi - 
\ds\int_{-1}^{+1} \mN_{\mrt,\xi}^\mrT\,\lambda\,\norm{\ba_\mrc}^{-3}\big(\ba_\mrc \otimes \ba_\mrc \big)\,\mN_{\mrt,\xi}\,\dif\xi
\label{e:Kintt}\eqe
needs to be added to the internal tangent matrix.

%
\vspace{5mm}
{\bf Conflict of Interest}\\
The authors declare that they have no conflict of interest.
%
\bibliographystyle{apalike}
\bibliography{SurfBib,bibliography}

\end{document}

%% file: neco_updated.tex
\newcommand{\bitm}{\begin{itemize}}
\newcommand{\eitm}{\end{itemize}}
\newcommand{\bnumr}{\begin{enumerate}}
\newcommand{\enumr}{\end{enumerate}}

\newcommand {\eqb}[1]{\begin{equation}\begin{array}{#1}}
\newcommand {\eqe}{\end{array}\end{equation}}

\newcommand {\esb}[1]{\begin{equation*}\begin{array}{#1}}
\newcommand {\ese}{\end{array}\end{equation*}}
\newcommand {\ds}{\displaystyle}

\newcommand {\pa}[2]{\frac{\partial{#1}}{\partial{#2}}}

\newcommand {\back}{\! \! \!}
\newcommand {\is}{\back &=& \back}
\newcommand {\dis}{\back &:=& \back}

\newcommand {\ais}{\back &\approx& \back}

\newcommand {\plus}{\back &+& \back}
\newcommand {\mi}{\back &-& \back}

\newcommand {\norm}[1]{\|#1\|}
\newcommand {\tr}{\mathrm{tr}\,}

\newcommand {\divz}{\mathrm{div}\,}

\newcommand {\dif}{\mathrm{d}}

\newcommand {\II}{{I\kern-.3em I}}
\newcommand {\III}{{I\kern-.3em I\kern-.3em I}}

\newcommand {\mra}{\mathrm{a}}

\newcommand {\mrc}{\mathrm{c}}

\newcommand {\mre}{\mathrm{e}}
\newcommand {\mrf}{\mathrm{f}}
\newcommand {\mrg}{\mathrm{g}}
\newcommand {\mrh}{\mathrm{h}}

\newcommand {\mrk}{\mathrm{k}}

\newcommand {\mrm}{\mathrm{m}}
\newcommand {\mrn}{\mathrm{n}}

\newcommand {\mrr}{\mathrm{r}}
\newcommand {\mrs}{\mathrm{s}}
\newcommand {\mrt}{\mathrm{t}}

\newcommand {\mrv}{\mathrm{v}}

\newcommand {\mrx}{\mathrm{x}}

\newcommand {\mrC}{\mathrm{C}}

\newcommand {\mrN}{\mathrm{N}}

\newcommand {\mrT}{\mathrm{T}}

\newcommand {\mf}{\mathbf{f}}

\newcommand {\mh}{\mathbf{h}}

\newcommand {\mk}{\mathbf{k}}
\newcommand {\ml}{\mathbf{l}}

\newcommand {\mw}{\mathbf{w}}
\newcommand {\mx}{\mathbf{x}}

\newcommand {\ba}{\boldsymbol{a}}
\newcommand {\bb}{\boldsymbol{b}}

\newcommand {\bff}{\boldsymbol{f}}
\newcommand {\bg}{\boldsymbol{g}}

\newcommand {\bm}{\boldsymbol{m}}
\newcommand {\bn}{\boldsymbol{n}}

\newcommand {\bq}{\boldsymbol{q}}
\newcommand {\br}{\boldsymbol{r}}

\newcommand {\bt}{\boldsymbol{t}}
\newcommand {\bu}{\boldsymbol{u}}
\newcommand {\bv}{\boldsymbol{v}}
\newcommand {\bw}{\boldsymbol{w}}
\newcommand {\bx}{\boldsymbol{x}}

\newcommand {\mF}{\mathbf{F}}

\newcommand {\mK}{\mathbf{K}}

\newcommand {\mN}{\mathbf{N}}

\newcommand {\mU}{\mathbf{U}}

\newcommand {\mX}{\mathbf{X}}

\newcommand {\bX}{\boldsymbol{X}}

\newcommand {\sig}{\sigma}

\newcommand {\bsig}{\mbox{\boldmath$\sigma$}}

\newcommand {\bone}{\mathbf{1}}

\newcommand {\IR}{{\rm\kern.24em
   \vrule width.02em height1.53ex depth-.05ex
   \kern-.3em R}}
\newcommand {\ic}{{\rm\kern.20em
   \vrule width.02em height1.0ex depth-.05ex
   \kern-.22em c}}
\newcommand {\ia}{{\rm\kern.20em
   \vrule width.02em height1.05ex depth-.0ex
   \kern-.25em a}}
\newcommand {\IC}{{\rm\kern.24em
   \vrule width.02em height1.4ex depth-.05ex
   \kern-.26em C}}
\newcommand {\ID}{{\rm\kern.34em
   \vrule width.02em height1.5ex depth-.05ex
   \kern-.36em D}}
\newcommand {\IS}{{\rm\kern.24em
   \vrule width.02em height1.6ex depth.05ex
   \kern-.26em S}}
\newcommand {\IT}{{\rm\kern.50em
   \vrule width.02em height1.55ex depth-.05ex
   \kern-.52em T}}

\newcommand {\IE}{{\rm\kern.24em
   \vrule width.02em height1.55ex depth-.05ex
   \kern-.33em E}}
\newcommand {\IEa}{{\rm\kern.24em
   \vrule width.02em height1.55ex depth-.05ex
   \kern-.33em E}^{1}_{ijkl}}
\newcommand {\IEb}{{\rm\kern.24em
   \vrule width.02em height1.55ex depth-.05ex
   \kern-.33em E}^{2}_{ijkl}}

\newcommand {\sB}{\mathcal{B}}
\newcommand {\sC}{\mathcal{C}}
\newcommand {\sD}{\mathcal{D}}

\newcommand {\sP}{\mathcal{P}}

\newcommand {\sS}{\mathcal{S}}

\newcommand {\sW}{\mathcal{W}}

\newcommand {\Ass}[2]{\kern 0.9ex \vrule width0.45em height0.2ex depth0ex \kern -2.1ex \bigwedge_{#1}^{#2}}
\newcommand {\ASS}[2]{\kern 1.45ex \vrule width0.5em height0.2ex depth0ex \kern -2.65ex \bigwedge_{#1}^{#2}}

%% file: surfactant_v12.bbl
\begin{thebibliography}{}

\bibitem[Alonso et~al., 2004]{alonso04}
Alonso, C., Alig, T., Yoon, J., Bringezu, F., Warriner, H., and Zasadzinski,
  J.~A. (2004).
\newblock More than a monolayer: {Relating} lung surfactant structure and
  mechanics to composition.
\newblock {\em Biophys. J.}, {\bf 87}(6):4188--4202.

\bibitem[Archie, 1973]{archie73}
Archie, J. (1973).
\newblock A mathematical model for pulmonary mechanics: The alveolar surface
  contribution.
\newblock {\em Int. J. Engrg. Sci.}, {\bf 11}(6):659--671.

\bibitem[Bachofen and Sch{\"u}rch, 2001]{bachofen01}
Bachofen, H. and Sch{\"u}rch, S. (2001).
\newblock Alveolar surface forces and lung architecture.
\newblock {\em Compar. Biochem. Physiol. A Molec. Integ. Physiol.}, {\bf
  129}(1):183--193.

\bibitem[Bangyozova et~al., 2017]{bangyozova17}
Bangyozova, M., Jordanova, A., Tsanova, A., Stoyanova, V., Tasheva, E.,
  Ivanova, K., Todorov, R., Hristova, E., and Lalchev, Z. (2017).
\newblock Application of axisymmetric drop shape analysis and brewster angle
  microscopy for assessment of clinical samples from prematurely born infants
  with {NRDS}.
\newblock {\em Coll. Surf. A Physicochem. Engrg. Asp.}, {\bf 519}:187--191.

\bibitem[Borden et~al., 2011]{borden11}
Borden, M.~J., Scott, M.~A., Evans, J.~A., and Hughes, T. J.~R. (2011).
\newblock Isogeometric finite element data structures based on {B}ezier
  extraction of {NURBS}.
\newblock {\em Int. J. Numer. Meth. Engng.}, {\bf 87}:15--47.

\bibitem[Butt et~al., 2006]{butt06}
Butt, H.-J., Graf, K., and Kappl, M. (2006).
\newblock {\em Physics and chemistry of interfaces}.
\newblock John Wiley \& Sons, Weinheim.

\bibitem[Clements, 1957]{clements57}
Clements, J.~A. (1957).
\newblock Surface tension of lung extracts.
\newblock {\em Exper. Bio. Med.}, {\bf 95}(1):170--172.

\bibitem[Dale et~al., 1980]{dale80}
Dale, P.~J., Matthews, F.~L., and Schroter, R.~C. (1980).
\newblock Finite element analysis of lung alveolus.
\newblock {\em J. Biomech.}, {\bf 13}(10):865--873.

\bibitem[Denny and Schroter, 1995]{denny95}
Denny, E. and Schroter, R. (1995).
\newblock The mechanical behavior of a mammalian lung alveolar duct model.
\newblock {\em J. Biomech. Engrg.}, {\bf 117}(3):254--261.

\bibitem[Denny and Schroter, 1997]{denny97}
Denny, E. and Schroter, R. (1997).
\newblock Relationships between alveolar size and fibre distribution in a
  mammalian lung alveolar duct model.
\newblock {\em J. Biomech. Engrg.}, {\bf 119}(3):289--297.

\bibitem[Denny and Schroter, 2000]{denny00}
Denny, E. and Schroter, R. (2000).
\newblock {Viscoelastic Behavior of a Lung Alveolar Duct Model}.
\newblock {\em J. Biomech. Engrg.}, {\bf 122}(2):143--151.

\bibitem[Denny and Schroter, 2006]{denny06}
Denny, E. and Schroter, R. (2006).
\newblock A model of non-uniform lung parenchyma distortion.
\newblock {\em J. Biomech.}, {\bf 39}(4):652--663.

\bibitem[Duncan et~al., 1995]{duncan95}
Duncan, D., Li, D., Gaydos, J., and Neumann, A. (1995).
\newblock Correlation of line tension and solid-liquid interfacial tension from
  the measurement of drop size dependence of contact angles.
\newblock {\em J. Coll. Interf. Sci.}, {\bf 169}(2):256--261.

\bibitem[Enhorning, 1977]{enhorning77}
Enhorning, G. (1977).
\newblock Pulsating bubble technique for evaluating pulmonary surfactant.
\newblock {\em J. Appl. Physiol.}, {\bf 43}(2):198--203.

\bibitem[Franses et~al., 1996]{franses96}
Franses, E.~I., Basaran, O.~A., and Chang, C.-H. (1996).
\newblock Techniques to measure dynamic surface tension.
\newblock {\em Curr. Opin. Coll. Interf. Sci.}, {\bf 1}(2):296--303.

\bibitem[Ganesan and Tobiska, 2009]{ganesan09}
Ganesan, S. and Tobiska, L. (2009).
\newblock A coupled arbitrary {Lagrangian--Eulerian} and {Lagrangian} method
  for computation of free surface flows with insoluble surfactants.
\newblock {\em J. Comput. Phys.}, {\bf 228}(8):2859--2873.

\bibitem[Goerke, 1998]{goerke98}
Goerke, J. (1998).
\newblock Pulmonary surfactant: functions and molecular composition.
\newblock {\em Biochim. Biophys. Acta Molec. Basis Dis.}, {\bf
  1408}(2-3):79--89.

\bibitem[Gregory et~al., 1991]{gregory91}
Gregory, T., Longmore, W., Moxley, M., Whitsett, J., Reed, C., Fowler, A.,
  Hudson, L., Maunder, R., Crim, C., and Hyers, T. (1991).
\newblock Surfactant chemical composition and biophysical activity in acute
  respiratory distress syndrome.
\newblock {\em J. Clinic. Invest.}, {\bf 88}(6):1976--1981.

\bibitem[Hallman et~al., 2001]{hallman01}
Hallman, M., Glumoff, V., and R{\"a}met, M. (2001).
\newblock Surfactant in respiratory distress syndrome and lung injury.
\newblock {\em Compar. Biochem. Physiol. A Molec. Integ. Physiol.},
  129(1):287--294.

\bibitem[Han and Mallampalli, 2015]{han15}
Han, S. and Mallampalli, R.~K. (2015).
\newblock The role of surfactant in lung disease and host defense against
  pulmonary infections.
\newblock {\em Ann. Amer. Thorac. Soc.}, {\bf 12}(5):765--774.

\bibitem[Hermans et~al., 2015]{hermans15}
Hermans, E., Bhamla, M.~S., Kao, P., Fuller, G.~G., and Vermant, J. (2015).
\newblock Lung surfactants and different contributions to thin film stability.
\newblock {\em Soft Matter}, {\bf 11}(41):8048--8057.

\bibitem[Hildebran et~al., 1979]{hildebran79}
Hildebran, J., Goerke, J., and Clements, J. (1979).
\newblock Pulmonary surface film stability and composition.
\newblock {\em J. Appl. Physiol.}, {\bf 47}(3):604--611.

\bibitem[Hills, 1985]{hills85}
Hills, B.~A. (1985).
\newblock Alveolar liquid lining: Langmuir method used to measure surface
  tension in bovine and canine lung extracts.
\newblock {\em J. Physiol.}, {\bf 359}(1):65--79.

\bibitem[Horn and Davis, 1975]{horn75}
Horn, L.~W. and Davis, S.~H. (1975).
\newblock Apparent surface tension hysteresis of a dynamical system.
\newblock {\em J. Colloid Interface Sci.}, {\bf 51}(3):459--476.

\bibitem[Ingenito et~al., 1999]{ingenito99}
Ingenito, E., Mark, L., Morris, J., Espinosa, F., Kamm, R., and Johnson, M.
  (1999).
\newblock Biophysical characterization and modeling of lung surfactant
  components.
\newblock {\em J. Appl. Physiol.}, {\bf 86}(5):1702--1714.

\bibitem[Karakaplan et~al., 1980]{karakaplan80}
Karakaplan, A., Bieniek, M., and Skalak, R. (1980).
\newblock A mathematical model of lung parenchyma.
\newblock {\em J. Biomech. Engrg.}, {\bf 102}(2):124--136.

\bibitem[Koji{\'c} et~al., 2009]{kojic09}
Koji{\'c}, M., Filipovi{\'c}, N., Stojanovi{\'c}, B., and Koji{\'c}, N. (2009).
\newblock {\em Biological Soft Tissue}, chapter~11, pages 201--225.
\newblock John Wiley \& Sons, Chichester, England Hoboken, NJ.

\bibitem[Koji{\'c} et~al., 2006]{kojic06}
Koji{\'c}, M., Vlastelica, I., Stojanovi{\'c}, B., Rankovi{\'c}, V., and Tsuda,
  A. (2006).
\newblock Stress integration procedures for a biaxial isotropic material model
  of biological membranes and for hysteretic models of muscle fibres and
  surfactant.
\newblock {\em Int. J. Numer. Meth. Engrg.}, {\bf 68}(8):893--909.

\bibitem[Kowe et~al., 1986]{kowe86}
Kowe, R., Schroter, R., Matthews, F., and Hitchings, D. (1986).
\newblock Analysis of elastic and surface tension effects in the lung alveolus
  using finite element methods.
\newblock {\em J. Biomech.}, {\bf 19}(7):541--549.

\bibitem[Krueger and Gaver, 2000]{krueger00}
Krueger, M.~A. and Gaver, D.~P. (2000).
\newblock A theoretical model of pulmonary surfactant multilayer collapse under
  oscillating area conditions.
\newblock {\em J. Colloid Interface Sci.}, {\bf 229}(2):353--364.

\bibitem[Lewis and Jobe, 1993]{lewis93}
Lewis, J.~F. and Jobe, A.~H. (1993).
\newblock Surfactant and the adult respiratory distress syndrome.
\newblock {\em Amer. Rev. Respir. Disease}, {\bf 147}:218--233.

\bibitem[Loglio et~al., 1991]{loglio91}
Loglio, G., Tesei, U., Innocenti, N., Miller, R., and Cini, R. (1991).
\newblock Non-equilibrium surface thermodynamics. {Measurement} of transient
  dynamic surface tension for fluid--fluid interfaces by the trapezoidal pulse
  technique.
\newblock {\em Colloids and Surfaces}, {\bf 57}(2):335--342.

\bibitem[Luginsland and Sauer, 2017]{luginsland17}
Luginsland, T. and Sauer, R.~A. (2017).
\newblock A computational study of wetting on chemically contaminated
  substrates.
\newblock {\em Coll. Surf. A Physicochem. Engrg. Asp.}, {\bf 531}:81--92.

\bibitem[Ma and Ma, 2012]{ma12}
Ma, C. C.-H. and Ma, S. (2012).
\newblock The role of surfactant in respiratory distress syndrome.
\newblock {\em Open Respir. Medic. J.}, {\bf 6}:44--53.

\bibitem[Marmur, 1997]{marmur97}
Marmur, A. (1997).
\newblock Line tension and the intrinsic contact angle in solid--liquid--fluid
  systems.
\newblock {\em J. Coll. Interf. Sci.}, {\bf 186}(2):462--466.

\bibitem[Miller et~al., 1994]{miller94}
Miller, R., Joos, P., and Fainerman, V.~B. (1994).
\newblock Dynamic surface and interfacial tensions of surfactant and polymer
  solutions.
\newblock {\em Adv. Coll. Interf. Sci.}, {\bf 49}:249--302.

\bibitem[Morris et~al., 2001]{moris01}
Morris, J., Ingenito, E., Mark, L., Kamm, R., and Johnson, M. (2001).
\newblock Dynamic behavior of lung surfactant.
\newblock {\em J. Biomech. Engrg.}, {\bf 123}(1):106--113.

\bibitem[Nkadi et~al., 2009]{nkadi09}
Nkadi, P.~O., Merritt, T.~A., and Pillers, D.-A.~M. (2009).
\newblock An overview of pulmonary surfactant in the neonate: {Genetics},
  metabolism, and the role of surfactant in health and disease.
\newblock {\em Molec. Genet. Metabol.}, {\bf 97}(2):95--101.

\bibitem[Olmeda et~al., 2010]{olmeda10}
Olmeda, B., Vill{\'e}n, L., Cruz, A., Orellana, G., and Perez-Gil, J. (2010).
\newblock Pulmonary surfactant layers accelerate {O2} diffusion through the
  air-water interface.
\newblock {\em Biochim. Biophys. Acta Biomembr.}, {\bf 1798}(6):1281--1284.

\bibitem[Otis et~al., 1994]{otis94}
Otis, D.~R., Johnson, M., and Kamm, R.~D. (1994).
\newblock Dynamic surface tension of surfactant {TA}: experiments and theory.
\newblock {\em J. Appl. Physiol.}, {\bf 77}(6):2681--2688.

\bibitem[Possmayer et~al., 2001]{possmayer01}
Possmayer, F., Nag, K., Rodriguez, K., Qanbar, R., and Sch{\"u}rch, S. (2001).
\newblock Surface activity in vitro: role of surfactant proteins.
\newblock {\em Compar. Biochem. Physiol. A Molec. Integ. Physiol.}, {\bf
  129}(1):209--220.

\bibitem[Roohbakhshan, 2018]{roohbakhshan18thesis}
Roohbakhshan, F. (2018).
\newblock {\em {Membrane and Shell Formulations for Biological Materials such
  as Arteries and Lung Tissue}}.
\newblock PhD thesis, RWTH Aachen University.

\bibitem[Roohbakhshan et~al., 2016]{biomembrane}
Roohbakhshan, F., Duong, T.~X., and Sauer, R.~A. (2016).
\newblock A projection method to extract biological membrane models from {3D}
  material models.
\newblock {\em J. Mech. Behav. Biomed. Mater.}, {\bf 58}:90--104.

\bibitem[Rotenberg et~al., 1983]{rotenberg83}
Rotenberg, Y., Boruvka, L., and Neumann, A. (1983).
\newblock Determination of surface tension and contact angle from the shapes of
  axisymmetric fluid interfaces.
\newblock {\em J. Coll. Interf. Sci.}, {\bf 93}(1):169--183.

\bibitem[Rudiger et~al., 2005]{rudiger05}
Rudiger, M., Tolle, A., Meier, W., and Rustow, B. (2005).
\newblock Naturally derived commercial surfactants differ in composition of
  surfactant lipids and in surface viscosity.
\newblock {\em Amer. J. Physiol. Lung Cell Molec. Physiol.}, {\bf
  288}(2):L379--L383.

\bibitem[Saad, 2011]{saad11thesis}
Saad, S.~M. (2011).
\newblock {\em {Axisymmetric Drop Shape Analysis (ADSA) and Lung Surfactant}}.
\newblock PhD thesis, University of Toronto.

\bibitem[Saad et~al., 2010]{saad10}
Saad, S.~M., Neumann, A., and Acosta, E.~J. (2010).
\newblock {A dynamic compression--relaxation model for lung surfactants}.
\newblock {\em Coll. Surf. A Physicochem. Engrg. Asp.}, {\bf 354}(1-3):34--44.

\bibitem[Saad and Neumann, 2016]{saad16}
Saad, S.~M. and Neumann, A.~W. (2016).
\newblock Axisymmetric drop shape analysis {(ADSA): An} outline.
\newblock {\em Adv. Coll. Interf. Sci.}, {\bf 238}:62--87.

\bibitem[Saad et~al., 2012]{saad12}
Saad, S.~M., Policova, Z., Acosta, E.~J., and Neumann, A.~W. (2012).
\newblock Effect of surfactant concentration, compression ratio and compression
  rate on the surface activity and dynamic properties of a lung surfactant.
\newblock {\em Biochim. Biophys. Acta Biomembr.}, {\bf 1818}(1):103--116.

\bibitem[Saad et~al., 2011]{saad11design}
Saad, S.~M., Policova, Z., and Neumann, A.~W. (2011).
\newblock Design and accuracy of pendant drop methods for surface tension
  measurement.
\newblock {\em Coll. Surf. A Physicochem. Engrg. Asp.}, {\bf
  384}(1-3):442--452.

\bibitem[Sahu et~al., 2017]{sahu17}
Sahu, A., Sauer, R.~A., and Mandadapu, K.~K. (2017).
\newblock Irreversible thermodynamics of curved lipid membranes.
\newblock {\em Physic. Rev. E}, {\bf 96}(4):042409.

\bibitem[Sauer, 2014]{droplet}
Sauer, R.~A. (2014).
\newblock Stabilized finite element formulations for liquid membranes and their
  application to droplet contact.
\newblock {\em Int. J. Numer. Meth. Fluids}, {\bf 75}(7):519--545.

\bibitem[Sauer, 2016a]{memtheo}
Sauer, R.~A. (2016a).
\newblock A contact theory for surface tension driven systems.
\newblock {\em Math. Mech. Solids}, {\bf 21}(3):305--325.

\bibitem[Sauer, 2016b]{frictdroplet}
Sauer, R.~A. (2016b).
\newblock A frictional sliding algorithm for liquid droplets.
\newblock {\em Comput. Mech.}, {\bf 58}:937--956.

\bibitem[Sauer, 2018]{sauer18CISM}
Sauer, R.~A. (2018).
\newblock On the computational modeling of lipid bilayers using thin-shell
  theory.
\newblock In {\em The Role of Mechanics in the Study of Lipid Bilayers}, pages
  221--286. Springer International Publishing, Cham.

\bibitem[Sauer and Duong, 2017]{shelltheo}
Sauer, R.~A. and Duong, T.~X. (2017).
\newblock On the theoretical foundations of solid and liquid shells.
\newblock {\em Math. Mech. Solids}, {\bf 22}(3):343--371.

\bibitem[Sauer et~al., 2014]{membrane}
Sauer, R.~A., Duong, T.~X., and Corbett, C.~J. (2014).
\newblock A computational formulation for constrained solid and liquid
  membranes considering isogeometric finite elements.
\newblock {\em Comput. Methods Appl. Mech. Engrg.}, {\bf 271}:48--68.

\bibitem[Sauer et~al., 2017]{liquidshell}
Sauer, R.~A., Duong, T.~X., Mandadapu, K.~K., and Steigmann, D.~J. (2017).
\newblock A stabilized finite element formulation for liquid shells and its
  application to lipid bilayers.
\newblock {\em J. Comput. Phys.}, {\bf 330}:436--466.

\bibitem[Sauer et~al., 2018]{sauer18multi}
Sauer, R.~A., Ghaffari, R., and Gupta, A. (2018).
\newblock The multiplicative deformation split for shells with application to
  growth, chemical swelling, thermoelasticity, viscoelasticity and
  elastoplasticity.
\newblock available online, arXiv:1810.10384.

\bibitem[Sauer and Luginsland, 2018]{sauer18mono}
Sauer, R.~A. and Luginsland, T. (2018).
\newblock A monolithic fluid--structure interaction formulation for solid and
  liquid membranes including free-surface contact.
\newblock {\em Comput. Methods Appl. Mech. Engrg.}, {\bf 341}:1--31.

\bibitem[Sch{\" u}rch et~al., 1989]{schurch89}
Sch{\" u}rch, S., Bachofen, H., Goerke, J., and Possmayer, F. (1989).
\newblock A captive bubble method reproduces the in situ behavior of lung
  surfactant monolayers.
\newblock {\em J. Appl. Physiol.}, {\bf 67}(6):2389--2396.

\bibitem[Sch{\" u}rch et~al., 2001]{schurch01}
Sch{\" u}rch, S., Bachofen, H., and Possmayer, F. (2001).
\newblock Surface activity in situ, in vivo, and in the captive bubble
  surfactometer.
\newblock {\em Compar. Biochem. Physiol. A Molec. Integ. Physiol.}, {\bf
  129}(1):195 -- 207.

\bibitem[Scriven and Sternling, 1960]{scriven60}
Scriven, L. and Sternling, C. (1960).
\newblock {The Marangoni effects}.
\newblock {\em Nature}, {\bf 187}(4733):186--188.

\bibitem[Sharma et~al., 2017]{sharma17}
Sharma, R., Corcoran, T.~E., Garoff, S., Przybycien, T.~M., and Tilton, R.~D.
  (2017).
\newblock Transport of a partially wetted particle at the liquid/vapor
  interface under the influence of an externally imposed surfactant generated
  {Marangoni} stress.
\newblock {\em Coll. Surf. A Physicochem. Engrg. Asp.}, {\bf 521}:49--60.

\bibitem[Sosnowski et~al., 2017]{sosnowski17}
Sosnowski, T.~R., Kubski, P., and Wojciechowski, K. (2017).
\newblock New experimental model of pulmonary surfactant for biophysical
  studies.
\newblock {\em Coll. Surf. A Physicochem. Engrg. Asp.}, {\bf 519}:27--33.

\bibitem[Stone, 1990]{stone90}
Stone, H. (1990).
\newblock A simple derivation of the time--dependent convective--diffusion
  equation for surfactant transport along a deforming interface.
\newblock {\em Phys. Fluid. A Fluid. Dyn.}, {\bf 2}(1):111--112.

\bibitem[Todorov et~al., 2017]{todorov17}
Todorov, R., Exerowa, D., Alexandrova, L., Platikanov, D., Terziyski, I.,
  Nedyalkov, M., Pelizzi, N., and Salomone, F. (2017).
\newblock Behavior of thin liquid films from aqueous solutions of a pulmonary
  surfactant in presence of corticosteroids.
\newblock {\em Coll. Surf. A Physicochem. Engrg. Asp.}, {\bf 521}:105--111.

\bibitem[Velarde and Zeytounian, 2002]{velarde02}
Velarde, M.~G. and Zeytounian, R.~K. (2002).
\newblock {\em Interfacial phenomena and the Marangoni effect}.
\newblock Springer, Wien.

\bibitem[Veldhuizen and Haagsman, 2000]{veldhuizen00}
Veldhuizen, E.~J. and Haagsman, H.~P. (2000).
\newblock Role of pulmonary surfactant components in surface film formation and
  dynamics.
\newblock {\em Biochim. Biophys. Acta Biomembr.}, {\bf 1467}(2):255--270.

\bibitem[Wiechert, 2011]{wiechert11thesis}
Wiechert, L. (2011).
\newblock {\em {Computational Modeling of Multi-field and Multi-scale Phenomena
  in Respiratory Mechanics}}.
\newblock PhD thesis, Technical University of Munich.

\bibitem[Wiechert et~al., 2009]{wiechert09}
Wiechert, L., Metzke, R., and Wall, W. (2009).
\newblock {Modeling the Mechanical Behavior of Lung Tissue at the Microlevel}.
\newblock {\em J. Engrg. Mech.}, {\bf 135}(5):434--438.

\bibitem[Yang and James, 2007]{yang07}
Yang, X. and James, A.~J. (2007).
\newblock An arbitrary {Lagrangian--Eulerian (ALE)} method for interfacial
  flows with insoluble surfactants.
\newblock {\em FDMP}, {\bf 3}:65--96.

\end{thebibliography}
